\documentstyle[aps,epsfig]{revtex}

\begin{document}
\title{Medium effects in high energy heavy-ion collisions}
\author{C. M. Ko and G. Q. Li} 
\address{Cyclotron Institute and Physics Department\\
Texas A\&M University, College Station, Texas 77843, USA}
 
\maketitle
 
\begin{abstract} 
 
The change of hadron properties in dense matter based on various 
theoretical approaches are reviewed. Incorporating these
medium effects in the relativistic transport model, which
treats consistently the change of hadron masses and energies in dense 
matter via the scalar and vector fields, heavy-ion collisions at 
energies available from SIS/GSI, AGS/BNL, and SPS/CERN are studied. 
This model is seen to provide satisfactory explanations for the 
observed enhancement of kaon, antikaon, and antiproton yields as 
well as soft pions in the transverse direction from the SIS experiments.  
In the AGS heavy-ion experiments, it can account for the enhanced 
$K^+/\pi^+$ ratio, the difference in the slope parameters of the 
$K^+$ and $K^-$ transverse kinetic energy spectra, and the lower 
apparent temperature of antiprotons than that of protons. This model 
also provides possible explanations for the observed enhancement of 
low-mass dileptons, phi mesons, and antilambdas in heavy-ion collisions 
at SPS energies. Furthermore, the change of hadron properties in hot 
dense matter leads to new signatures of the quark-gluon plasma to 
hadronic matter transition in future ultrarelativistic heavy-ion 
collisions at RHIC/BNL.
 
\end{abstract}
 
\newpage
 
\section{introduction}
 
An important feature of quantum chromodynamics (QCD) is its approximate
$SU(3)_L\times SU(3)_R$ chiral symmetry as a result of small quark 
masses \cite{HOLS}. This symmetry is, however, spontaneously broken, 
leading to a finite quark condensate in vacuum, $\langle\bar qq\rangle_0
\approx\langle\bar uu\rangle_0\approx \langle\bar dd\rangle_0$, as given 
by the Gell$-$Mann-Oaks-Renner relation, 
\begin{equation}
m_\pi^2f_\pi^2=-2{m_q}\langle\bar qq\rangle_0.
\end{equation}
In the above, $m_\pi\approx 138$ MeV and $f_\pi\approx a_\mu93$ MeV are the pion 
mass and decay constant, respectively; and $m_q=(m_u+m_d)/2\approx 5.5$ MeV
is the average up and down quark masses. The quark condensate in vacuum 
thus has a value $\langle\bar qq\rangle_0\approx -(245\,{\rm MeV})^3$. 

As the density and/or temperature of a hadronic system increase, this 
spontaneously broken symmetry is expected to be partially restored, 
so the quark condensate would decrease with increasing density and/or 
temperature. At zero baryon chemical potential, the decrease of quark 
condensate with increasing temperature has been observed in lattice QCD 
simulations \cite{LQCD} as well as in calculations based on the chiral 
perturbation theory \cite{LEUT89} and the interacting pion gas model 
\cite{WAM95}.  At finite baryon density, model-independent studies 
using the Feynman-Hellmann theorem \cite{LEV90,COHEN92} have shown that
the ratio of the quark condensate in medium to its value in vacuum
is given by 
\begin{equation}
\frac{\langle\bar qq\rangle_\rho}{\langle\bar qq\rangle_0}
\approx 1-\frac{\Sigma_{\pi N}}{f_\pi^2m_\pi^2}\rho_N,
\end{equation}
where $\rho_N$ is the nuclear density and $\Sigma_{\pi N}\approx 45$ MeV 
is the ${\pi}N$ sigma term.  At normal nuclear matter density 
$\rho_0\approx 0.16$ fm$^{-3}$, the condensate is seen to decrease 
already by $\approx 1/3$.  Higher-order contributions to the quark 
condensate in nuclear matter due to nucleon-nucleon (NN) interactions 
have been studied using various models 
\cite{COHEN92,BIRSE93,ERIC93,LI94A,DEY95,BROCK96}.
It is found that at normal nuclear matter density they change the 
leading-order result by only about 5\% \cite{LI94A,BROCK96}. The 
decrease of quark condensate in medium may lead to reduced hadron masses 
as shown in QCD sum-rule studies \cite{COHEN91,HAT92}. The study of chiral 
symmetry breaking and partial restoration in hot dense matter is a topic 
of great current interest in nuclear physics \cite{WEIS93,BIRSE94,BR95A}.
 
In high energy heavy-ion collisions, theoretical simulations based on
transport models have shown that a hot dense hadronic matter
is formed in the initial stage of the collisions.  This provides thus 
the possibility of studying the partial restoration of chiral symmetry 
through the changes of hadron properties in medium
\cite{BROWN91A,KO95,KOLI95}.  Unfortunately, the dynamics of heavy-ion 
collisions is very complex, involving a violent initial compression, 
which is then followed by a relatively slow expansion and finally reaches 
the freeze out when particle interactions become unimportant.
The entire reaction typically lasts for about a few tens fm/c. The 
interesting physics of hadron in-medium properties and chiral symmetry 
restoration can only be studied for a few fm/c during the early part 
of the expansion stage when both the density and temperature of the 
hadronic matter are high. This stage of the collision can in principle 
be probed by detecting the emitted electromagnetic radiation such as the 
real and virtual (dilepton) photons. However, both are not easy to 
measure due to their small rates. Furthermore, they can also be produced 
from initial hard collisions and final hadron decays, and this makes
it difficult to extract the signals from the hot dense matter.  What are 
usually measured in heavy-ion experiments are instead the momentum 
distributions of hadrons, such as the nucleon, nuclear clusters, pion, 
kaon, etc., which are mostly ejected from the colliding system at freeze 
out.  To infer what have happened in the initial hot dense matter 
from the final hadron phase space distributions requires thus a model that 
can describe the whole collision process. Indeed, various 
transport models have been developed during the past ten years for this 
purpose. We shall review briefly the history of the development of 
transport models.  We note that a number of good review articles
are available on both relativistic and nonrelativistic transport models 
for heavy-ion collisions at various energies
\cite{STOCK86,BERT88,MAL90,MOS90,AICH91,MOS93,PEIL94,BONA95}.  

Since the reduced hadron masses in medium can be consistently included 
in the relativistic transport model \cite{KO87,ELZE87,BLAT88} 
we shall concentrate mostly on the results obtained from this model.
In the relativistic transport model, which is derived from the  
Walecka model \cite{QHD1,QHD2}, the effective hadron mass is 
connected to the scalar field while its energy is shifted by the vector 
potential.  Because of dropping hadron masses due to the attractive 
scalar field, particle production is enhanced as a result of reduced 
threshold and increased phase space.   Furthermore, the dense nuclear 
matter also provides a strong mean-field vector potential, which then
affects the momentum distributions of hadrons. The relation of the 
Walecka model to the chiral effective field theory has recently been 
explored in Refs. \cite{GEL95,BR96A,FURN}.  The attractive scalar and 
repulsive vector potentials for a nucleon in nuclear matter have also 
been obtained in the Dirac-Brueckner-Hartree-Fock (DBHF) approach
\cite{MACH87,MACH89,ANA83,HM87,MACH90,LI92A} as well as the QCD sum-rule 
studies \cite{COHEN91,FURN92,JIN93,JIN94A}.
 
For heavy-ion collisions at energies available from the SIS at GSI, 
the relativistic transport model gives a satisfactory explanation for 
the observed enhancement of kaon \cite{FANG94,LI95A,LI95B,KAOSE}, 
antikaon \cite{LI94B,KIEN}, and antiproton yields \cite{LI94C,antip}
as well as low energy pions in the transverse direction 
\cite{XIONG93,SPE,TAPS}. Experiments on kaon flow have also provided 
important information on the kaon potential in dense matter 
\cite{LI95C,FOPI}.  Furthermore, the study of dilepton production from 
these collisions is expected to allow us to probe directly the in-medium 
properties of vector mesons in dense matter \cite{LI95D,HADES}.
 
The relativistic transport model has also been used to describe the
expansion stage of heavy-ion collisions at energies available from the 
AGS at BNL. It can account for the enhanced $K^+/\pi^+$ ratio 
\cite{KO91}, the difference between the slope parameters of the 
$K^+$ and $K^-$ transverse mass spectra \cite{FANG93A}, and the 
lower apparent temperature of antiprotons than that of protons
\cite{KOCH91} as observed in experiments \cite{AGS1,AGS2}.
 
For the expansion stage of heavy-ion collisions at energies available 
from the SPS at CERN, it has been found that the relativistic transport 
model can explain quantitatively the observed enhancement of low-mass 
dileptons if one includes the decrease of vector meson masses in hot 
dense matter \cite{LI95E,LI95F,CERES,HELIOS,SUMDI}.  In a simplified 
hydrochemical model, medium effects also lead to enhanced production 
of phi mesons \cite{sa91,guil91} and antilambdas \cite{leva92,bart90} 
in these collisions.
 
The success of theoretical studies including the medium effects
in explaining a large body of experimental data from high energy 
heavy-ion collisions thus provides the possible evidence for the 
partial restoration of chiral symmetry in hot dense matter
formed in high energy heavy-ion collisions. The purpose of this article
is to review the theoretical predictions for hadron properties in dense
matter, the consistent incorporation of these medium effects 
in the relativistic transport model, and the observable consequences in 
heavy-ion collisions from SIS to SPS energies.
 
In Section II, hadron properties in dense matter are discussed
using various theoretical approaches, such as the Walecka model, the 
DBHF approach,  the QCD sum-rule approach, and the effective hadronic 
model. The relativistic transport model based on the nonlinear
$\sigma$-$\omega$ model is then described in Section III, together 
with a brief review of the development of transport models.
In Section IV, the results obtained from the relativistic transport model 
for particle production and  flow in heavy-ion collisions at SIS energies, 
the strangeness enhancement in heavy-ion collisions at AGS and SPS 
energies, and dilepton production in heavy-ion collisions are then 
presented and compared with the experimental data.  The relevance of 
these experimental observables as the probes of hadron properties in 
dense matter is emphasized.  A summary and outlook is then given in 
Section V.
 
\section{hadrons in dense matter}
 
Theoretically, lattice QCD simulations should provide us with
the most reliable information about the temperature and/or density 
dependence of hadron properties.  Currently, these simulations can be 
carried out only for finite temperature systems with zero baryon chemical 
potential. Furthermore, the quenched approximation is usually introduced 
for quarks, so their dynamical role is only crudely included.  From the 
hadronic correlation functions at large spatial separations the hadron 
screening masses have been obtained in the lattice QCD 
\cite{detar87,hash93,boyd95}. In a recent paper, Boyd {\it et al}
\cite{boyd95} have shown that, up to about 0.92$T_c$, where $T_c$ is
the critical temperature for the quark-gluon to hadronic matter phase 
transition, there is no significant change of the rho-meson screening mass.  
They have also found that the quark condensate does not change very much 
until one is extremely close to the critical temperature. This is in 
contrast with the result from chiral perturbation theory which shows a more 
appreciable decrease of the quark condensate with temperature \cite{LEUT89}.   
Therefore, the study of hadron properties in dense nuclear matter at 
present time has to rely on theoretical models.  In this section, we 
shall review some of these approaches.  In particular, the properties 
of baryons (mainly nucleon), pseudoscalar mesons (mainly pion 
and kaon), and vector mesons (mainly rho, omega, and phi mesons) 
will be discussed.
 
\subsection{baryons}
 
A simple model for nuclear matter at high density is the Walecka model,
also known as the quantum hadrodynamics (QHD-I) or linear 
$\sigma$-$\omega$ model, in which nucleons interact with each other 
through the exchange of a scalar sigma and a vector omega meson 
\cite{QHD1,QHD2}. The sigma meson gives rise to an intermediate-range 
attraction while the omega exchange leads to a short-range repulsion, 
both are the essential properties of nuclear force. In this 
phenomenological model, the nucleon mass is reduced by the attractive 
scalar interaction, while its energy is shifted by the repulsive vector 
interaction. The parameters in this model are adjusted to fit the nuclear 
matter properties at saturation density using the mean-field approximation.
 
This model gives, however, too small a nucleon effective mass and too 
large a compression modulus at saturation density compared to what 
are known empirically. To correct for these deficiencies, self interactions 
of cubic and quartic forms have been introduced for the scalar field 
\cite{NL1,NL2}.  The Lagrangian for this nonlinear
$\sigma$-$\omega$ model is given by
\begin{eqnarray}\label{lag}
{\cal L}& =&\bar N[\gamma_\mu (i\partial ^\mu-g_\omega\omega^\mu)
-(m_N-g_\sigma N)]N +{1\over 2}(\partial_\mu\sigma\phi\partial^\mu\sigma-
m_\sigma^2\sigma^2)
-{1\over 3}b\sigma^3-{1\over 4}c\sigma^4\nonumber\\
&-&{1\over 4} (\partial _\mu\omega_\nu-\partial_\nu\omega_\mu )
(\partial ^\mu\omega^\nu -\partial ^\nu\omega^\mu)+{1\over 2}m_\omega^2
\omega_\mu\omega^\mu ,
\end{eqnarray}
where $N$ is the nucleon field with mass $m_N$, while $\sigma$ and
$\omega_\mu$ are the fields for the scalar and vector mesons with mass 
$m_\sigma$ and $m_\omega$, respectively. The parameters $b$ and $c$ 
determine the strength of scalar field self-interaction. With $b=c=0$ 
the nonlinear $\sigma$-$\omega$ model reduces to the original Walecka 
model. The coupling constants between a nucleon and the scalar and 
vector fields are denoted by $g_\sigma$ and  $g_\omega$, respectively.  
In a static nuclear matter, meson fields in the mean-field approximation 
become constants and are independent of the spatial coordinates. 
Furthermore, the vector field has a nonvanishing value only in its time-like 
component due to translational invariance in infinite nuclear matter. The 
equation of motion for a nucleon with momentum {\bf p} in a nuclear 
matter of density $\rho_N$ is then given by the Dirac equation
\begin{eqnarray}\label{dirac}
({\bbox \alpha}\cdot {\bf p}+\beta m_N^*)N =\epsilon ^* ({\bf p})N,
\end{eqnarray}
where the nucleon effective mass is related to the scalar potential
$\Sigma_S=-g_\sigma\langle\sigma\rangle$, i.e., 
\begin{eqnarray}\label{effms}
m_N^*=m_N+\Sigma _S,
\end{eqnarray}
and its energy is shifted by the vector potential 
$\Sigma_V^0=\frac{g_\omega^2}{m_\omega^2}\rho_N$, i.e., 
\begin{eqnarray}
\epsilon^*= (m_N^{*2}+{\bf p}^2)^{1/2}+\Sigma _V^0.
\end{eqnarray}
 
While the vector potential is simply proportional to the nuclear density
$\rho_N $, the scalar potential depends on the scalar density $\rho_S$ 
via the following self-consistent equation, obtained by minimizing the 
energy,
\begin{equation}\label{nlsom}
m_\sigma ^2 \langle\sigma\rangle +b \langle\sigma \rangle^2 +c\langle\sigma 
\rangle^3=g_{\sigma} \rho_S.
\end{equation}
 
\vskip 0.5cm
{\bf Table I} Parameters of the nonlinear $\sigma$-$\omega$ model
corresponding to the soft and stiff equations of state and the nuclear
matter properties at the saturation density $\rho _0=0.16 ~$fm$^{-3}$.
\vskip 0.5cm
\begin{center}
\begin{tabular}{ccccccccc}
\hline\hline
 & $C_\sigma=\frac{g_\sigma m_N}{m_\sigma}$& 
$C_\omega=\frac{g_\omega m_N}{m_\omega}$& 
$B=\frac{b}{g_\sigma^3m_N}$& $C=\frac{c}{g_\sigma^4}$ & 
$\frac{\cal E}{A} ~({\rm MeV})$& 
$\frac{m_N^*}{m_N}$ & $K ~({\rm MeV})$\\
\hline
soft  & 13.95 & 8.498 & 0.0199 & -0.00296  & -15.96 & 0.83 & 200.0 \\
stiff & 15.94 & 12.92 & $8.0\times 10^{-4}$  
&$2.26\times 10^{-3}$ & -15.96 & 0.68 & 380.0\\
\hline
\end{tabular}
\end{center}
\vskip 0.5cm 

The four parameters in the nonlinear $\sigma$-$\omega$ model are
determined by the nuclear matter saturation density ($\rho _0$=0.16 
fm$^{-3}$), binding energy (B=15.96 MeV), compression modulus (K),
and the nucleon effective mass ($m^*$).  In this review, only two sets 
of parameters, corresponding to the soft and stiff equations of state 
(EOS), are considered.  At $\rho_0$, the soft EOS 
corresponds to $m_N^*/m_N=0.83$ and K=200 MeV, while the stiff EOS 
has $m_N^*/m_N=0.68$ and K=380 MeV. The corresponding parameters are 
listed in Table 1.  The two nuclear equations of state, together with 
the nucleon effective mass, are shown in Fig. 1 by solid curves. 
Both the soft and the stiff equation of state in this relativistic 
formulation are very close to that from the Skyrme interaction 
\cite{AICH91} shown in the right panel of Fig. 1 by dotted curves.
 
It is worthy to mention that in the literature the expression
`effective mass' has been used to denote different quantities.
A nice clarification of the relationship between these various
quantities can be found in Ref. \cite{MAH89}. In nonrelativistic
models, the nucleon effective mass is related to the nonlocality,
or momentum dependence, of the single-particle potential. This quantity 
should be identified with the Lorentz mass \cite{MAH89} defined in terms of 
the Schr\"odinger-equivalent potential in relativistic models, rather
than the Dirac mass defined in terms of the scalar potential 
as in Eq. (\ref{effms}).
 
Extensions of the Walecka-type model from zero to finite temperature
have been studied in Refs. \cite{SAITO89,SAITO90,FURN90}. It is found 
that at finite density the nucleon effective mass first increases 
slightly at low temperature and then decreases rapidly at high 
temperature.  A similar temperature dependence of the constituent 
quark mass has been found in studies based on the Nambu$-$Jona-Lasinio 
model \cite{KLIM90}.
 
\begin{figure}
\epsfig{file=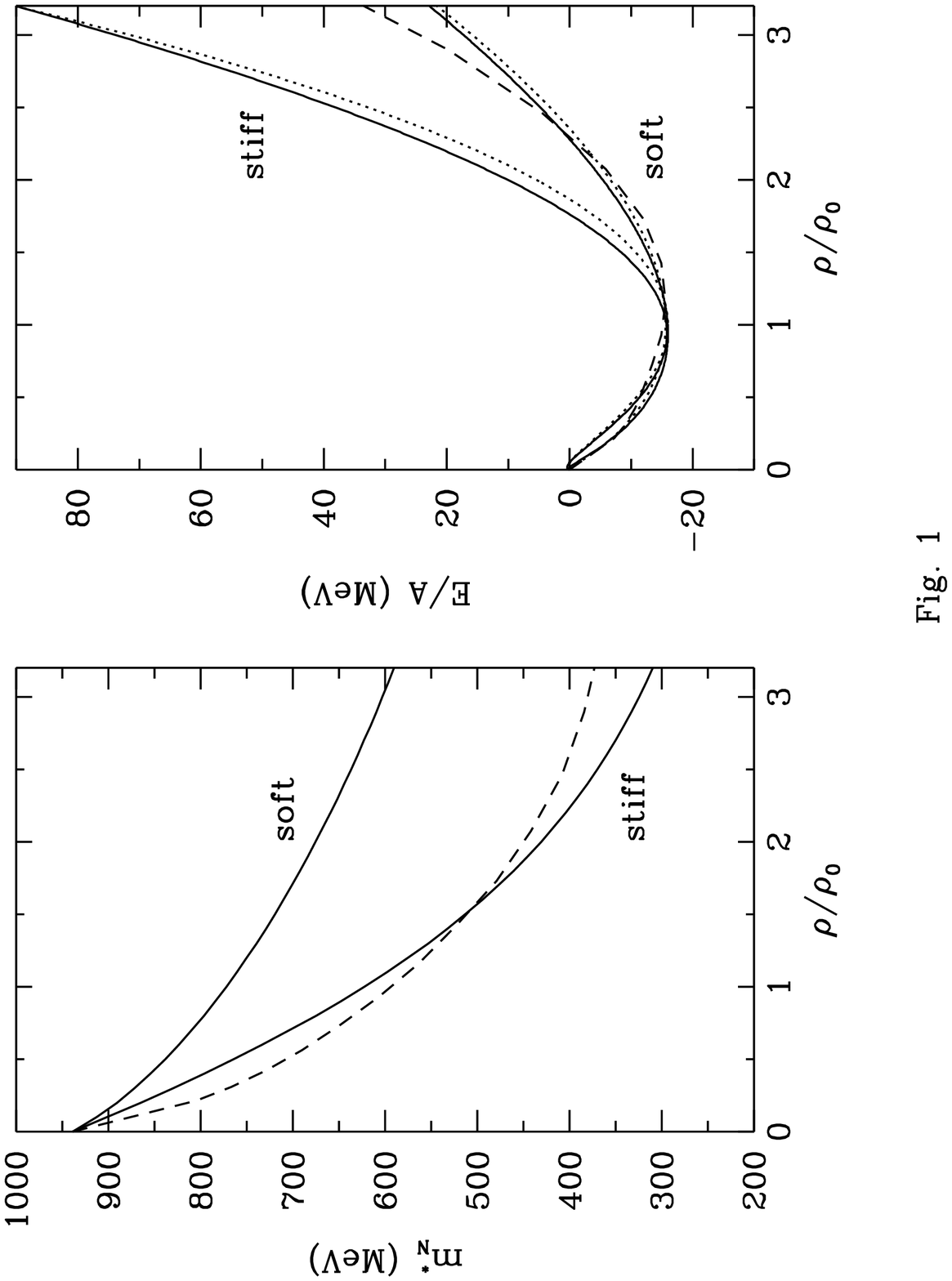,height=3in,width=5.5in}
\vskip 0.4cm
Fig. 1 ~Left panel: The nucleon effective mass in medium.
Solid and dashed curves are from the nonlinear $\sigma$-$\omega$
model and DBHF theory, respectively. Right panel: Nuclear equation
of state.  Dotted curves are from the Skyrme interaction.
\end{figure}

A more microscopic approach to nucleon properties in nuclear matter 
is provided by the DBHF calculation \cite{MACH89,ANA83,HM87,MACH90,LI92A}
using, e.g., the Bonn one-boson-exchange potential \cite{MACH87,MACH89}.
As in conventional Brueckner approach, the basic quantity in DBHF  
approach is the $G$-matrix which satisfies the in-medium Thompson 
equation \cite{MACH89,MACH90,LI92A},
\begin{eqnarray}\label{bethe}
 G({\bf q',q; P}) =&& V({\bf q',q})\nonumber\\
&+&\int {d^3k\over (2\pi )^3} V({\bf q',k})\frac{m_N^{*2}}
{E^{*2}_{{\bf P/2}+{\bf k}}}\frac{Q({\bf k,P})}
{2 E^*_{{\bf P}/2+{\bf q}}-2E^*_{{\bf P}/2+{\bf k}}}G({\bf k,q;P}),
\end{eqnarray}
with $E^*_{\bf k}= (m_N^{*2}+{\bf k}^2)^{1/2}.$  The momenta ${\bf P}$ 
and ${\bf q}$ denote, respectively, the center-of-mass and the relative 
momentum of two interacting nucleons.  The kernel of this integral 
equation is given by the sum of the one-meson-exchange amplitudes from 
six nonstrange  bosons ($\pi, \sigma, \rho, \omega, \eta, \delta$). 
A form factor of monopole type is introduced at each meson-nucleon 
vertex to take into account the short-range dynamics due to quarks 
and gluons.  The $G$-matrix becomes density-dependent through the 
Pauli projection operator $Q$ and the scalar field $\Sigma _S$, which 
appears in $m_N^*$ as shown by Eq. (\ref{effms}).
 
The Dirac equation and the in-medium Thompson equation are then solved 
self-consistently to determine the $G$-matrix, from which the nuclear 
matter properties can be derived.  For example, the nuclear equation 
of state, or the energy-per-nucleon, is given by
\begin{eqnarray}
{{\cal E}\over A}={1\over A}\sum _{i\le k_F} {m_Nm^*_N+{\bf p}_i^2\over E^*_i}
+{1\over 2A}\sum _{i,j\le k_F}{m_N^{*2}\over E^*_iE^*_j}\langle
ij|G|ij-ji\rangle -m_N,
\end{eqnarray}
where  the nucleon bare mass has been subtracted out.
 
The $G$-matrix also allows one to determine the nucleon scalar and vector 
potentials in nuclear matter, i.e., 
\begin{eqnarray}
{m^*_N\over E^*_i}\Sigma _S+\Sigma _V^0 ={\rm Re }\sum _{j\le k_F}
{{m_N^*}^2\over E^*_iE^*_j}\langle ij|G|ij-ji\rangle .
\end{eqnarray}
The nuclear equation of state and the nucleon effective mass obtained from 
the DBHF calculation with the Bonn A potential are shown in Figs. 1 by 
dashed curves.
 
Nucleon properties in dense matter have also been studied in approaches 
based on the QCD. In the QCD sum-rule approach 
\cite{LEV90,COHEN91,FURN92,JIN93,JIN94A,HAT90,ADAMI91}, 
hadron properties are related to a number of condensates, which are 
the vacuum expectation values of quark and gluon fields and describe 
the nonperturbative aspects of the QCD vacuum.  The QCD sum rules have 
been quite successful in describing the hadron properties in free space, 
e.g, the nucleon mass is quantitatively reproduced in this approach 
\cite{SHIF79,IOFF81,REIN85,SHIF92}.  The extension of QCD sum rules to 
hadrons in nuclear matter is achieved by using in-medium condensates.
For a nucleon, it has been shown that the change of scalar quark 
condensate in medium leads to an attractive scalar potential which
reduces its mass, while the change of vector quark condensate
leads to a repulsive vector potential which shifts its energy 
\cite{COHEN91,FURN92,JIN93,JIN94A}. The nucleon scalar and vector 
self-energies in this study are given, respectively, by
\begin{eqnarray}\label{cohen}
\Sigma_S&\approx& -\frac{8\pi^2}{M_B^2}(\langle\bar qq\rangle_\rho-
\langle\bar qq\rangle_0)
\approx -\frac{8\pi^2}{M_B^2}\frac{\Sigma_{\pi N}}{m_u+m_d}
\rho_N,\nonumber\\
\Sigma_V&\approx& \frac{64\pi^2}{3M_B^2}\langle q^+q\rangle_\rho=
\frac{32\pi^2}{M_B^2}\rho_N,
\end{eqnarray}
where the Borel mass $M_B$ is an arbitrary parameter. With $M_B\approx m_N$, 
and $m_u+m_d\approx 11$ MeV, these self-energies have values which are 
similar to those determined from both the Walecka model and the DBHF 
approach.
 
Experimental evidences for these strong scalar and vector potentials
have been inferred from the proton-nucleus scattering at intermediate energies 
\cite{RAY92} via the Dirac phenomenology \cite{CLARK,COOP85} in which 
the Dirac equation with scalar and vector potentials is solved.
The proton-nucleus scattering has also been successfully described using
realistic NN interactions in the impulsive approximation \cite{WAL,HORO87}
or the local density approximation \cite{LI93A}.
 
In heavy-ion collisions at energies above the pion production threshold,
baryon resonances can be produced. Transport model simulations have shown 
that in central heavy-ion collisions at incident energies around 1 
GeV/nucleon, the baryon resonances to nucleon ratio can be as large as 0.2 
\cite{FANG94}. This ratio approaches 1 when the incident energy is 
increased to about 10 GeV/nucleon at AGS \cite{SORGE95A,BALI}. The 
mean-field potentials for resonances are thus needed in transport models, 
and up to now have been assumed to be the same as those for a nucleon.
This assumption for a delta particle seems to be supported by analyses of 
the pion-nucleus scattering data \cite{CHEN93}. However, recent QCD 
sum-rule studies \cite{JIN95A} have shown that the delta vector potential 
is considerably weaker than that for a nucleon, while its scalar 
potential is somewhat stronger.  This would lead to a larger net 
attraction for a delta resonance than for a nucleon in nuclear medium.
 
Also of interest is the potentials for hyperons in nuclear matter.
As for nucleon, most of the empirical information about hyperons
in nuclear matter have been extracted from the structure of hypernuclei.
The potential for the lambda  particle is relatively well determined
to be about -30 MeV \cite{DOV88}. The separation of this 
nonrelativistic potential into the scalar and vector potentials 
in the relativistic approach is, however, not without ambiguities
\cite{SCH92}.
 
The lambda potential in nuclear matter has also been studied using
the DBHF approach \cite{SPETH93}. The result ranges from -25 to -40
MeV, depending on the input boson-exchange models for the $\Lambda N$
interaction. There have also been various attempts to generalize the
Walecka-type model from SU(2) to SU(3) to include the hyperon degrees
of freedom \cite{BROCK77,BOUY77,RUFA90,GLEN91}. In the naive SU(3) 
quark model, the hyperon potential is about 2/3 of the nucleon 
potential, as there are only two light quarks in a hyperon
instead of three light quarks in a nucleon.  Recently, hyperon properties 
in nuclear matter have also been studied using the QCD sum-rule approach. 
It has been found that both lambda scalar and vector potentials are 
significantly weaker than the prediction from the naive quark model 
\cite{JIN95B}, while those of the sigma hyperon are somewhat stronger
and are close to the ones for the nucleon \cite{JIN94B}. The accuracy of
these findings are, however, limited by uncertainties in the nucleon
strangeness content and certain in-medium four-quark condensates.
 
\subsection{pseudoscalar mesons}
 
Pseudoscalar mesons play a special role in nuclear physics. On the one 
hand, they are the  Goldstone bosons of spontaneously broken chiral
symmetry and are thus closely related to chiral symmetry restoration
in nuclear medium. Furthermore, their small masses make the chiral 
perturbation theory applicable for low energy phenomena 
\cite{MEIS93,RHO94A,RHO94B,RHO95B}.  On the other hand, in heavy-ion 
collisions, pions are the most copiously produced particles; e.g., 
at SPS energies, the pion/nucleon ratio approaches 10 in midrapidity
region. Knowledge on the properties of pseudoscalar mesons in medium 
is thus very important for understanding both chiral symmetry restoration 
and the dynamics of high energy heavy-ion collisions.
 
\subsubsection{pion}
 
Pion properties in dense matter have been a subject of extensive studies 
over many years \cite{WEISE}.  The strong p-wave pion-nucleon interaction 
through the delta resonance is known to modify appreciably the pion 
dispersion relation at finite density \cite{WEISE,BROWN75,PAND81}.
In the delta-hole model, the pion dispersion relation in nuclear medium
can be written as 
\begin{eqnarray}
\omega ({\bf k}, \rho_B) =m_\pi^2+{\bf k}^2+\Pi (\omega, {\bf k}),
\end{eqnarray}
where the pion self-energy is given by
\begin{eqnarray}
\Pi (\omega, {\bf k})={{\bf k}^2 \chi(\omega,{\bf k})\over 1-g^\prime\chi
(\omega,{\bf k})},
\end{eqnarray}
with $g^\prime\approx 0.6$ being the Migdal parameter due to 
short-range correlations. The pion susceptibility $\chi$ is given by
\begin{eqnarray}
\chi(\omega , {\bf k}) \approx {8\over 9}\Big({f_{\pi N\Delta} \over m_\pi
}\Big)^2 {\omega_R \over \omega ^2-\omega^2_R}
{\rm exp}\Big(-2{\bf k}^2/b^2\Big) \rho _N,
\end{eqnarray}
where $f_{\pi N\Delta}\approx 2$ is the pion-nucleon-delta coupling 
constant, $b\approx 7m_\pi$ is the range of the form factor, and
$\omega_R\approx {{\bf k}^2\over 2m_\Delta } +m_\Delta -m_N$.
 
The pion dispersion relation obtained from this model is shown in 
Fig. 2. The pion branch in the lower part of the figure is seen to 
become softened, while the delta-hole branch in the upper part of 
the figure is stiffened.  These results are, however, modified by the 
imaginary part of the pion self-energy due to the delta decay width 
and pion absorption \cite{kxs89,BROWN89,xss94,henn,km95}. As a result, 
the strength of the delta-hole branch is significantly reduced, and 
the peak in the spectral function is less shifted in energy 
than the pion branch obtained without the imaginary pion self-energy.
 
Pion properties in dense matter have also been studied using the chiral 
perturbation theory \cite{BROWN91B,THOR95}.  In effective chiral 
Lagrangians a pion has a negligible vector interaction with a nucleon. 
The s-wave $\pi N$ scalar interaction, proportional to the $\pi N$ 
sigma term $\Sigma _{\pi N}$, would lead to a reduction of the pion 
mass \cite{BROWN91B}. This effect is, however, almost completely canceled 
by the range term from the energy dependence of the $\pi N$ 
scattering amplitude, so the net mean-field effects are quite small
\cite{ERICSON}, and the pion mass is found to increase only slightly
with increasing density. This is consistent with the results determined
from the pionic atom data \cite{WEISE}.
 
The Nambu$-$Jona-Lasinio model has also been used to study pion
(or in general Goldstone boson) properties at both finite density and 
temperature \cite{MEIS88,LUTZ92}. In this model, pseudoscalar mesons 
are treated as quark-antiquark collective modes. Because of the finite 
scalar mean squared radius of a constituent quark, the s-wave $\pi N$ 
attraction is found to be screened, leading to a slow increase of 
the pion mass with increasing density and/or temperature \cite{LUTZ92}. 
On the other hand, the pion mass is found to decrease slightly with 
temperature in an effective Lagrangian that includes its interactions 
in a pion gas via the rho and $a_1$ mesons \cite{SONG94A}. 

\begin{figure}
\epsfig{file=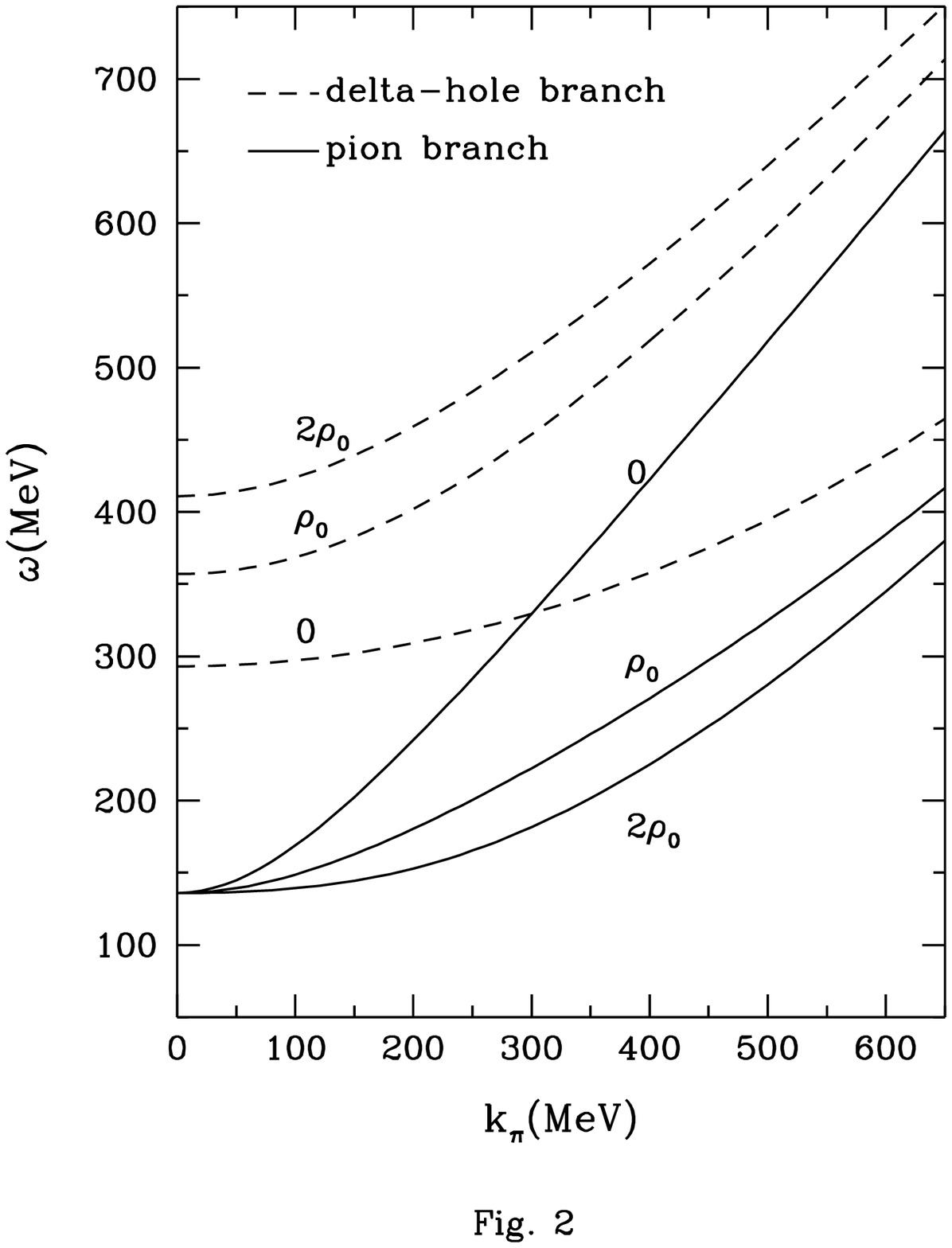,height=4in,width=4in}
\vskip 0.4cm
Fig. 2 ~The pion dispersion relation from the delta-hole model.
(From Ref. \cite{PAND81})
\end{figure} 

\subsubsection{kaon}
 
The study of kaon properties in dense matter has recently attracted
a lot of attention as it is relevant for understanding not only
the strangeness enhancement in heavy-ion collisions, which has been 
suggested as a possible signature of quark-gluon plasma formation
in these reactions \cite{RAF}, but also the kaon condensation in 
neutron stars \cite{BROWN88A} and the possible formation of mini-black 
holes in galaxies \cite{BROWN94}.
 
The properties of kaon in dense matter were first studied in chiral
Lagrangian by Kaplan and Nelson \cite{KL86}, and further pursued by 
many others
\cite{RHO94A,RHO94B,RHO95B,BRO87,PW91,LUTZ94,LYNN90,MU92,BRO92,YA93,SCH94}. 
The SU(3)$_L\times$SU(3)$_R$ nonlinear chiral Lagrangian used
by Kaplan and Nelson is written as \cite{KL86,LYNN90,BRO92}
\begin{eqnarray}
{\cal L}&=&{1\over 4}f^2{\rm Tr}\partial^\mu\Sigma\partial_\mu\Sigma^+
+{1\over 2}f^2\Lambda[{\rm Tr}M_q(\Sigma-1)+{\rm h.c.}]
+{\rm Tr}{\bar B}(i\gamma^\mu\partial_\mu-m_B)B\nonumber\\
&+&i{\rm Tr}{\bar B}\gamma^\mu[V_\mu, B]
+D{\rm Tr}{\bar B}\gamma^\mu\gamma^5\{A_\mu, B\}
+F{\rm Tr}{\bar B}\gamma^\mu\gamma^5[A_\mu, B]\nonumber\\
&+&a_1{\rm Tr}{\bar B}(\xi M_q\xi+{\rm h.c.})B
+a_2{\rm Tr}{\bar B}B(\xi M_q\xi+{\rm h.c.})\nonumber\\
&+&a_3[{\rm Tr}M_q\Sigma+{\rm h.c.}]{\rm Tr}{\bar B}B.\label{LAG}
\end{eqnarray}
In the above, $B$ is the baryon octet with a degenerate mass $m_B$, and
\begin{equation}
\Sigma=\exp (2i\pi/f),~~{\rm and}~~
\xi=\sqrt\Sigma=\exp(i\pi/f),
\end{equation}
with $\pi$ being the pseudoscalar meson octet.  The pseudoscalar meson 
decay constants are equal in the SU(3)$_V$ limit and are denoted by 
$f=f_\pi \simeq 93$ MeV.  The meson vector $V_\mu$ and axial vector 
$A_\mu$ currents are defined as
\begin{equation}
V_\mu={1\over 2}(\xi^+\partial_\mu\xi+\xi\partial_\mu\xi^+)~~{\rm and}~~
A_\mu={i\over 2}(\xi^+\partial_\mu\xi-\xi\partial_\mu\xi^+),
\end{equation}
respectively. The current quark mass matrix is given by 
$M_q={\rm diag}\{m_q,m_q,m_s\}$, if we neglect the small difference 
between the up and down quark masses.
 
Terms involving the axial vector current can be ignored as they have 
no effects on the kaon mass. Expanding $\Sigma$ to order of $1/f^2$ and
keeping explicitly only the kaon field, the first two terms in Eq. 
(\ref{LAG}) can be written as
\begin{equation}
\partial^\mu\bar K\partial_\mu K-\Lambda(m_q+m_s)\bar KK+\cdots,
\end{equation}
where
\begin{equation}
K=\left(\matrix{K^+ \cr
                K^0 \cr}\right)~~
{\rm and} ~~\bar K=(K^- ~~\bar {K^0}),
\end{equation}
and the ellipsis denotes terms containing other mesons.
 
Keeping explicitly only the nucleon and kaon, the third and fourth terms in 
Eq. (\ref{LAG}) become
\begin{equation}
{\bar N}(i\gamma^\mu\partial_\mu-m_B)N
-{3i\over 8f^2}{\bar N}\gamma^0 N 
\bar K \buildrel \leftrightarrow\over \partial_t K+\cdots,
\end{equation}
where
\begin{equation}
N=\left(\matrix{p \cr
                n \cr}\right)~~{\rm and}~~\bar N=(\bar p~~ \bar n),
\end{equation}
and the ellipsis denotes terms involving other baryons and mesons.
 
The last three terms in Eq. (\ref{LAG}) can be similarly worked out, and
the results are
\begin{eqnarray}
{\rm Tr}{\bar B}(\xi M_q\xi+{\rm h.c.})B&=&2m_q{\bar N}N-{{\bar N}N\over 2f^2}
(m_q+m_s){\bar K}K+\cdots,\nonumber\\
{\rm Tr}{\bar B}B(\xi M_q\xi+{\rm h.c.})&=&2m_s{\bar N}N-{{\bar N}N\over f^2}
(m_q+m_s){\bar K}K+\cdots,\nonumber\\
\,[{\rm Tr}M_q\Sigma+{\rm h.c.}]{\rm Tr}{\bar B}B&=&2(2m_q+m_s){\bar N}N
-{2{\bar N}N\over f^2}(m_q+m_s){\bar K}K+\cdots.
\end{eqnarray}
Combining above expressions, one arrives at the following Lagrangian,
\begin{eqnarray}
{\cal L}&=&{\bar N}(i\gamma^\mu\partial_\mu-m_B)N
+\partial^\mu{\bar K}\partial_\mu K-\Lambda(m_q+m_s){\bar K}K\nonumber\\
&-&{3i\over 8f^2}{\bar N}\gamma^0 N
\bar K \buildrel \leftrightarrow\over \partial_t K
+[2m_qa_1+2m_sa_2+2(2m_q+m_s)a_3]{\bar N}N\nonumber\\
&-&{{\bar N}N{\bar K}K\over 2f^2}(m_q+m_s)(a_1+2a_2+4a_3)
+\cdots.\label{LAG1}
\end{eqnarray}
 
From this equation, one can identify the kaon mass by
\begin{equation}
m_K^2=\Lambda(m_q+m_s)
\end{equation}
and the nucleon mass by
\begin{eqnarray}\label{mn}
m_N=m_B-2[a_1m_q+a_2m_s+a_3(2m_q+m_s)].
\end{eqnarray}
Also, the $KN$ sigma term can be expressed as
\begin{eqnarray}
\Sigma_{KN}&\equiv&{1\over 2}(m_q+m_s)\langle N|{\bar u}u+{\bar s}s|
N\rangle\nonumber\\
&=&{1\over 2}(m_q+m_s)\Big[{1\over 2}\frac{\partial m_N}{\partial m_q}+
\frac{\partial m_N}{\partial m_s}\Big]\nonumber\\
&=&-{1\over 2}(m_q+m_s)(a_1+2a_2+4a_3).
\end{eqnarray}
The second line in the above follows from explicit chiral symmetry
breaking in the QCD Lagrangian, and the last step is obtained using 
Eq. (\ref{mn}).   Then Eq. (\ref{LAG1}) can be rewritten as
\begin{eqnarray}
{\cal L}&=&{\bar N}(i\gamma^\mu\partial_\mu-m_N)N
+\partial^\mu{\bar K}\partial_\mu K
-(m_K^2-{\Sigma_{KN}\over f^2}{\bar N}N){\bar K}K\nonumber\\
&-&{3i\over 8f^2}{\bar N}\gamma^0 N
\bar K \buildrel \leftrightarrow\over \partial_t K
+\cdots.\label{LAG2}
\end{eqnarray}
 
The above Lagrangian does not describe properly the nuclear matter 
properties. In Ref. \cite{LYNN90}, an attempt has been made to combine 
this SU(3)$_L\times$SU(3)$_R$ chiral Lagrangian with the Walecka model
\cite{QHD1,QHD2} by allowing the nucleon to also couple with the 
phenomenological scalar ($\sigma$) and vector ($\omega$) fields so 
that the saturation properties of nuclear matter can be obtained in 
mean-field level. Possible justifications of this approach have been 
further studied in Ref. \cite{LYNN93} based on the chiral perturbation 
theory. Following this approach, one modifies Eq. (\ref{LAG2}) to
\begin{eqnarray}
{\cal L}&=&{\bar N}(i\gamma^\mu\partial_\mu-m_N+g_\sigma\sigma)N
-g_\omega\bar N\gamma^\mu N\omega_\mu+{\cal L}_0(\sigma,\omega_\mu)\nonumber\\
&+&\partial^\mu{\bar K}\partial_\mu K
-(m_K^2-{\Sigma_{KN}\over f^2}{\bar N}N){\bar K}K
-{3i\over 8f^2}{\bar N}\gamma^0 N
\bar K \buildrel \leftrightarrow\over \partial_t K+\cdots.\label{LAG3}
\end{eqnarray}
In the above, ${\cal L}_0$ contains the kinetic and mass terms of the
$\sigma$ and $\omega$ fields as well as possible cubic and quartic
$\sigma$ self-interaction terms.
 
From the Euler-Lagrange equation and using the mean-field approximation 
for the nucleon field, i.e., replacing $\langle{\bar N}\gamma ^0 N\rangle$ 
by the nuclear density $\rho_N$ and $\langle{\bar N}N\rangle$ by the scalar 
density $\rho_S$, one obtains the following Klein-Gordon equation for 
a kaon in nuclear medium,
\begin{eqnarray}
\Big[\partial _\mu\partial ^\mu+{3i\over 4f^2} \rho_N \partial _t
+\big(m_K^2-{\Sigma _{KN}\over f^2}\rho _S\big)\Big]K=0.\label{KG}
\end{eqnarray}
 
The kaon dispersion relation in nuclear matter is then given by
\begin{eqnarray}
\omega^2({\bf k},\rho _B) =m_K^2+{\bf k}^2 -{\Sigma_{KN}\over f^2}
\rho_S +{3\over 4}{\omega\over f^2}\rho_N,\label{DIS}
\end{eqnarray}
where {\bf k} is the three-momentum of the kaon. The third term in the 
above equation is from the attractive scalar interaction due to explicit
chiral symmetry breaking and depends on the kaon-nucleon sigma term 
$\Sigma _{KN}$. With a strangeness content of the nucleon, defined by 
$y=2\langle N|\bar ss|N\rangle /\langle N|\bar uu+\bar dd|N\rangle 
\approx 0.1-0.2$ as normally used, its value is $370<\Sigma_{KN}<405$ 
MeV if one takes the strange quark to the light quark mass ratio to be 
$m_s/m\approx 29$. On the other hand, recent lattice gauge calculations 
\cite{DONG95,fuku} show that $y\approx 0.33$ which would give 
$\Sigma_{KN}\approx 450$ MeV.  The last term in Eq. (\ref{DIS}) is from 
the repulsive vector interaction and is proportional to the nuclear 
density $\rho _N$. For an antikaon this term becomes attractive due 
to G parity.
 
From the dispersion relation, the kaon energy in medium can be 
obtained, i.e.,
\begin{eqnarray}\label{omek}
\omega({\bf k}, \rho_N)=\left[m_K^2+{\bf k}^2-{\Sigma_{KN}\over f^2}\rho_S
+\left({3\over 8}{\rho_N\over f^2}\right)^2\right]^{1/2}
+{3\over 8}{\rho_N\over f^2}.\label{DIS1}
\end{eqnarray}
Using the KFSR relation ($m_\rho=2\sqrt{2}fg_\rho$) and the SU(3) 
relation ($g_\omega=3g_\rho$) \cite{bhaduri}, the kaon vector potential 
in the last term can be written as $(1/3)(g_\omega/m_\omega)^2\rho_N$ 
which is just 1/3 of the nucleon vector potential. This can be understood 
in the constituent quark model as the kaon contains only one light quark 
as compared to three light quarks in a nucleon. Since an antikaon has 
one light antiquark, its vector potential becomes attractive, leading to 
a reduction of its energy in medium. 
 
Corrections to the mean-field results of chiral Lagrangian have been 
studied in Refs. \cite{RHO94A,RHO94B,RHO95B}. The term which contributes 
at the same order in chiral perturbation theory as the Kaplan-Nelson 
term is the energy-dependent range term, which can be included
by multiplying the Kaplan-Nelson term by a factor $1-0.37\omega^2/m_K^2$ 
\cite{BR96A}. In Ref. \cite{FANG94}, a smaller value of $\Sigma_{KN}
\approx 350$ MeV has been used in evaluating the kaon scalar potential 
so that the effect of range term is effectively included. Also, 
the Nambu$-$Jona-Lasinio model has been used to estimate these 
corrections, and they are found to cancel off almost completely the 
attractive scalar potential \cite{WEIS93}. One notes, however, that 
the former is based on a systematic expansion while the latter is 
model-dependent.
 
The in-medium kaon mass is normally given by the kaon energy at zero 
three-momentum. With $\Sigma_{KN}\approx 350$ MeV, the kaon mass increases 
by about 10 MeV at normal nuclear matter density as shown in Fig. 3.
The antikaon mass is also shown and is seen to decrease with density
as its vector potential is also attractive.

Because of baryon number conservation, vector potentials in the initial 
and final states of a hadron-hadron reaction are similar and can thus 
be neglected in estimating the threshold for particle production in 
medium.  It is then sometimes useful to consider the kaon effective mass 
by including only effects due to the scalar field, i.e.,
\begin{equation}\label{kmass}
m_K^*\approx m_K\left[1-{\Sigma_{KN}\over f^2m_K^2}\rho_S\right]^{1/2}, 
\end{equation}
for both kaon and antikaon. In Ref. \cite{BRO92}, the quadratic 
vector term, $\left({3\over 8}{\rho_B\over f^2}\right)^2$, in Eq. 
(\ref{DIS1}) was discarded based on the expectation that it would be 
cancelled by higher order terms not included in the calculation.  Then, 
the kaon ``effective" mass should involve only the scalar field, while 
the vector mean field shifts the kaon chemical potential or energy, as 
in Walecka mean-field theory. 
 
From Eq. (\ref{omek}) one can define the kaon potential as the difference
between its energies in the medium and in free space \cite{LI95C,SHU92}. 
In the mean-field approximation this potential at normal nuclear matter
density is about 10 MeV for a kaon at rest. This is smaller than what 
is expected from the impulse approximation using the kaon-nucleon 
(KN) scattering length in free space which gives about 30 MeV. The 
inclusion of higher-order terms \cite{RHO94A,RHO94B,RHO95B} and possible 
scaling of the pion decay constant $f$ in medium \cite{BR96A,BROWN96B,LI96A} 
would bring the chiral perturbation results in agreement with that obtained
from the KN scattering length.
 
\begin{figure}
\epsfig{file=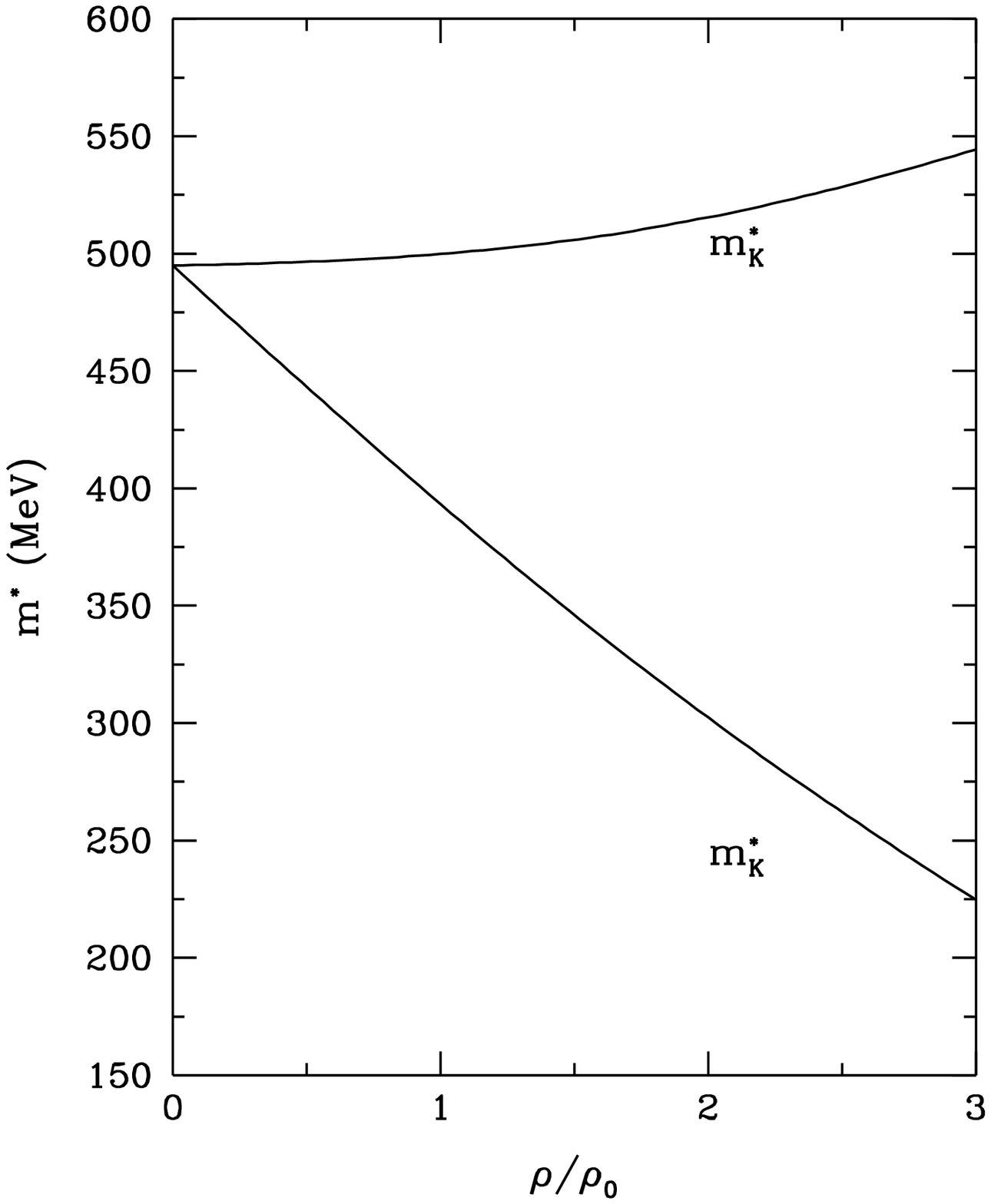,height=3in,width=4in}
\vskip 0.4cm
Fig. 3 ~Kaon and antikaon in-medium masses as functions of 
density from the mean-field approximation to the chiral Lagrangian.
\end{figure}

Antikaon properties in nuclear matter deserve some further discussions.
The possibility of $K^-$ condensation in neutron stars has been the 
focus of many recent studies \cite{RHO94A,MARU94A,MARU96,PRAK95,PAND95}. 
The isospin averaged $K^-N$ scattering length is negative in free space 
\cite{MARTIN81}, implying a repulsive $K^-$ optical potential in the 
simple impulse approximation. However, a systematic analysis of the 
kaonic atom data shows that the $K^-$ optical potential is deeply 
attractive, with a value of about 200$\pm$ 20 MeV at normal nuclear 
matter density \cite{BATTY93}. Unlike the kaon which interacts with 
nucleons relatively weakly so that the impulse approximation is reasonable, 
the antikaon interacts strongly with nucleons so we do not expect the 
impulse approximation to be reliable. The antikaon-nucleon (${\bar K}N$) 
interaction at low-energy is strongly affected by the $\Lambda (1405)$ 
which is a quasi bound state of an antikaon and a nucleon in the isospin 
$I=0$ channel and can decay into $\Sigma \pi$ channel \cite{WEISE95,WEISE96}. 
Thus, in principle one needs to carry out a coupled-channel calculation 
for ${\bar K}N$, $\Lambda \pi$ and $\Sigma \pi$ including the effects 
of $\Lambda (1405)$ in both free space and in nuclear medium 
\cite{WEISE95,WEISE96,BROCK78,TOKI94,KOCH94A,RHO96}.  Because of Pauli 
blocking effects on the intermediate states, the possibility of forming 
a bound ${\bar K}N$ state ($\Lambda (1405)$) decreases with increasing 
density, leading to a dissociation of $\Lambda(1405)$ in nuclear medium.  
This induces a transition of the $K^-$ potential from repulsion at very 
low densities to attraction at higher densities \cite{WEISE96,KOCH94A}. 
It was shown in Ref. \cite{RHO96}, that, because of the diminishing 
role of $\Lambda (1405)$ in dense matter, the prediction for the kaon 
condensation threshold is relatively robust with respect to different 
treatments of $\Lambda (1405)$. Also, the strong attraction found in 
kaonic atoms can be understood in chiral perturbation theory if the 
possible scaling of the pion decay constant $f$ in nuclear medium is 
included \cite{BR96A,BROWN96B,LI96A}.
 
\subsection{vector mesons}
 
Based on the restoration of scale invariance of QCD, Brown and Rho
have argued that masses of nonstrange vector mesons would be reduced 
in dense matter \cite{BR91,ADAMI93}. This is supported by studies based 
on the QCD sum-rule approach \cite{HAT92}, the effective hadronic model 
including $\bar NN$ vacuum polarization \cite{JEAN94,HAT94,SONG95}, 
and the quark-meson coupling model \cite{THOMAS}. 
 
In the QCD sum-rule study of vector meson masses in nuclear medium, 
one considers the correlation function of a vector current,
\begin{equation}
\Pi_{\mu\nu}(q) = i\int e^{iqx} \langle T J_{\mu}(x)J_{\nu}(0) 
\rangle_\rho d^4x, 
\end{equation}
In nuclear medium, it can be written as 
\begin{eqnarray}
\Pi_{00} & = & {\bf q}^2 \Pi_L,\nonumber \\
\Pi_{ij} & = & \left ( \delta_{ij}-\frac{q_i q_j}{{\bf q} ^2} \right ) 
\Pi_{T}+\frac{q_i q_j \omega ^2 }{{\bf q} ^2}\Pi_L,
\end{eqnarray}
where $\Pi_T$ and $\Pi_L$ denote the transverse and longitudinal parts 
of the polarization tensor, respectively. At zero momentum, ${\bf q} = 
{\bf 0}$, there is no distinction between the transverse and longitudinal
directions, and $\Pi_L$ is thus related to $\Pi_T$, i.e., 
\begin{eqnarray}
\quad \Pi_{T} & = & (\omega ^2 - {\bf q} ^2 ) \Pi _L,
\nonumber \\
\Pi_L & = & -\frac{1}{3 \omega ^2} g^{\mu\nu} \left. \Pi_{\mu\nu} \right |
_{\bf q \rightarrow {\bf 0}}.
\end{eqnarray}
In this case, only the longitudinal correlation function is needed.
 
For a rho meson, the current is taken to be
\begin{equation}
J_{\mu}^{(\rho)}=\frac{\bar{u}\gamma_{\mu}u-\bar{d}\gamma_{\mu}d}{2}.
\end{equation}
For large Euclidean four momenta, $Q^2 (=-q^2=-s) \rightarrow \infty $, $%
\Pi_L(Q^2)$ can be evaluated perturbatively by the operator product expansion
(OPE) \cite{REIN85}. Including operators up to dimension 6, one has
\begin{eqnarray}
\label{corr}
\Pi_L(Q^2 )&=& -\frac{1}{8\pi^2} \left ( 1+ \frac{\alpha_s}{\pi} \right )
\log \frac{Q^2}{Q_0^2} 
+\frac{m_q}{Q^4}\langle \bar{q}q\rangle_\rho
+\frac{\alpha_s}{24\pi Q^4}\langle G_{\mu\nu}G^{\mu\nu}\rangle_\rho
\nonumber\\
&&-\frac{\pi\alpha_s}{Q^6}\langle (\bar{q}\gamma_\mu \gamma_5 \lambda^a q)
(\bar{q}\gamma^\mu \gamma_5 \lambda^a q)\rangle_\rho
-\frac{2\pi\alpha_s}{9 Q^6}
\sum_{q'=u,d}\langle (\bar{q}\gamma_\mu \lambda^a q)
(\bar{q'} \gamma^\mu \lambda^a q')\rangle_\rho.
\end{eqnarray}
In the above, $\alpha_s$ is the QCD coupling constant while $Q_0$ is an
arbitrary scale parameter. Both light quark masses and their expectation
values are taken to be the same, i.e., $m_q=m_u\approx m_d$ and 
$\langle\bar{q}q\rangle_\rho=\langle\bar{u}u\rangle_\rho\approx
\langle\bar{d}d\rangle_\rho $.
 
The imaginary part of the correlation function at $s > 0$ is related
to the rho meson spectral function in medium and is normally parameterized
phenomenologically by a contribution from the rho meson pole and a
continuum, i.e., 
\begin{equation}\label{spec}
8\pi {\rm Im}\Pi_L(s) = F \delta ( s -{m^*_\rho}^2 ) + \left (1 + 
\frac{\alpha_s}{\pi} \right ) \theta (s- s_0),
\end{equation}
where $F$ is a constant.
 
The theoretical side ($Q^2 > 0 $) and the phenomenological side ($s > 0 $)
are related through the dispersion relation, 
\begin{equation}
\label{disp} {\rm Re} \Pi_L(Q^2 ) = \frac{{\rm P}}{\pi} \int_{0}^{\infty}ds 
\frac{{\rm Im}\Pi_L(s)}{s + Q^2 } + {\rm ~subtractions}. 
\end{equation}
From the above equation, the rho meson mass $m_\rho^*$ and the continuum
threshold $s_0$ can then be determined.
 
To suppress contributions from the higher order operators, one usually
introduces the Borel transform defined by 
\begin{equation}
L_{M_B}=\lim_{Q^2,n\to\infty; \, Q^2 \! /n=M_B^2}\,\frac{1}{(n-1)!}(Q^2 )^n
\left ( -\frac{d}{dQ^2} \right ) ^n, 
\end{equation}
where $M_B$ is the Borel mass in Eq. (\ref{cohen}). The Borel transform 
also removes the need for subtractions in the dispersion relation.
 
Carrying out the Borel transform of both sides of eq. (\ref{disp}) and
taking the ratio of the resulting equation to its derivative with respect 
to $-1/M_B^2$, one obtains
\begin{equation}\label{sumrule}
\frac{{m^*_\rho}}{M_B^2} = \frac{(1+\frac{\alpha_s}{\pi})[1-(1+\frac{s_0}
{M_B^2})e^{-s_0/M_B^2}]-\frac{8\pi^2 m_q}{M_B^4}\langle \bar{q}q 
\rangle_\rho-\frac{\alpha_s \pi }{3 M_B^4} \langle G_{\mu\nu}G^{\mu\nu} 
\rangle_\rho +\frac{896\pi^3\alpha_s}{81 M_B^6} \langle \bar{q}q 
\rangle_\rho^2}{(1+\frac{\alpha_s}{\pi})(1-e^{-s_0/M_B^2})+\frac{8\pi^2 m_q}
{M_B^4}\langle \bar{q}q \rangle_\rho+\frac{\alpha_s \pi }{3 M_B^4} 
\langle G_{\mu\nu}G^{\mu\nu} \rangle_\rho-\frac{448\pi^3\alpha_s}{81 M_B^6} 
\langle \bar{q}q \rangle_\rho^2}. 
\end{equation}
In deriving eq. (\ref{sumrule}), one has made use of the mean-field
approximation, 
\begin{eqnarray}\label{fac}
\langle (\bar{q}\gamma_{\mu}\gamma_{5}\lambda^{a}q)
(\bar{q}\gamma^{\mu}\gamma_{5}\lambda^{a}q)\rangle_\rho& \approx& \frac{16}{9}
\langle \bar{q} q \rangle_\rho^2, \nonumber \\
\langle (\bar{q}\gamma_{\mu}\lambda^{a}q)
(\bar{q'}\gamma^{\mu}\lambda^{a}q')\rangle _\rho&\approx& -\frac{16}{9}
\langle \bar{q} q \rangle_\rho^2\delta_{qq'}.
\end{eqnarray}
Also, terms involving nonscalar operators due to lack of Lorentz 
invariance in a medium are not shown. Similar sum rules can be derived 
for the omega and phi meson masses, using the currents 
$J_{\mu}^{(\omega)}=(\bar{u}\gamma_{\mu}u+\bar{d}\gamma_{\mu}d)/2$
and $J_{\mu}^{(\phi)} =\bar{s}\gamma_{\mu}s$, respectively.
 
To determine the rho meson mass, one minimizes the sum rule, Eq. 
(\ref{sumrule}), with respect to the Borel mass $M_B$ at an optimal $s_0$.  
In free space, using the commonly adopted values for the vacuum 
condensates ($\langle \bar qq\rangle _0=-(245\,{\rm MeV})^3$, 
$\frac{\alpha _s}\pi \langle G_{\mu \nu }G^{\mu \nu }\rangle_0
= (330\,{\rm MeV})^4$), quark mass ($m_q=5.5\,$MeV), and QCD coupling 
constant ($\alpha _s=0.3$), the empirical rho meson mass (776 MeV) can 
be reproduced with $s_0\approx 1.77\,{\rm GeV}^2$ in the range of
Borel masses $0.55\leq M_B^2\leq 0.75\,{\rm GeV}^2$. 
 
In nuclear medium, vector meson masses depend on in-medium condensates. 
Because of small light quark masses, terms in Eq. (\ref{sumrule}) 
linear in quark condensates can be neglected. The results for $\rho$ and 
$\omega$ mesons in nuclear medium thus depend on the density dependence 
of four-quark condensates. For phi meson, $\langle\bar ss\rangle_\rho$ 
dominates due to the relatively large strange quark mass $m_s\approx 
100\sim 200$ MeV, its in-medium mass thus depends on the nucleon 
strangeness content as $\langle\bar ss\rangle_\rho\approx \langle\bar 
ss\rangle_0+\langle N|\bar ss|N\rangle\rho_N$, where $\langle\bar 
ss\rangle\approx -0.8\langle\bar qq\rangle_0$ is the strange quark 
condensate in vacuum. Based on the factorization assumption for the 
four-quark condensate, i.e., Eq. (\ref{fac}), and using a simple 
delta-function plus continuum ansatz for the vector meson spectral 
function as shown in Eq. (\ref{spec}), Hatsuda and Lee have obtained the 
following results for the in-medium vector meson masses \cite{HAT92},
\begin{eqnarray}\label{rmass}
{m^*_{\rho ,\omega}\over m_{\rho ,\omega}}\approx 1-0.18 (\rho_N/\rho _0),
\end{eqnarray}
and
\begin{eqnarray}\label{pmass}
{m^*_{\phi}\over m_{\phi}}\approx 1-0.15y(\rho_N/\rho _0).
\end{eqnarray}
The density dependence of rho (omega) and phi meson masses is shown in 
Fig. 4, where the nucleon strangeness content is taken to be $y=0.17$. 
It is seen that the rho (omega) meson mass decreases significantly with 
density due to a strong density dependence of the light quark condensate. 
Since the strange quark condensate does not change much in nuclear 
medium as a result of the small nucleon strangeness content, the phi 
meson mass thus shows a weaker density dependence. However, the phi 
meson mass would decrease substantially if the temperature is high so
that many strange particles are present and reduce thus the strange quark 
condensate \cite{ASA94}.  The temperature dependence of phi meson mass 
is shown in Fig. 5 and is seen to decrease significantly at high 
temperatures.  We note that recent studies on phi meson mass at finite
temperature using the hidden gauge theory \cite{bhat,csong} show a much 
smaller reduction than that from the QCD sum rules. This may be due to the 
incomplete treatment of vacuum effects in the latter approach as it 
is carried out only at the one-loop level. 
 
\begin{figure}
\epsfig{file=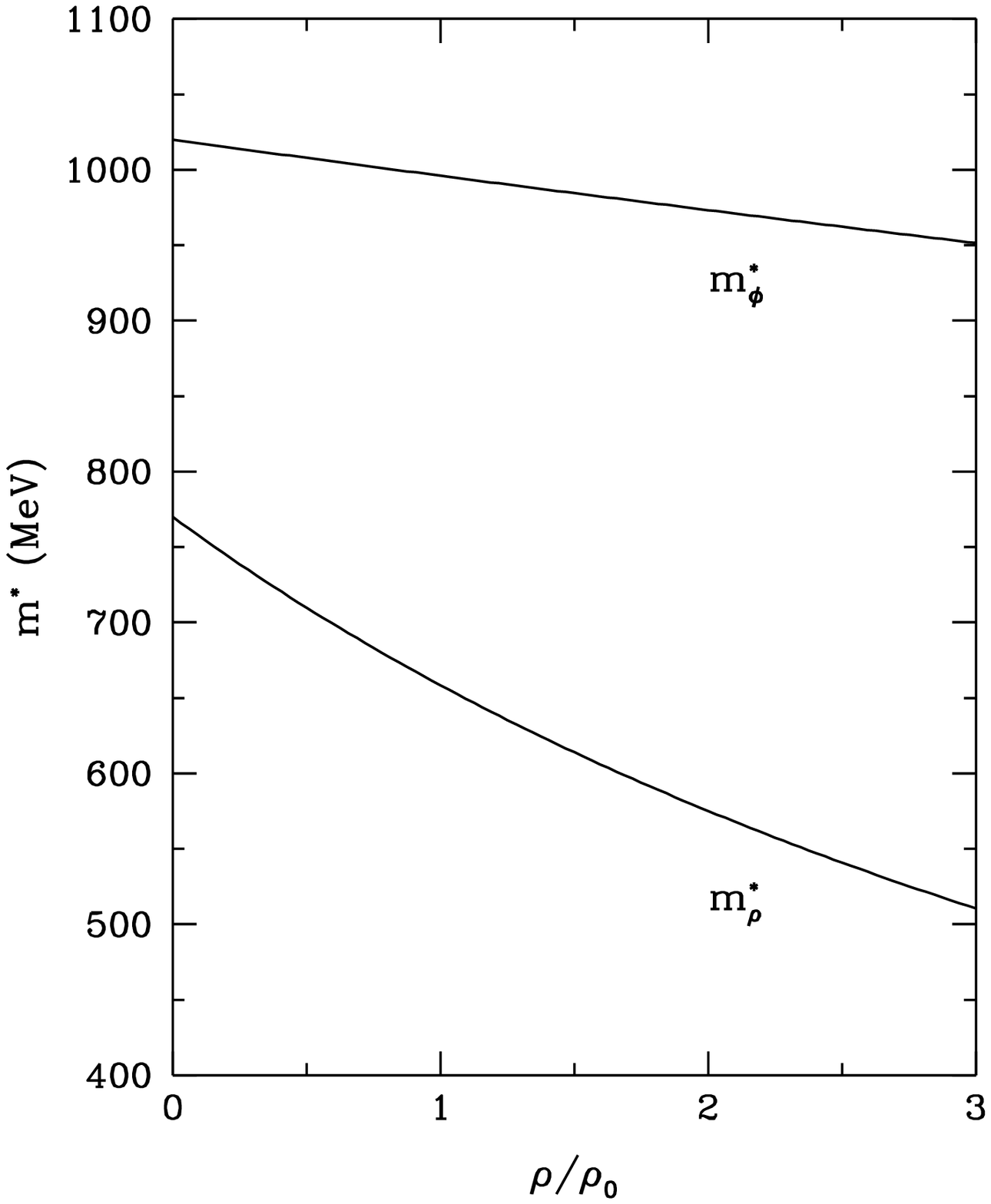,height=3in,width=4in}
\vskip 0.4cm
Fig. 4 ~Rho and phi meson masses as functions of density
from QCD sum rules. (from Ref. \cite{HAT92})
\end{figure}

A similar QCD sum-rule calculation of in-medium rho-meson mass has been 
carried out by Asakawa and Ko \cite{ASA93} using in the RHS of Eq. 
(\ref{disp}) a more realistic spectral function that is obtained by 
coupling the rho meson to pions via the vector dominance model and 
using the pion in-medium dispersion relation determined from the 
delta-hole model as discussed in Section II.B.1 \cite{ASA92,HERR93}.
A decrease of rho meson mass with increasing density as that of Ref. 
\cite{HAT92} has also been obtained. This is due to the stronger effect 
from reduced quark condensate in the presence of nucleons than that from 
the matter polarization.
 
Vector meson masses have also been studied using effective hadronic models
\cite{JEAN94,HAT94,SONG95}. Taking into account the change of $\bar NN$
vacuum polarization in medium as a result of reduced nucleon mass, the 
rho and omega meson masses have been found to decrease with density
and temperature as in QCD sum-rule approach.  If one includes only the 
matter effect, i.e., the polarization of the Fermi sea, the rho meson 
mass is seen to increase at high density. The inclusion of the vacuum 
effect, i.e., the polarization of the Dirac sea, however, brings the 
rho meson mass down in medium. The modification of the vacuum polarization 
in effective hadronic models is thus related to the change of quark 
condensate in medium in the QCD sum-rule approach.  Since the nucleon is
a composite particle, the large effect of $\bar NN$ vacuum polarization 
has, however, been questioned in Ref. \cite{cohen}. 
 
\begin{figure}
\epsfig{file=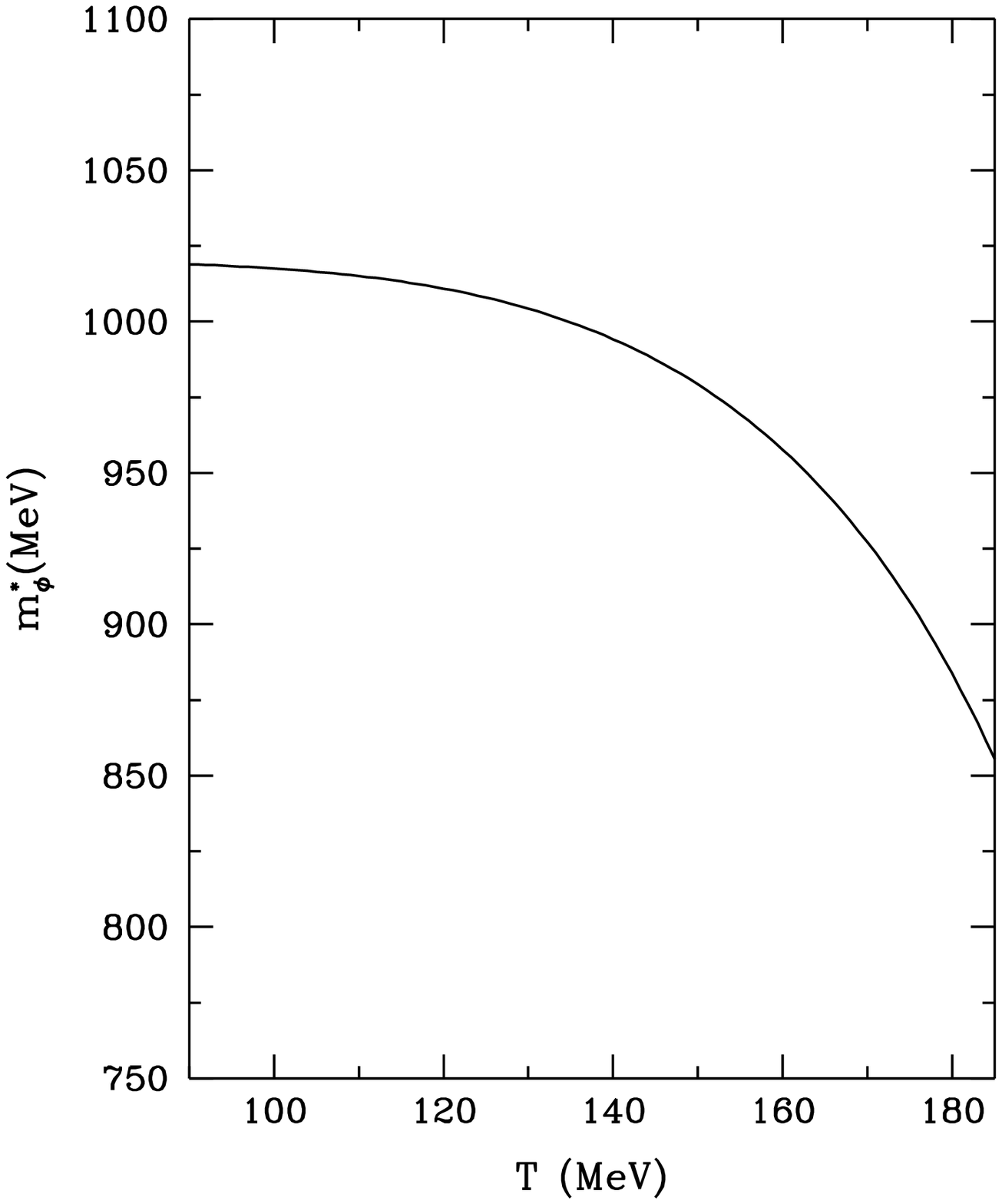,height=4in,width=4in}
\vskip 0.4cm
Fig. 5 ~Phi meson mass as a function of temperature from 
QCD sum rules. (from Ref. \cite{ASA94})
\end{figure} 

Experimentally, the quenching of the longitudinal response in quasielastic 
electron-nucleus scattering \cite{BR89}, and the enhancement of the 
$K^+$-nucleus scattering cross section \cite{BROWN88B,CHEN92} have been 
considered as indirect indication of the decrease of omega meson mass 
in nuclear medium. More direct observation of the vector meson mass in 
dense matter can be provided by the dilepton invariant mass spectrum from 
heavy-ion collisions. Since low-mass dileptons are mainly produced from 
pion-pion annihilation that proceeds through a rho meson, a change in
the rho-meson mass in dense matter is expected to be reflected in the
dilepton spectrum as a shift of the rho-meson peak to a lower mass
\cite{LI95D}. Similarly, the in-medium properties of phi meson can be
studied from the dilepton spectrum via kaon-antikaon annihilation.
 
\section{The relativistic transport model} 
 
\subsection{a brief review of the development of transport models}
 
Heavy-ion collisions involve very complicated nonequilibrium dynamics
and exhibit different features at different incident energies. At low 
energies, because of Pauli blocking of two-body collisions, a suitable
approach is the mean-field theory, such as the time-dependent Hartree-Fock 
(TDHF) theory and its semiclassical approximation 
\cite{BON76,CUS79,KOON81,NEG85,CUS85}. At high energies, the reaction 
dynamics at the initial stage is mainly governed by two-body collisions, 
so the mean-field can be safely neglected. Cascade-type models, in which 
heavy-ion reactions are visualized as a sequence of independent two-body 
collisions, have been developed for this purpose, first for heavy-ion 
collisions at Bevalac energies \cite{YARIV79,CUG80}, and recently for 
heavy-ion collisions at AGS \cite{ARC} and SPS energies \cite{RQMD}.
 
At energies between these two limits, and for the expansion stage of 
heavy-ion collisions at high energies, both the mean field and two-body 
collisions are important and need to be included in the model.  There 
have been attempts to extend the TDHF approach to include the effects of 
two-body collisions. In this way, one obtains the extended time-dependent
Hartree-Fock (ETDHF) equation, which were, however, found to be too 
complicated to be useful for heavy-ion collisions \cite{WONG83,KOH88}.  
On the other hand, semiclassical approximations, which make it possible 
to treat heavy-ion collisions more efficiently, have been shown to give 
results that are close to that of TDHF \cite{tang}.  This has thus led 
to the development of the well-known Boltzmann-Uehling-Uhlenbeck 
(BUU) equation, which is also known as the Vlasov-Uehling-Uhlenbeck 
(VUU), Landau-Vlasov (LV), and Boltzmann-Nordheim-Landau (BNL) equations
\cite{BERT84,KRUS85,MOLI85,AICH85A,GRE87,LENK89}. The BUU equation 
includes simultaneously the effects of mean field, two-body collisions, 
and the Pauli principle. Its solution then gives the time evolution of 
the one-body phase space distribution function.
 
The BUU equation is usually solved by the test-particle method
\cite{WONG82}, in which the propagation of particles in the mean field
(the Vlasov part) is given by the Hamilton's equation of motion, while
two-body collisions (the Uehling-Uhlenbeck part) are treated by 
Monte-Carlo procedure. The introduction of test particles is necessary
to remove numerical fluctuations associated with the evaluation of
the mean field, but at the same time they also suppress correlations that
are important for describing the instability and fragmentation
of the system at low densities. A similar approach, the quantum molecular 
dynamics (QMD) \cite{AICH91,AICH86A,BOAL88}, has thus been developed. 
In the QMD model each particle is represented by a Gaussian wave packet 
in both space and momentum.  In this way, one can obtain a smooth mean 
field without introducing test particles (in other words, one test 
particle for one physical particle). The QMD model seems to provide a 
reasonable description of the nuclear fragmentation phenomena \cite{AICH91}. 
Similar but improved models, such as the fermionic molecular dynamics 
(FMD) \cite{FELD90} and the antisymmetrized molecular dynamics (AMD) 
\cite{MARU90,ONO92}, have also been developed.
 
The two important ingredients in these transport models, namely, the 
mean-field potential in the Hamilton's equation of motion and the 
in-medium cross sections for two-body scattering, are usually introduced 
separately. For example, the Skyrme parameterization \cite{BERT88,AICH91}
for the mean field and the Cugnon parameterization for the NN cross 
sections \cite{BERT88,CUG81} are often used in transport models. Also, 
the momentum dependence of the nuclear optical potential has been introduced 
as it has been found to be important in describing the nucleon collective
flow in heavy-ion collisions \cite{GALE87A,AICH87A,DAN93,ZHANG94}. Since 
the mean field and in-medium two-body cross sections are related through
the in-medium effective NN interactions (the $G$ matrix), using separate 
parameterization inevitably introduces ambiguities in the transport models.  
There have been attempts to derive consistently the mean field and the 
in-medium two-body cross sections from the same underlying NN interactions 
\cite{BOH89,FAE89,PURI94}.
 
Also, relativistic effects are expected to become important in heavy-ion 
collisions at high energies. Although relativistic kinematics have been 
incorporated in most transport models, the relativistic covariance, or 
the frame independence, of the transport model has only been addressed 
in the relativistic quantum molecular dynamics (RQMD) \cite{RQMD} and 
similar models \cite{MARU91} by the use of Hamiltonian dynamics that 
is constrained by Poincare invariants. The meson-exchange nature of 
NN interactions, by which the nucleon mean-field potential can be 
separated into different Lorentz components (e.g., scalar and vector
potentials), has been included in the relativistic 
Boltzmann/Vlasov-Uehling-Uhlenbeck (RBUU/RVUU) approach
\cite{KO87,ELZE87,BLAT88}. The relativistic transport model allows one 
to investigate consistently the medium effects on hadron properties through
the change of the scalar and vector potentials. We shall discuss briefly 
in the next subsection the derivation of the relativistic transport model
from Walecka-type models.
 
\subsection{the relativistic transport model}
 
Based on the Lagrangian of nonlinear $\sigma$-$\omega$ model, Eq. 
(\ref{lag}), in which the nuclear matter is treated as a system of 
interacting nucleons and mesons, a transport equation for the phase 
space distribution function of nucleons has been derived in Refs. 
\cite{MAL90,MOS93,KO87,NORE90,NORE94,WA91,HEINZ} using the mean-field
approximation.. 
 
An important quantity in these derivations is the nucleon Green's 
function, defined by 
\begin{eqnarray}
iG(x_1,x_{1'})&=&\langle |T[N(x_1){\bar N}(x_{1'})]|\rangle\nonumber\\
&=&\theta(t_1-t_{1'})iG^>(x_1,x_{1'})+
\theta(t_{1'}-t_1)iG^<(x_1,x_{1'}),
\end{eqnarray}
where $\langle\cdots\rangle$ denotes the expectation value in the nuclear 
many-body state and $T$ is the time-ordering operator defined on a contour 
in the complex time plane \cite{daniel,kada}. It satisfies the following 
equation 
\begin{equation}
(i\gamma_\mu\partial^\mu_{x_1}-m_N)G(x_1,x_{1'})=\delta(x_1-x_{1'})
+\int d^4u\Sigma(x_1,u)G(u,x_{1'}),
\end{equation}
where $\Sigma$ is the nucleon self-energy and can be evaluated via
the perturbative expansion. The first-order nucleon self-energy is normally 
written as $\Sigma(x_1,x_2)\approx\Sigma_{HF}(x_1)\delta(x_1-x_2)$ with
\begin{equation}
\Sigma_{HF}(x)=\Sigma_S+i\gamma_\mu\Sigma_V^\mu(x).
\end{equation}
The scalar self-energy can be absorbed into the nucleon effective mass 
as in Eq. (\ref{effms}) while the vector self-energy leads to an effective 
momentum $p_\mu^*=p_\mu+\Sigma_{V\mu}$.  In this limit, the nucleon 
Green's function satisfies the equation,
\begin{equation}\label{gdirac}
\{i\gamma_\mu[\partial_{x_1}^\mu-\Sigma_V^\mu(x_1)]-m_N^*(x_1)\}G^<
(x_1,x_{1'})=0.
\end{equation}
 
Introducing the Fourier transform of the nucleon Green's function
\begin{equation}
G^<(x,p^*)=\int{d^4y}e^{ipx}G^<(x+y/2,x-y/2),
\end{equation}
then in local density and semi-classical approximations, it is related
to the seven dimensional on-shell nucleon's phase-space distribution 
function $f(x,{\bf p}^*)$, i.e., 
\begin{equation}
{\rm Tr}[G^<(x,p^*)]=i16\pi\delta({p^*}^2-{m^*_N}^2)m_N^*f(x,p^*).
\end{equation}
From Eq. (\ref{gdirac}), a relativistic Vlasov equation follows and has the 
form
\begin{equation}\label{vuu} 
\{[\partial _x^\mu -(\partial _x^\mu \Sigma_v^\nu -\partial _x^\nu 
\Sigma _v^\mu )\partial _\nu ^{p^{*}}]p_\mu ^{*}+m_N^*
({\partial _x^\mu } m_N^{*})\partial _\mu ^{p^{*}}\}f(x,{\bf p}^{*})=0,
\end{equation}
This equation can be more conveniently written as \cite{lk}
\begin{equation}
\frac{\partial}{\partial t}f+{\bf v}\cdot\nabla_xf-\nabla_xU\cdot\nabla_pf=0,
\end{equation}
where 
\begin{equation}
{\bf v}={\bf p^*}/E^*,\qquad\qquad U=E^*+(g_\omega/m_\omega)^2\rho_N,
\end{equation}
with $E^*=({\bf p^*}^2+{m^*}^2)^{1/2}$. 
 
In terms of the phase-space distribution function, the scalar and current
densities in the local-density approximation can be expressed, respectively,
as 
\begin{equation}
\rho_s=\,4\int \frac{d^3{\bf p}^*}{(2\pi)^3}f(x,{\bf p}^*)m^*/E^*, ~~~~
\rho_\mu=\,4\int \frac{d^3{\bf p}^*}{(2\pi)^3}f(x,{\bf p}^*)p_\mu^*/E^*.
\end{equation}
In heavy-ion collisions
the nucleon vector potential has both spatial-like and time-like components
and are related to the nuclear current density $\rho _\mu$. 
 
The relativistic transport equation is solved by the test-particle method 
\cite{WONG82}, so each nucleon is replaced by a collection of test particles.
The one-body phase-space distribution function is then given by the
distribution of these test particles in phase space. To solve the Vlasov
equation is thus equivalent to solving the following classical equations 
of motion for all test particles, 
\begin{equation}\label{beom}
\frac{d{\bf x}}{dt}=\,{\bf p}^*/E^*, \qquad
\frac{d{\bf p}}{dt}=\,-{\bf\nabla}_xU(x). 
\end{equation}
 
These equations of motion can be extended to other hadrons. For baryon 
resonances, they are the same as Eq. (\ref{beom}), as we have assumed 
that they have same mean-field potentials. For lambda and sigma hyperons, 
the corresponding mean-field potentials are taken to be 2/3 of those for 
the nucleon. The equations of motion for an antiproton are the same as 
those of the nucleon except that the sign of the vector potential is 
changed due to G parity. For a kaon, the equations of motion in the 
mean-field approximation to chiral Lagrangian \cite{LI95C} are obtained 
from Eq. (\ref{beom}) with $E^*=\left[m_K^2+{\bf p}^2-{\Sigma_{KN}\over 
f^2}\rho _S +\big({3\over 8}{\rho_B\over f^2}\big)^2\right]^{1/2}$
and $U=\omega({\bf p},\rho_B)-\omega_0({\bf p})$, where 
$\omega({\bf p},\rho_B)$ and $\omega_0({\bf p})$ are the kaon dispersion
relation in medium and in free space, respectively \cite{SHU92}.   
Again, the vector potential for an antikaon has an opposite sign
from that of the kaon \cite{LI94B}.
 
If one includes the second-order Born terms in the nucleon self-energy, 
then its imaginary part can be shown to lead to a collisional integral on 
the right hand side of the relativistic Vlasov equation
\cite{KO87,NORE94}, i.e.,
\begin{eqnarray}
I_c&=&\,\int{{d{\bf p}_2^*}\over(2\pi)^3}\int{{d{\bf p}_3^*}\over(2\pi)^3}\
\int{d\Omega}v\frac{d\sigma}{d\Omega}
\delta^3({\bf p}^*+{\bf p}_2^*-{\bf p}_3^*-{\bf p}_4^*)\nonumber\\
&&\cdot \{f(x,{\bf p}_3^*)f(x,{\bf p}_4^*)[1-f(x,{\bf p}^*)]
[1-f(x,{\bf p}_2^*)]
-f(x,{\bf p}^*)f(x,{\bf p}_2^*)[1-f(x,{\bf p}_3^*)]\nonumber\\
&&\cdot [1-f(x,{\bf p}_4^*)]\},
\end{eqnarray}
where $v$ is the relative velocity between the colliding particles and 
$\frac{d\sigma}{d\Omega}$ is the two-body NN differential scattering cross 
section calculated from meson exchanges. The factor $(1-f)$ takes into 
account the Pauli-blocking of fermions in the final state of a scattering.
 
For heavy-ion collisions at a few GeV/nucleon, nucleons, delta resonances,
and pions are the most important degrees of freedom. As in normal BUU model 
\cite{BERT88}, the isospin-averaged cross sections measured in free space 
\cite{CUG81} are used for the elastic ($NN\rightarrow NN$) and the delta 
excitation ($NN\rightarrow N\Delta $) process. The cross section for the 
inverse process $N\Delta\to NN$ is determined from the detailed balance
relation \cite{DAN91}. Both the nucleon-delta ($N\Delta\rightarrow
N\Delta$) and the delta-delta ($\Delta\Delta\rightarrow\Delta\Delta$ )
elastic collision are also allowed, and their cross sections are assumed 
to be the same as that for the nucleon-nucleon elastic scattering at the 
same center-of-mass energy. 
 
When a delta is formed, its mass distribution is taken to be of the
Breit-Wigner form 
\begin{eqnarray}
P(m_\Delta )={(\Gamma  (q)/2)^2\over (m_\Delta -m_0)^2+(\Gamma (q)/2)^2},
\end{eqnarray}
where $ m _0=$ 1.232 GeV and the momentum-dependent delta
width \cite{TOKI86},
$\Gamma (q)$, is 
\begin{eqnarray}
\Gamma (q)={0.47q^3\over [1+0.6(q/m_\pi )^2]m_\pi ^2}.
\end{eqnarray}
In the above, $q$ is the pion momentum in the rest frame of a delta and 
is related to its mass $m_\Delta$ by $q=\sqrt {\big[ m_\Delta^2
-(m_N+m_\pi )^2\big] \big[m_\Delta ^2-(m_N-m_\pi )^2\big]}/(2m_\Delta)$.
 
The collision between two baryons is treated in the same way as in cascade 
model \cite{CUG81}. A collision occurs when the distance between them is 
less than $\sqrt{\sigma /\pi}$ with $\sigma $ being their interaction cross
section.  After the collision, directions of their momenta change in a 
statistical way according to the angular distribution. However, collisions 
are allowed only among particles in the same simulation but the mean nuclear 
density and current are computed with test particles from all simulations
in the ensemble.  

We note that a different treatment of two-body collisions has recently 
been introduced in Ref. \cite{kahana}. In this approach, the collision 
between two particles is treated classically so that not only the angular 
momentum is conserved but also the reaction plane is preserved. 
Furthermore, only repulsive orbits are allowed in the collision due to 
the largely repulsive force in nucleon-nucleon scattering below 1 
GeV/nucleon. An attempt to address the question of separating the nuclear
interaction into a mean-field and a two-body collision part in nuclear
collisions has recently been given in Ref. \cite{pawel}.
 
Pion production is included in the relativistic transport model through
delta decay, i.e., $\Delta\to\pi N$, using the momentum-dependent delta 
decay width. The inverse reaction $\pi N\to\Delta$ is also included to 
take into account pion absorption \cite{XIONG90,WOLF90,BALI91}. In most 
transport models for pions, one does not include the pion potential so 
they propagate as free particles in nuclear medium. However, medium 
effects on pions are included in Ref. \cite{XIONG93} by treating them 
as quasiparticles. It is found that softening of the pion dispersion 
relation might be responsible for the observed enhancement of low 
transverse energy pions in heavy-ion collisions at both Bevalac 
and SIS energies \cite{SPE,TAPS}.
 
Because of their small production probability in hadron-hadron collisions, 
other particles, such as the kaon, antikaon, antiproton, and dilepton, are
treated perturbatively, i.e., the collision dynamics is not affected by 
their presence \cite{RK80,AICH85}.  These particles may also scatter 
with other hadrons. A kaon interacts only elastically with 
a baryon due to strangeness conservation, and the kaon-nucleon scattering 
cross section at low energies is about 10 mb in free space \cite{DOVER}.
On the other hand, an antikaon, in addition to elastic scattering, can 
be absorbed by a nucleon through the strangeness-exchange process 
$\bar KN\rightarrow \pi Y$. These cross sections are also taken from Ref.
\cite{DOVER}. Similarly, antibaryons can be annihilated by baryons, and 
their cross sections are taken from Ref. \cite{CUG89}. To include the 
final-state interactions of these particles, a perturbative test particle 
method has been introduced in \cite{FA93}. In this method, one allows many
particles to be produced in a hadron-hadron interaction. Each produced 
particle is then assigned a production probability, which is given by 
the ratio of its production cross section to the hadron-hadron total 
cross section. Their motions are then followed by solving the classical 
equations of motion. The collisions of these particles with nucleons 
are treated as usual except that their effects on nucleons are neglected, 
i.e., one does not allow the nucleon momenta to change in such collisions.
 
With reduced hadron masses in medium, the above cross sections should 
be modified as well. For nucleon-nucleon elastic scattering, the in-medium
cross section has been evaluated in Ref. \cite{HAAR87,lm} in the DBHF
approach using the one-boson-exchange potentials. It is found that the 
magnitude of the cross section decreases with density and the differential 
cross section becomes more isotropic in medium. Experiments on proton-nucleus 
scattering \cite{SUE94,DUB96,TANA96} seem to be consistent with these 
predictions. The $NN\to N\Delta$ cross section in medium has been studied 
in Refs. \cite{bbkl,KO89,mlz} via the one-pion exchange model. Including 
the softening of the pion dispersion relation, its magnitude is found to 
increase with density. Recently, the same model has been used for
evaluating the $\Delta\Delta\to\Delta\Delta$ cross section in medium
\cite{mlzz}. Although the magnitude of the cross section is found to 
decrease with density, the density dependence is less pronounced than
that of the nucleon-nucleon elastic cross section. 
 
The $\Delta\leftrightarrow\pi N$ cross section in medium can be easily
evaluated using in-medium hadron masses if one assumes that the cross
section remains to be of the Breit-Wigner form. No theoretical studies 
have been carried out for the kaon-nucleon elastic cross section in 
medium. It is thus of interest to pursue such a study following the 
method of Ref. \cite{lm} by using the meson-exchange model of Ref. 
\cite {ldhs}.  Similar calculations should also be done for the antikaon. 
 
The production cross sections for kaon, antikaon, antiproton, and dilepton 
from hadron-hadron interactions in both free space and medium will be 
discussed below.
 
\section{medium effects in heavy-ion collisions}
 
In this section, results obtained from the relativistic transport model 
are presented and compared with the experimental data from heavy-ion 
collisions. In particular, discussions will be given for particle 
production and flow at SIS/GSI, strangeness enhancement at AGS/BNL 
and SPS/CERN, and dilepton production in heavy-ion collisions. 
 
\subsection{heavy-ion collisions at SIS energies}
 
In heavy-ion collisions at SIS energies ($\sim 1$ GeV/nucleon), the 
colliding system consists mainly of nucleons, delta resonances, and pions. 
In addition, particles with small production probability, such as the eta, 
kaon, antikaon, antiproton, and dilepton, can be created from baryon-baryon, 
meson-baryon, and meson-meson interactions during the course of the 
collisions. Since the threshold energies for eta, kaon, antikaon, and 
antiproton production in NN collisions in free space are, respectively, 
1.26, 1.58, 2.5, and 5.6 GeV, their production in heavy-ion collisions 
at SIS energies is mostly below the production threshold in NN interactions
in free space, and is known as `subthreshold particle production'. 
Subthreshold particle production offers the possibility of exploring the 
extreme region of phase space, and thus may carry valuable information about 
the early, violent stage of heavy-ion collisions during which both density 
and temperature are high. In particular, the mean-field potential has been
found to play an important role in subthreshold particle production 
\cite{FANG94,LI95A,LI94B,LI94C,antip,RUSS92,HUANG93,AICH94,MARU94B,BALI94,MOS94,FAE94,HART94}.
This can be seen from Fig. 6, where we show the density dependence of 
the Q-value, which is defined by the difference between total hadron 
masses and potentials in the final and initial states, for the 
reaction $NN\to N\Lambda K$, $NN\to NNK\bar K$, and $NN\to NNN\bar N$. 
They are evaluated with the in-medium hadron masses and potentials 
discussed in Section II for the soft nuclear equation of state in Table I.
It is seen that the Q-value increases slightly for the reaction $NN\to 
N\Lambda K$ but decreases appreciably for the reactions $NN\to NNK\bar K$ 
and $NN\to NNN\bar N$. In the ideal case that the scalar and vector 
potentials on a hadron in nuclear medium is determined by the sum of 
scalar and vector potentials on its constituent quarks, the Q-value for 
the reaction $NN\to N\Lambda K$ is independent of density, while it is 
$2/3\Sigma_S$ and $2\Sigma_S$, respectively, below the values in free 
space for the reactions $NN\to NNK\bar K$ and $NN\to NNN\bar N$. At 
twice normal nuclear density, their values are -177 MeV and -530 MeV, 
respectively, which are slightly less than that shown in Fig. 6.

\begin{figure}
\epsfig{file=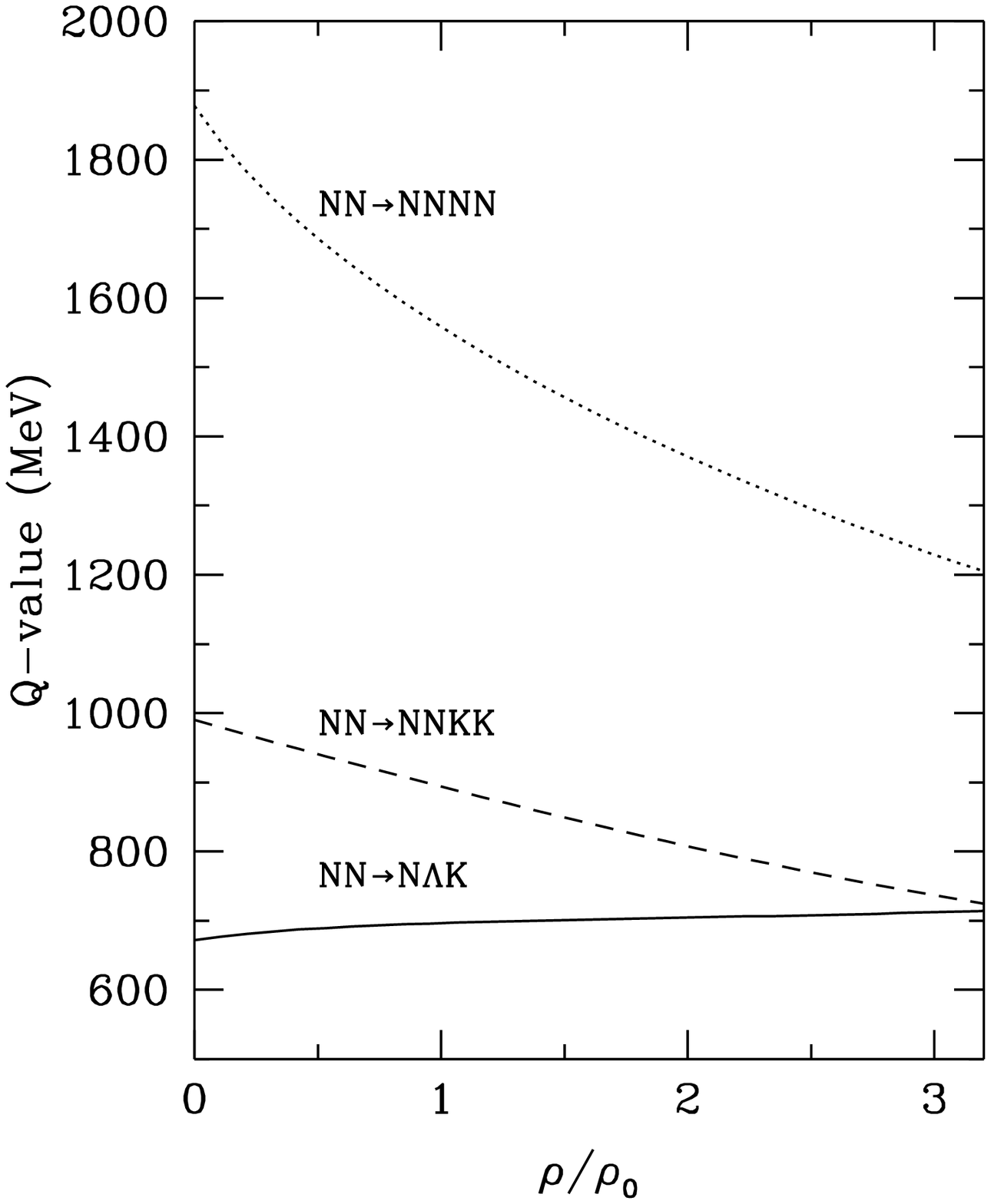,height=3in,width=4in}
\vskip 0.4cm
Fig. 6 ~The density dependence of the Q-value for the reactions
$NN\to N\Lambda K$ (solid curve), $NN\to NNK\bar K$ (dashed curve), and
$NN\to NNN\bar N$ (dotted curve).
\end{figure}

\subsubsection{subthreshold kaon production}
 
The first experiment on kaon production in heavy-ion collisions was 
carried out at Bevalac in the eighties at incident energies around 2 
GeV/nucleon \cite{NAGA81,LBL}.  These experiments had generated many 
theoretical studies based on both cascade and transport models 
\cite{AICH87A,RK80,AICH85,KO89,CUG84,SZ87,LI92B}. Recently, there are
new experiments on kaon production in heavy-ion collisions at incident 
energies around 1 GeV/nucleon by the KaoS collaboration at GSI \cite{KAOSE}. 
This has led to a resurgence of theoretical studies based on both 
relativistic and nonrelativistic transport models 
\cite{FANG94,LI95A,LI95B,HUANG93,AICH94,MARU94B,BALI94}. The study of 
kaon production in heavy-ion collisions at subthreshold energies allows 
us to investigate the properties of the dense matter formed in the initial 
stage of the collisions. These include the nuclear equation of state, the 
role of baryon resonances, and the kaon in-medium properties. 
 
To study the kaon production cross section in heavy-ion collisions,
we need its production cross section in baryon-baryon interactions
as the contribution from other processes is small. For example, 
it has been shown that in heavy-ion collisions the contribution to 
kaon production from pion-baryon interactions is only about 25\% 
\cite{XION90} while that due to many-body collisions is about 10\% 
\cite{BAT92}. Randrup and Ko \cite{RK80} have analyzed the available 
experimental data and proposed the following parameterization 
for the isospin-averaged kaon production cross section from the NN 
interaction in free space 
\begin{equation}\label{kcross}
\sigma _{NN\rightarrow BYK^+} (\sqrt s)\approx
36 ~{p_{\rm max}\over m_K} ~\mu {\rm b},
\end{equation}
where the kaon maximum momentum $p_{\rm max}$ is related to the NN
center-of-mass energy $\sqrt s$ by $p_{\rm max}=\sqrt {\big[ 
s-(m_B+m_Y+m_K)^2\big] \big[ s-(m_B+m_Y-m_K)^2\big] /4s}$.
 
Kaon production cross sections from the nucleon-delta and delta-delta 
interactions have also been analyzed in Ref. \cite{RK80} based mainly 
on isospin arguments. It has been found that the following scaling 
relations approximately hold,
\begin{eqnarray}
\sigma _{N\Delta\rightarrow BYK^+}(\sqrt s)&\approx& {3\over 4}
\sigma _{NN\rightarrow BYK^+}(\sqrt s),\nonumber\\
\sigma _{\Delta\Delta\rightarrow BYK^+}(\sqrt s)&\approx& {1\over 2}
\sigma _{NN\rightarrow BYK^+}(\sqrt s).
\end{eqnarray}
 
In addition to the total kaon production cross section, one also needs 
the kaon momentum distribution from baryon-baryon interaction in free space. 
This has been parameterized in Ref. \cite{RK80} based on phase space 
considerations, i.e.,
\begin{equation}
{d\sigma\over dp}\approx
\sigma _{K^+}(\sqrt s) {12\over p_{\rm max}}
\left ({p\over p_{\rm max}}\right )^2\left (1-{p\over p_{\rm max}}\right ),
\end{equation}
which reproduces reasonably the experimental data \cite{RK80}.
 
Using the one-pion plus one-kaon exchange model \cite{LAGET91}, it has been 
recently shown \cite{LI95B} that near kaon production threshold, where
no experimental data are available, the cross sections in the Randrup-Ko 
parameterization and scaling ansatz are overestimated for the nucleon-nucleon 
interaction but underestimated for the nucleon-delta interaction. However,
these discrepancy may be due to the neglect of final-state interactions, 
which have been shown to be important near the kaon production threshold
\cite{LAGET91}.
 
\begin{figure}
\epsfig{file=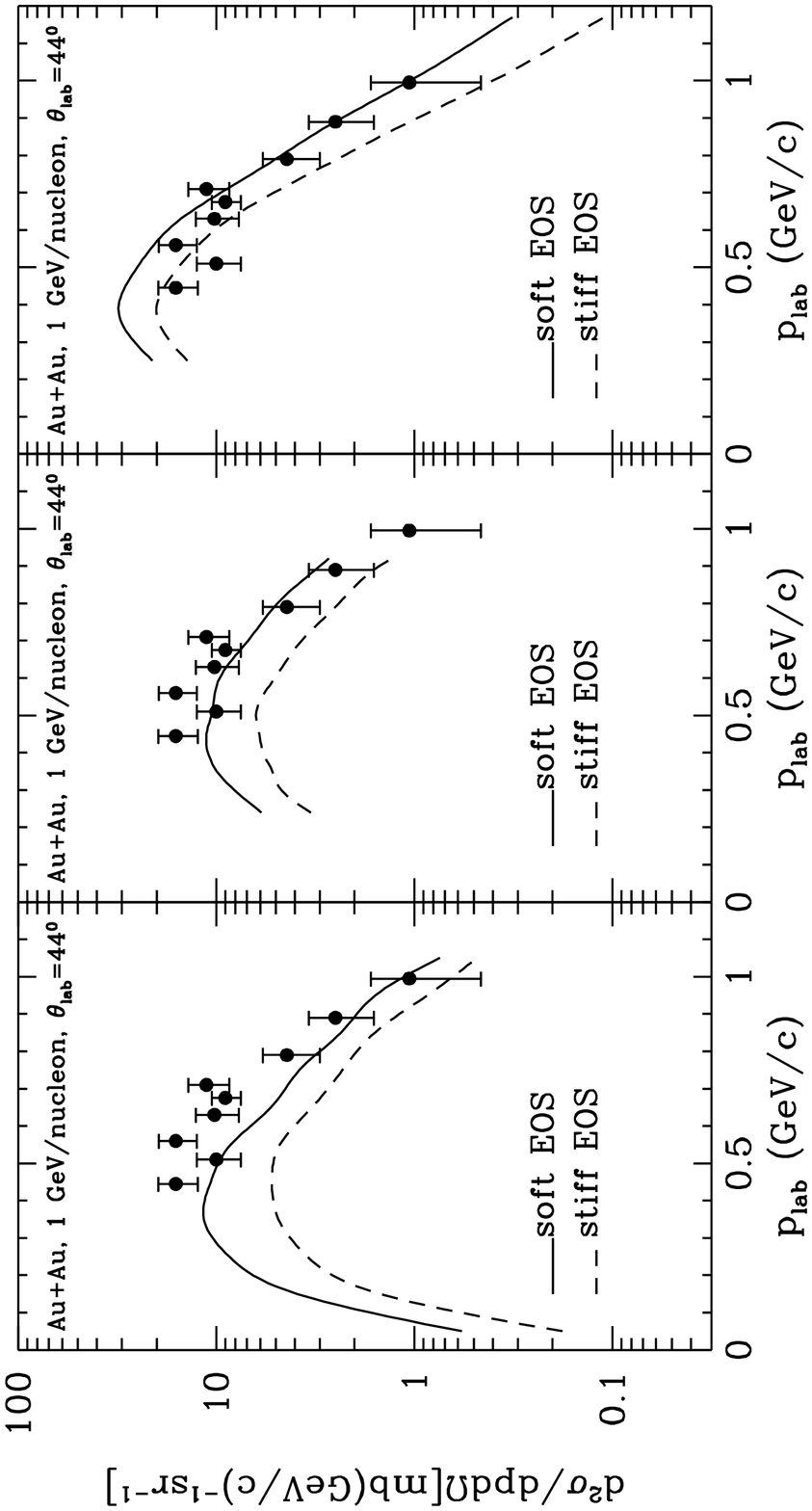,height=4in,width=6in}
\vskip 0.4cm
Fig. 7 ~Kaon momentum spectra from Au+Au collisions at 1
GeV/nucleon, obtained with the soft (solid) and stiff EOS (dashed),
respectively. The left, middle, and right panels are from
Refs. \cite{LI95A}, \cite{MARU94B}, and \cite{AICH94}, respectively.
The experimental data from the KaoS collaboration \cite{KAOSE} 
are also shown.
\end{figure}
 
As first pointed out by Aichelin and Ko \cite{AICH85}, the kaon yield 
from heavy-ion collisions at incident energies below the kaon production 
threshold in NN interaction in free space is sensitive to the nuclear 
equation of state at high densities. For Au+Au collisions at 1 GeV/nucleon, 
this dependence is shown in Fig. 7 in which the kaon momentum spectra 
obtained in three different calculations using the soft and the stiff 
equation of state are compared with experimental data from the KaoS 
collaboration \cite{KAOSE}.  The results in the left panel of Fig. 7 are 
from Ref. \cite{LI95A} based on the relativistic transport model including 
medium modifications of the kaon properties. In this calculation, the kaon 
production cross section in medium is obtained by using in-medium hadron 
masses to evaluate the $p_{\rm max}$ in Eq. (\ref{kcross}). The results in 
the middle panel are from Ref. \cite{MARU94B} based also on a relativistic 
transport model. The results in the right panel are from Ref. \cite{AICH94} 
using the nonrelativistic QMD model. Theoretical results from
relativistic transport models with the soft EOS are in reasonable 
agreement with the experimental data, while that with the stiff EOS 
are below the data by about a factor of two. However, the results from 
the QMD calculation seem to favor a stiff nuclear equation of state. 
As explained below, the QMD calculation of Ref. \cite{AICH94} has 
overestimated the contribution from delta particle as they are only 
allowed to decay at freeze out. Thus, the KaoS data from Au+Au collisions 
at 1 GeV/nucleon seem to be consistent with a soft EOS. 

\begin{figure}
\epsfig{file=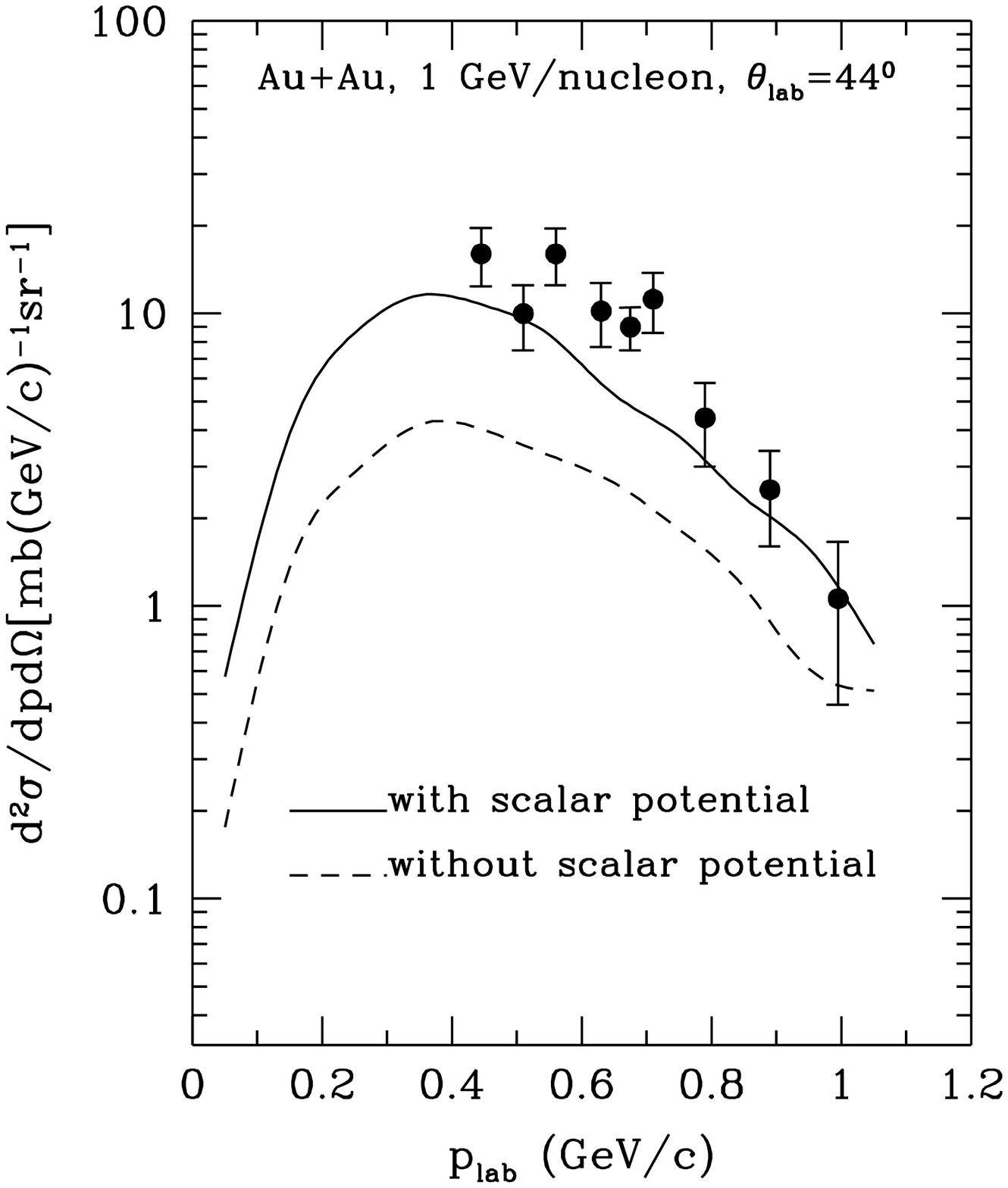,height=4in,width=4in}
\vskip 0.4cm
Fig. 8 ~Kaon momentum spectra from Au+Au collisions at 1
GeV/nucleon, obtained with (solid) and without (dashed) 
the kaon scalar potential. (from Ref. \cite{FANG94})
\end{figure} 

Subthreshold kaon production in heavy-ion collisions is also sensitive 
to the kaon scalar potential, and this is illustrated in Fig. 8. The 
solid and dashed curves correspond to results with and without kaon 
scalar potential, respectively. It is seen that without the attractive 
kaon scalar potential, the theoretical results are about a factor of 
3-4 below the experimental data. This can be easily understood from 
Fig. 6 in terms of the Q-value for the reaction $NN\to N\Lambda K$ 
in medium. At nuclear density $\rho=2\rho_0$, where most kaons 
are produced in heavy-ion collisions, the Q-value is about 33 MeV 
above that in free space. Without the attractive kaon scalar potential, 
which is about 90 MeV, the Q-value is increased by the same amount.  
This increases the kaon production threshold and thus reduces its yield.
 
In the calculation of Ref. \cite{MARU94B} based on the relativistic 
transport model, medium effects on kaons are not included so kaons
are treated as free particles as in the nonrelativistic QMD calculation 
\cite{HUANG93,AICH94}.  The results also agree with the experimental 
data and are thus comparable to that of Ref. \cite{FANG94}, as shown 
in the middle panel of Fig. 7. Although the kaon mean-field potential 
is taken to be zero in Ref. \cite{MARU94B}, the scalar and vector potentials 
for the hyperon are taken to have the same strength as that for a nucleon, 
which is stronger than the hyperon mean-field potential used in Ref. 
\cite{FANG94}, i.e., 2/3 of the nucleon mean-field potential. Thus, 
as far as the kaon production threshold is concerned, the treatment of 
Ref. \cite{MARU94B} is approximately equivalent to Ref. \cite{FANG94} 
in which both kaon scalar and vector potentials are included.

\begin{figure}
\epsfig{file=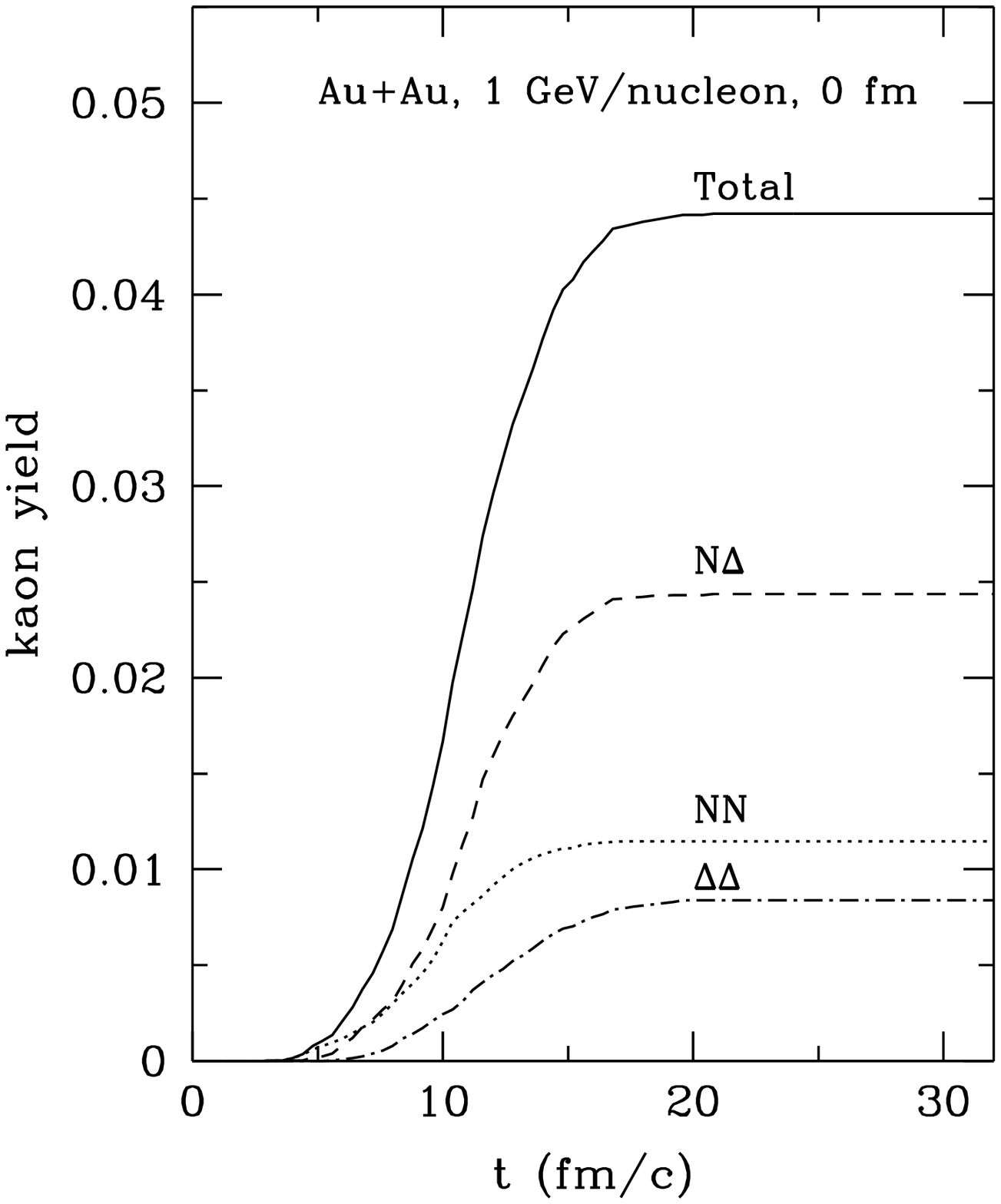,height=4in,width=4in}
\vskip 0.4cm
Fig. 9 ~Contributions to the total kaon production probability
from the NN, N$\Delta$ and $\Delta\Delta$ interactions in 
Au+Au collisions at 1 GeV/nucleon and impact parameter b=0 fm.
(from Ref. \cite{FANG94})
\end{figure}

Most kaons in heavy-ion collisions at subthreshold energies are produced 
from the two-step process in which a nucleon is first excited to a 
resonance and a kaon is then produced from the interaction of the 
resonance with another baryon. At incident energies around 1 GeV/nucleon, 
only the delta resonance is appreciably produced. Subthreshold kaon 
production can thus serve as a good probe of the properties of (delta) 
resonance matter. In Fig. 9, the total kaon production probability in 
Au+Au collisions at an incident energy of 1 GeV/nucleon and an impact 
parameter of b=0 fm from the relativistic transport model is separated 
into contributions from the nucleon-nucleon, the nucleon-delta and the 
delta-delta interaction. It is seen that the contribution from the
nucleon-delta interaction accounts for more than half of the total 
kaon yield, and the sum of contributions from the nucleon-delta and 
delta-delta interactions is about 75\% of the total kaon yield.  
Contrary to that of Refs. \cite{HUANG93,AICH94}, the contribution from 
the delta-delta interaction is slightly smaller than that from the 
nucleon-nucleon interaction. This is mainly due to following two 
reasons. First, in Refs. \cite{HUANG93,AICH94}, deltas are allowed
to decay only at final stage of the collision. This treatment certainly
overestimates the contribution from the  nucleon-delta and delta-delta 
interactions. Secondly, kaon production in Refs. \cite{HUANG93,AICH94} 
is calculated in the QMD with a Skyrme-type momentum-independent mean-field 
potential. It is well-known that introducing a momentum-dependence in 
the nucleon mean-field potential, which is automatically included in the 
relativistic transport model \cite{LI93B}, reduces the number of 
two-body collisions and hence the delta abundance, thus leading to a 
smaller contribution from the delta-delta interaction as compared to 
the case with a momentum-independent mean-field potential.
 
Although the interaction of a kaon with a nucleon is relatively weak 
as compared to other hadrons, they are still scattered by baryons 
in the dense medium \cite{FA93,RAN81}.  Furthermore, kaons propagate 
in both scalar and vector mean-field potentials. These final-state 
interactions change the kaon momentum spectra but not its total yield, 
leading to an increase of both its yield and the slope of its spectra 
at large angles. Actually, final-state interactions help to bring the 
theoretical results at $\theta _{\rm lab}=~44^o$ in better agreement 
with the experimental data. Since final-state interactions increase the 
kaon yield at large angles, it would be of interest to have 
experimental data to test this prediction.
 
\subsubsection{subthreshold antikaon production}
 
Effects of the attractive scalar potential can be more clearly seen 
in antikaon production than in kaon production as it leads to a significant
reduction of the Q-value for the reaction $NN\rightarrow NNK^+K^-$ in 
nuclear medium as shown in Fig. 6. From the analysis of available 
experimental data, Zwermann and Sch\"urmann \cite{ZS84} have proposed 
the following parameterization for the antikaon production cross section 
from nucleon-nucleon interaction in free space,
\begin{eqnarray}
{d\sigma\over dp}\approx
0.75 \left ({p\over p_{\rm max}}\right )^2
\left (1-{p\over p_{\rm max}}\right )^2~{\rm mb/GeV},
\end{eqnarray}
where $p_{\rm max}=\sqrt{[s-(2m+2m_K)^2](s-4m^2)/(4s)}$. It has been 
usually assumed that the same cross section is applicable for antikaon 
production from the delta-nucleon and the delta-delta interaction.
 
Including all final-state interactions (propagation in mean fields,
elastic scattering, and absorption) for antikaons, the results from 
the relativistic transport model for the antikaon production cross section 
from $^{58}$Ni+$^{59}$Ni collisions at 1.85 GeV/nucleon are shown in Fig. 10
\cite{LI94B}. The dashed curve gives the results using free kaon and antikaon
masses, while the results with in-medium kaon and antikaon masses are shown 
by the solid curve. Also shown in the figure by solid circles are the 
experimental data from the SIS at GSI \cite{KIEN}.
 
Using free kaon and antikaon masses, the theoretical results are about 
a factor of 5-10 smaller than the experimental values.  The antikaon yield 
is enhanced by a factor of 4-5 when in-medium kaon and antikaon masses 
are used.  The theoretical results are now in reasonable agreement with 
the experimental data, except at $p_{lab}^{\bar K}$=1.0 GeV/c where they 
are still below the data by about a factor of two. Including 
possible decrease of the pion decay constant in medium, which leads to 
a stronger reduction in antikaon mass, is expected to further increase
the calculated antikaon yield.
 
Subthreshold antikaon production in heavy-ion collisions has also been
measured earlier at Bevalac for Si+Si collisions at 1.55 and 2.1 GeV/nucleon
\cite{LBL}. Theoretical calculations using the QMD model without medium 
effects give results that are also smaller than the experimental data by 
about a factor of 3-4 \cite{LI92B,HUANG92}.  Again, this discrepancy 
is likely explained by the modifications of kaon and antikaon masses 
in medium.
 
\begin{figure}
\epsfig{file=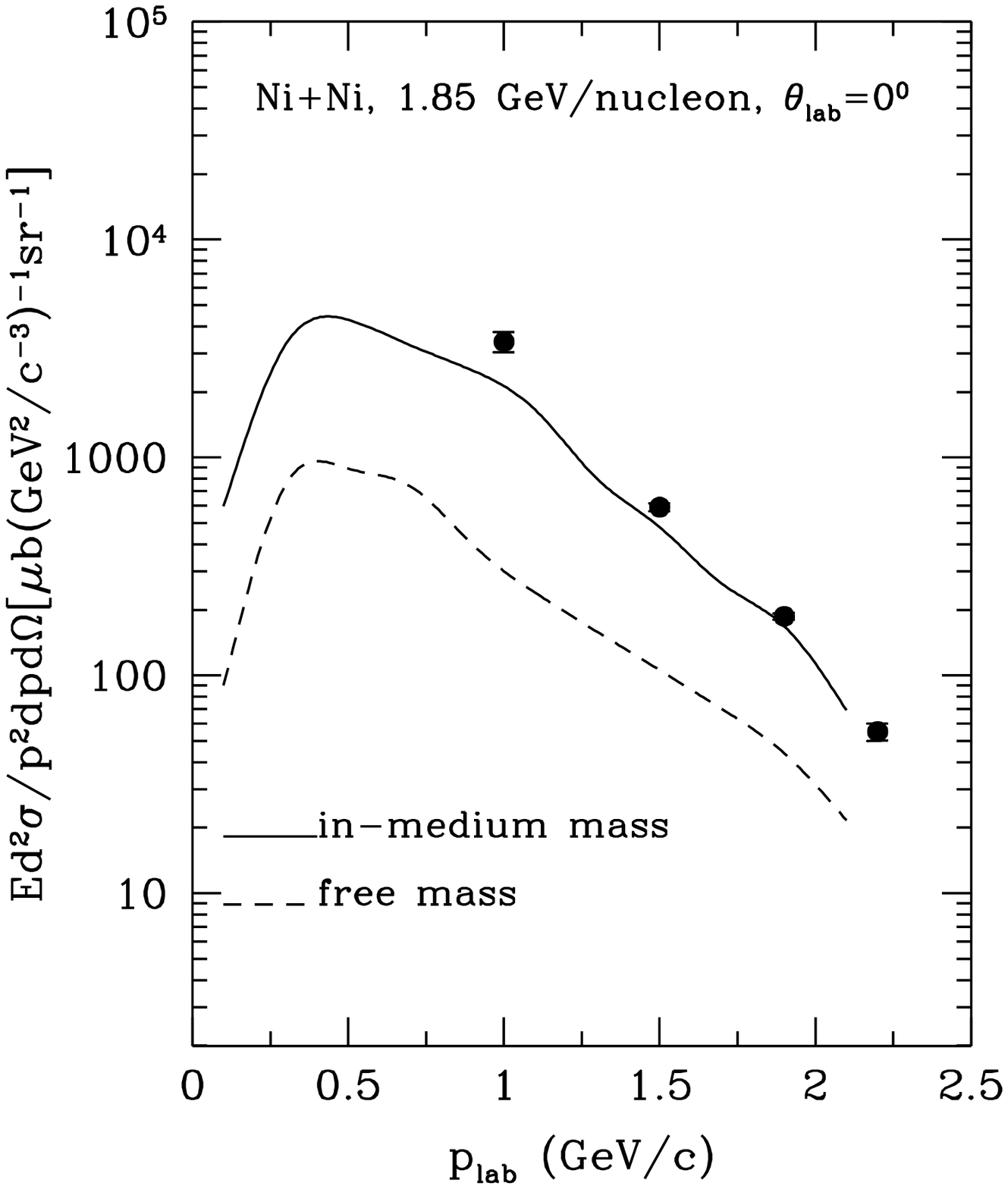,height=4in,width=4in}
\vskip 0.4cm
Fig. 10 ~Antikaon momentum spectra from Ni+Ni collisions at 1.85
GeV/nucleon, obtained with free (dashed) and in-medium (solid) kaon and 
antikaon masses. The experimental data from Ref. \cite{KIEN} are
shown by solid circles. (from Ref. \cite{LI94B})
\end{figure} 

It should be noted that an antikaon can also be produced from the process
$\pi Y\to\bar KN$ between a pion and a hyperon $Y$ produced from the 
reaction $BB\to NYK$. However, with the cross section of Ref. \cite{ko84} 
obtained from the K-matrix analysis of Ref. \cite{DOVER}, it has been 
found that the antikaon yield from $\pi Y\to \bar KN$ in heavy-ion 
collisions is only about 25\% of that from $BB\to NNK\bar K$. This is 
less than that from calculations based on the kinetic model \cite{ko83} 
and the cascade model \cite{barz85}, where more than 50\% of the antikaons 
are from this process. These earlier calculations are, however, less 
accurate than the one based on the transport model as they do not 
treat properly the collision dynamics and antikaon absorption. Similarly, 
antikaon production from $\pi B\rightarrow N K {\bar K}$ has been found 
to be much smaller than that from BB interactions.  Furthermore, the 
reaction $NY\to NNK\bar K$ is unimportant due to the small cross section 
\cite{ko84}. We note that these estimates were all based on calculations
without including medium effects and need to be further investigated. 
 
Antikaon production from $\phi$ decay could also be included in the
transport model calculation. Unfortunately, the elementary $\phi$ 
production cross section in nucleon-nucleon interaction is only
scarcely available at high beam momenta. At a  beam momentum of 10.0 GeV/c,  
the $\phi$ production cross section is 1.0$\pm$0.1 $\mu b$ \cite{PHI} 
in proton-proton interaction. This is to be compared with the antikaon 
production cross section of 33$\pm$16 $\mu b$ at the same beam momentum 
\cite{ANTIK}. Also, recent experiments at the AGS indicate that, at an 
incident energy of 14.6 GeV/c per nucleon, the ratio of $\phi$ to $K^-$
in a Si+Au collision is about 12\% \cite{AGS}. We expect an even smaller
ratio in heavy-ion collisions at beam energies of about 1-2 GeV/nucleon, 
and the contribution from phi decay to subthreshold antikaon production
can thus be neglected \cite{CHUNG96}.
 
\subsubsection{subthreshold antiproton production}
 
With a threshold at 5.6 GeV in nucleon-nucleon interaction in free space, 
antiproton production in nucleus-nucleus collisions at energies of a 
few GeV/nucleon is clearly the most extreme subthreshold process in 
particle production from heavy-ion collisions. The first observation 
of subthreshold antiproton production in proton-nucleus collisions dated 
back to the fifties \cite{CHAM} and sixties \cite{ELIO,DORF}. The 
experiments at the Bevalac \cite{LBL} and the JINR \cite{JINR} in the 
eighties provided the first evidence of subthreshold antiproton production 
in nucleus-nucleus collisions. Recently, new measurements of subthreshold 
antiproton production have been carried out at both KEK \cite{KEK} 
and GSI \cite{KIEN}.
 
The antiproton momentum spectrum in proton-proton interaction in free 
space has been parameterized as
\cite{CHAM,ELIO,DANI}
\begin{eqnarray}\label{antip}
{d\sigma \over dp}\approx
\sigma _{\bar p}(\sqrt s) \frac{105}{8p_{\rm max}}
\left (\frac{p}{p_{\rm max}}\right )^2 
\Big [1-\left (\frac{p}{p_{\rm max}}\right )^2\Big ]^2,
\end{eqnarray}
where the maximum momentum of the antiproton, $p_{\rm max}$, is related 
to the available center-of-mass energy $\sqrt s$  of the proton pair
by $p_{\rm max}=\sqrt {(s-16m_N^2)(s-4m_N^2)/4s}.$
 
This parameterization is based on phase space arguments and describes 
reasonably well the measured antiproton momentum spectrum in proton-proton 
collisions \cite{CHAM,ELIO,DANI}. The total antiproton production cross
section in Eq. (\ref{antip}) is fitted to the available experimental data, 
i.e., 
\begin{eqnarray}
\sigma _{\bar p}(\sqrt s)\approx 0.012 (\sqrt s-4m_N)^{1.846} ~
{\rm mb}.
\end{eqnarray}
 
The first investigation of subthreshold antiproton production using the 
transport  model has been carried out by Batko {\it et al.} \cite{MOS91}. 
In this study, the $\Delta $(1232) degree of freedom is included and is 
found to play an important role.  Neglecting the antiproton mean 
field and assuming that 90\% of the primordial antiprotons are absorbed, 
Batko {\it et al.} \cite{MOS91} have achieved a reasonable description 
of antiproton production in proton-nucleus collisions, but their results 
for nucleus-nucleus collisions are about a factor of five too small 
compared with the experimental data. A similar calculation has been 
carried out by Huang {\it et al.} \cite{HUANG92} using the QMD. Again, 
the antiproton yield in nucleus-nucleus collisions is less than the 
experimental data by about the same factor as in Ref. \cite{MOS91}. 
In both calculations \cite{HUANG92,MOS91}, antiproton annihilation is 
treated schematically, and the propagation as well as elastic scattering 
of antiprotons in medium are neglected. 
 
\begin{figure}
\epsfig{file=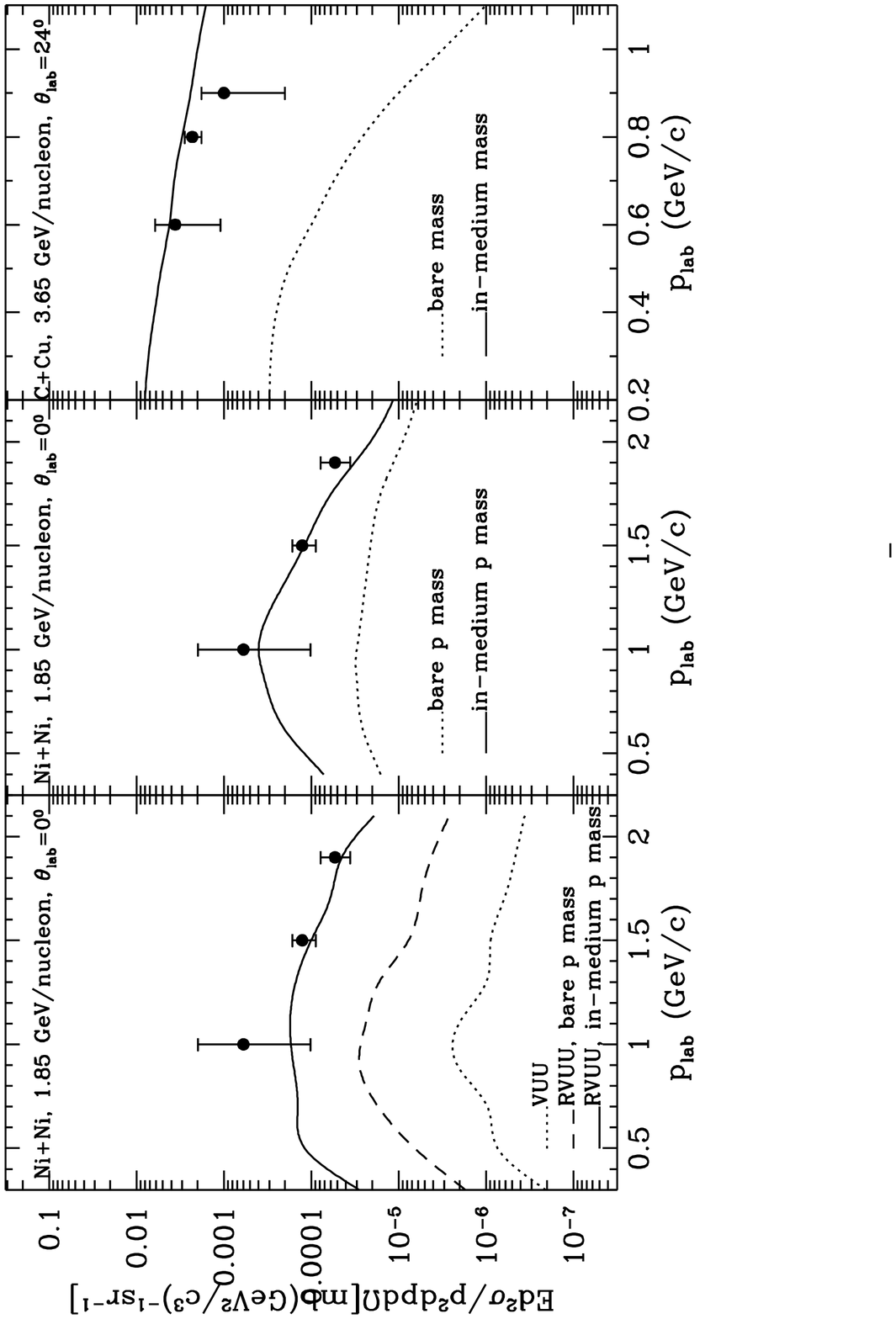,height=4in,width=6in}
\vskip 0.4cm
Fig. 11 ~Antiproton momentum spectra from Ni+Ni collisions 
at 1.85 GeV/nucleon, and C+Cu collisions at 3.85 GeV/nucleon.
The left, middle, and right panels are from Refs. \cite{antip},
\cite{MOS94}, and \cite{FAE94}, respectively. 
The experimental data from Ref. \cite{KIEN} for Ni+Ni collisions
and from Ref. \cite{JINR} for C+Cu collisions are shown 
by solid circles.
\end{figure} 

The importance of medium effects can be seen from the significant
decrease of the Q-value shown in Fig. 6 for the reaction $NN\to NNN\bar 
N$ in nuclear medium as a result of the attractive scalar potential.
This effect has been included in a number of studies based on transport
models.  In Fig. 11, theoretical results from these calculations for
the antiproton differential cross section in Ni+Ni collisions at 
1.85 GeV/nucleon and C+Cu collisions at 3.65 GeV/nucleon are compared 
with the experimental data from both GSI \cite{KIEN} and Dubna \cite{JINR}. 
In the left panel, the results from Ref. \cite{antip} based on the 
relativistic transport model are shown. The antiproton mean-field potential 
is obtained from the nucleon one by G parity and is thus very attractive.  
It is indeed seen that the experimental data can only be accounted for 
if in-medium baryon masses (solid curve) are used. The results with 
dropping only the nucleon mass but not the antiproton mass are shown 
by the dashed curve and are a factor of 8 smaller.  When free baryon 
masses are used in the calculation, the antiproton yield (dotted curve) 
is further reduced by a factor of 12. Overall, the antiproton production 
cross section is enhanced by about two orders of magnitude due to the 
reduction of baryon masses in medium. Similar results \cite{LI94C} have 
been obtained for antiproton production from Si+Si at 2.1 GeV/nucleon 
\cite{LBL}.
 
The results in the middle panel of Fig. 11 are from Ref. \cite{MOS94}
based also on the relativistic transport model. The results shown by 
the solid curve for the case with an in-medium antiproton mass are in good 
agreement with the data, while that with a bare antiproton mass (but still 
an in-medium nucleon mass) are below the data by about a factor of 5-10. 
We note that the theoretical results of Refs. \cite{antip,MOS94} agree 
with each other within a factor of two, and the antiproton potential 
at normal nuclear matter density extracted from these calculations is 
in the range of -150 to -250 MeV.
 
The results in the right panel of Fig. 11 are from Ref. \cite{FAE94}
based on the nonrelativistic QMD model. Medium effects on the produced 
nucleon and antinucleon masses are taken from the Nambu$-$Jona-Lasinio model.
It is again seen that without medium effects the theoretical results
are much below the experimental data.  These studies thus show that 
in order to describe the antiproton data from heavy-ion collisions 
at subthreshold energies it is necessary to include the reduction of 
both nucleon and antinucleon masses in nuclear medium.  Even at AGS 
energies, which are above the antiproton production threshold in NN 
interaction, a recent calculation using the RQMD shows that medium 
modifications of the antiproton properties are important for a 
quantitative description of the experimental data \cite{RQMD95}.
 
On the other hand, there have been suggestions based on schematic
considerations that the contributions from quasi-coherent multi-particle 
collisions \cite{DANI} and meson-meson interactions (e.g. $\rho\rho
\rightarrow p\overline p$) \cite{ko2} might be significant.  More accurate 
estimates of these contributions using the transport model have not 
been done and are needed.
 
\subsubsection{soft pions}
 
Transport models have also been used to study pion production from 
heavy-ion collisions. Of particular interest is the experimental
observation of enhanced low (transverse) momentum pions in these
collisions at various energies \cite{SPE,TAPS,LOVE91,STROB88}. Different 
mechanisms have been proposed to explain this phenomenon, including the 
collective flow \cite{LEE89}, the resonance decay at freeze-out 
\cite{BALI91,SOLL90,BARZ91,BROWN91}, the finite pion chemical potential
at freeze-out \cite{RUUS90,YANG91}, and the medium modifications of the 
pion dispersion relation \cite{XIONG93,mosel,SHUR90,SHUR92}. In this 
subsection, we discuss mainly the results for heavy-ion collisions 
at Bevalac and SIS energies.
 
Several nonrelativistic BUU calculations have been carried out
by different groups for the pion spectra from heavy-ion collisions at
1-2 GeV/nucleon \cite{XIONG93,BALI91,mosel,DAN95}. The results of 
Ref. \cite{XIONG93} for the pion kinetic energy spectra from La+La
collisions at 1.35 GeV/nucleon are shown in the left panel of Fig. 12.
The dashed histogram is obtained using the free-space pion dispersion 
relation, and for low energy pions it differs significantly from the 
experimental data shown by solid circles with error bars \cite{SPE}. 
With a softened pion in-medium dispersion relation described in Section 
II.B.1., there is an enhancement of low energy pions as shown by 
the solid histogram. This enhancement is due to several mechanisms. 
From delta decays in dense medium, quasipions in the pion branch 
generally carry less energy than free pions.  With their increased 
widths, it is more probable to form small mass deltas, so that more soft 
pions are created from the decay of these deltas. Also, as quasipions 
propagate from a more dense region to a less dense region, as most 
likely happened during heavy-ion collisions, their momenta should decrease 
since the p-wave interaction reduces. Enhanced soft pions are also seen 
in more recent experiments at SIS \cite{TAPS}. Calculations within 
transport models with a free pion dispersion relation fail to explain 
this enhancement \cite{mosel,DAN95}. A similar enhancement of low 
energy etas has also been observed \cite{BERG94}, and this may again 
be due to the change of the eta dispersion relation in medium as a 
result of the $N^*(1535)$-hole polarization.
 
\begin{figure}
\epsfig{file=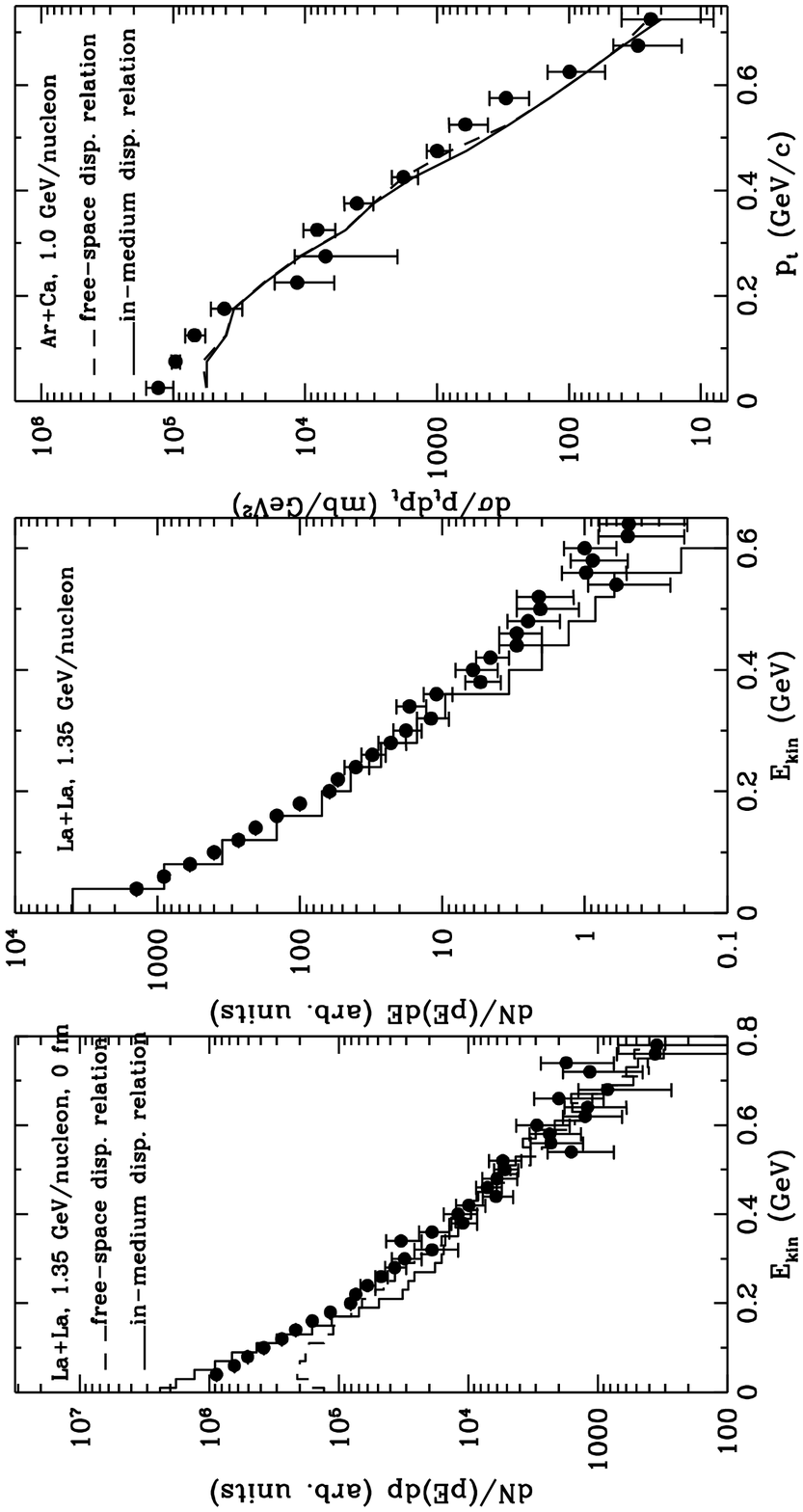,height=4in,width=6in}
\vskip 0.4cm
Fig. 12 ~Pion transverse kinetic energy spectra in La+La collisions
at 1.35 GeV/nucleon and in Ar+Ca collisions at 1 GeV/nucleon. The left, 
middle, and right panels are from Refs. \cite{XIONG93}, \cite{BALI91}, 
and \cite{mosel}, respectively. The experimental data from Ref. 
\cite{SPE} for La+La collisions and from Ref. \cite{TAPS} for Ar+Ca 
collisions are shown by solid circles.
\end{figure} 

It should be mentioned that there are still some controversies regarding
transport model calculations of pion spectra in heavy-ion collisions. 
For example, in Ref. \cite{BALI91} enhanced soft pions can also be 
obtained by treating pions as free particles. This is shown in the 
middle panel of Fig. 12 for La+La collisions at 1.35 GeV. The 
low-momentum pions were found to come mainly from the decay of delta 
resonances at freeze-out. However, this was not confirmed in similar 
transport model calculations by other groups \cite{XIONG93,mosel,DAN95}.
 
Furthermore, the effects of a modified pion dispersion relation in medium
on pion spectra do not all agree. In Ref. \cite{mosel}, this effect is
found to be very small due to a different treatment of the in-medium
pion dispersion relation in the transport model. This is shown 
in the right panel of Fig. 12 where the solid and dashed curves
are theoretical results obtained with the in-medium and the free-space
pion dispersion relation, respectively. Therefore, further work on 
quasipions in transport models are required, and this has recently 
been taken up in Ref. \cite{hel}.
 
Medium effects on pion (also kaon) momentum spectra in hot hadronic
matter have been investigated in Refs. \cite{SHUR90,SHUR92}
based on the `optical potential' constructed from the forward
scattering amplitude. The attractive pion collective potential
was proposed in Ref. \cite{SHUR90} as a possible explanation for
the observed enhancement of low transverse momentum pions (the
soft pion puzzle) in heavy-ion collisions at SPS energies \cite{STROB88}, 
in much the same way the attractive pion potential leads to enhanced 
soft pions in heavy-ion collisions at SIS energies. However, it was 
shown in Ref. \cite{KOCH93} that in order for this attractive potential 
to be effective in cooling down pions, the colliding system has to 
expand slower than the speed of these low-momentum pions. Since at 
SPS energies, the system and hence the source of the pion potential
is made of pions as well, it is unlikely that the above condition 
can be satisfied.
 
\subsubsection{collective flows}
 
Besides particle spectra, the collective motion of particles and fragments, 
both in and out-of the reaction plane, has been extensively studied in 
heavy-ion collisions at SIS energies
\cite{LI95C,FOPI,DAN93,ZHANG94,GUT89,GUT90,EOS,DAN85,LIBA91,BASS93,LIBA94,GRE94}.
It has been shown that the proton (and fragment) flow is sensitive to both
the density (related to the nuclear equation of state) and momentum 
dependence of the nuclear potential \cite{DAN93}, as well as the in-medium 
NN cross sections \cite{XU91,KROF92,ZHOU94}. Pion flow has also been 
identified in heavy-ion collisions, and is seen to undergo a transition 
from flow at small impact parameters to antiflow at large impact 
parameters as a result of the shadowing effects from the spectator
nucleons \cite{DAN95,BASS93,LIBA94}. Antiproton and antikaon flows
in heavy-ion collisions have been seen in both the RQMD \cite{GRE94} 
and the relativistic hadronic cascade (ARC) \cite{kahana} calculations. 
Because of their large annihilation cross section, both antiproton and 
antikaon flows are found to be opposite to that of nucleons. Kaon flow 
has recently been proposed as another observable for studying the kaon 
properties in dense matter \cite{LI95C}, and is being studied 
experimentally by both the FOPI \cite{FOPI} and the EOS \cite{EOS}
collaboration in heavy-ion collisions at a few GeV/nucleon.
 
\begin{figure}
\epsfig{file=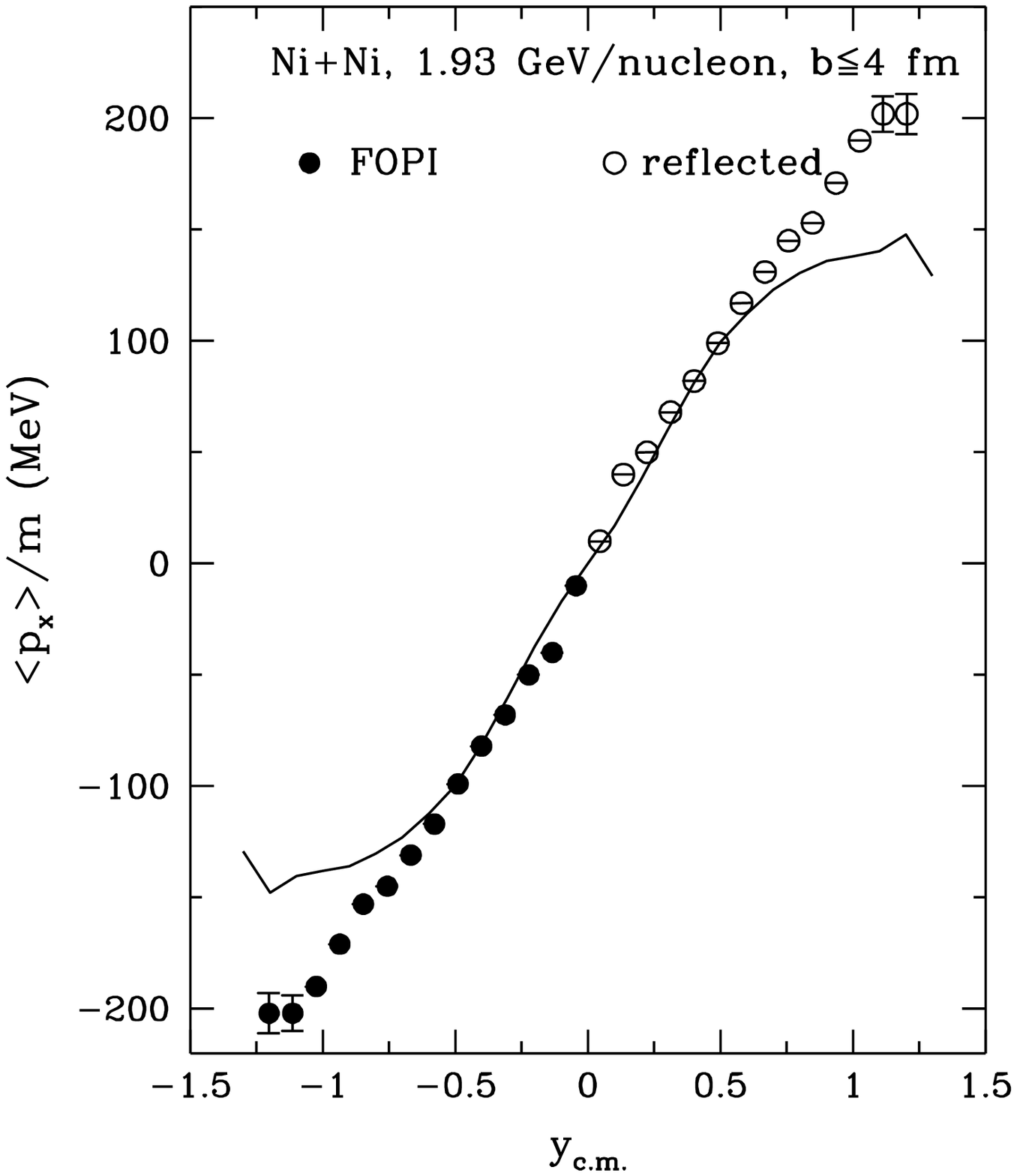,height=4in,width=4in}
\vskip 0.4cm
Fig. 13 ~Proton average in-plane transverse momentum from Ni+Ni 
collisions at 1.93 GeV/nucleon. The experimental data from the FOPI 
collaboration \cite{FOPI} are shown by circles.
\end{figure}

In-plane particle flow is usually shown as the average in-plane 
transverse momentum $\langle p_x \rangle$ as a function of rapidity 
$y$. The results for proton flow in Ni+Ni collisions at 1.93 GeV/nucleon 
using the relativistic transport model with a soft EOS are shown 
in Fig. 13. They are obtained after taking the impact-parameter-weighted 
average over b$\le$ 4 fm, corresponding approximately to the
centrality selection in experiments by the FOPI collaboration
\cite{FOPI}. Both the experimental data, shown in the figure
by solid circles, and the theoretical results include a low transverse
momentum cut of $p_t>0.5m_N$. It is seen that the theoretical 
predictions are in good agreement with the data except at 
the projectile and target rapidities where there are
some differences, which are not understood at present.
 
To extract nuclear equation of state at high densities, proton flow has 
also been studied extensively in nonrelativistic transport models with 
the inclusion of a momentum-dependence in the nuclear potential 
\cite{DAN93,ZHANG94}. The results of Refs. \cite{DAN93} and \cite{ZHANG94} 
are shown in the left and right panels of Fig. 14, respectively. The 
flow parameter $F$ in these work is defined by the average transverse 
velocity $\langle p_x \rangle /m$ at $y=0$. The momentum-dependence 
in the nuclear mean field was found to be important in reproducing 
correctly the dependence of the flow parameter on the incident energy 
and the colliding system. It is seen that the soft EOS with a 
compression modulus of about 200 MeV gives a good fit to the 
experimental data from both Nb+Nb and Au+Au collisions.
 
In Fig. 15, pion flow in Au+Au collisions at 1 GeV/nucleon and 
impact parameters b =3, 6, 9 fm is shown. In central collisions 
(dotted curve), pions are seen to follow the flow of nucleons, while 
in mid-central (dashed curve) and peripheral (solid curve) collisions,
the pion flow direction is opposite to that of nucleons. This transition 
from flow in central collisions to antiflow in mid-central and peripheral 
collisions has also been predicted in Refs. \cite{DAN95,LIBA94}. The
`anticorrelation'  of pions to nucleons at large impact parameters
is mainly due to the rescattering and absorption of pions by the
spectator nucleons.

\begin{figure}
\epsfig{file=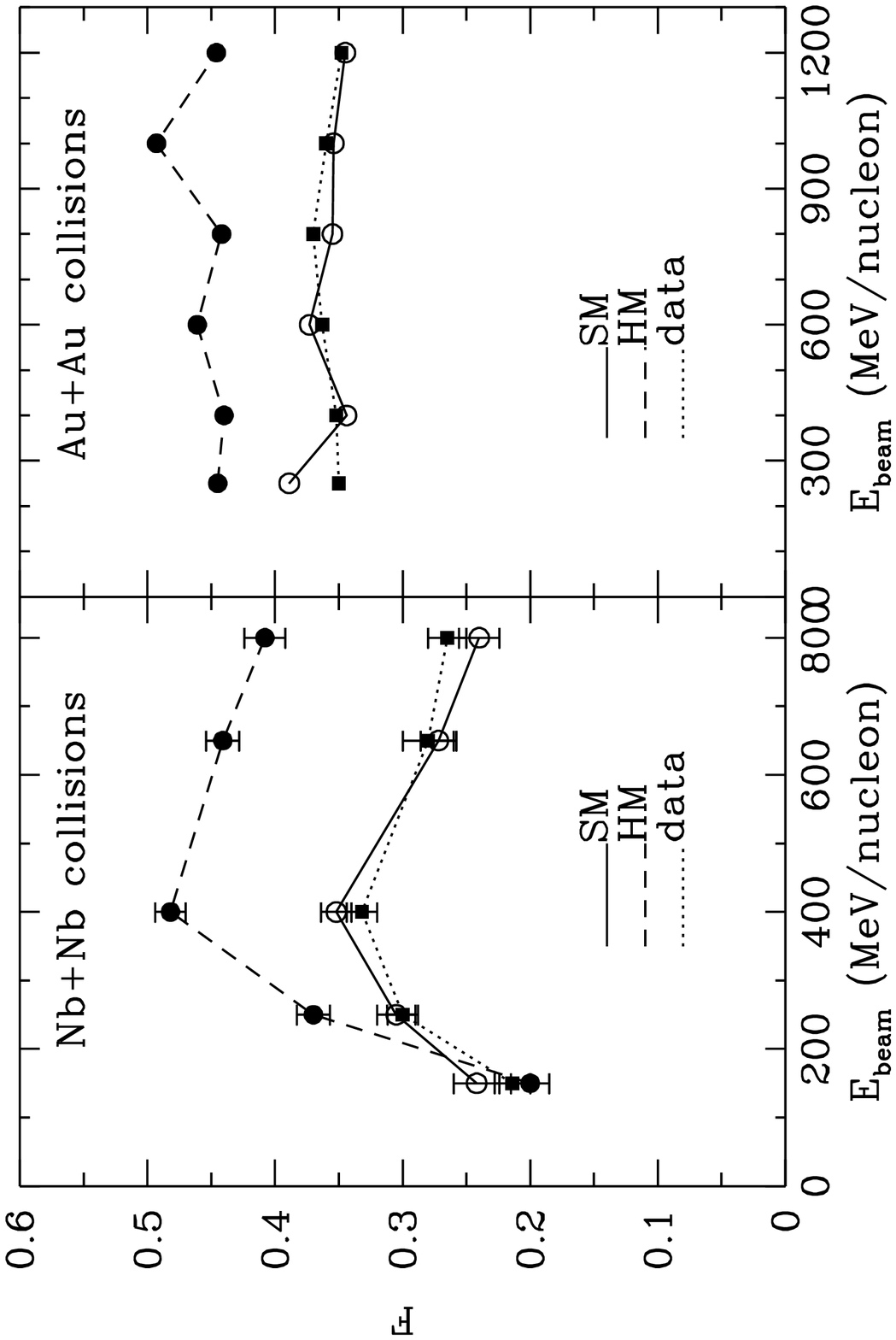,height=4in,width=5in}
\vskip 0.4cm
Fig. 14 ~Proton flow parameter $F$ as a function of the incident
energy for Nb+Nb (from Ref. \cite{DAN93}) and Au+Au collisions 
(from Ref. \cite{ZHANG94})
\end{figure}

Fig. 16 shows the average transverse momentum of kaons in the reaction 
plane as a function of the center-of-mass rapidity $y_{cm}$ in Ni+Ni 
collisions at 1.93 GeV/nucleon and for impact parameters b$\leq$4 fm. 
The dotted curve, corresponding to the case without kaon potential, shows
that kaons flow in the same direction as nucleons but with a smaller flow
velocity. The results with only the kaon vector potential are shown by 
the dashed curve.  The kaon flow in this case is opposite to that of 
nucleons, i.e., the appearance of an `antiflow' of kaons with respect 
to nucleons. With a weak repulsion due to both scalar and vector 
potentials, it is seen that the kaon flow, shown by the solid curve, 
is in the same direction as that of nucleons but its strength is 
significantly reduced.  Thus, the repulsive kaon potential tends to
make kaons flow away from nucleons, and its effect depends sensitively 
on the strength of the kaon potential. It is therefore possible to 
study the kaon potential in nuclear medium by measuring kaon flow in 
heavy-ion collisions. Preliminary data from the FOPI collaboration 
\cite{FOPI}, shown by the solid circles with error bars, seem to be 
consistent with the existence of both scalar and vector potentials 
for a kaon in nuclear medium.

\begin{figure}
\epsfig{file=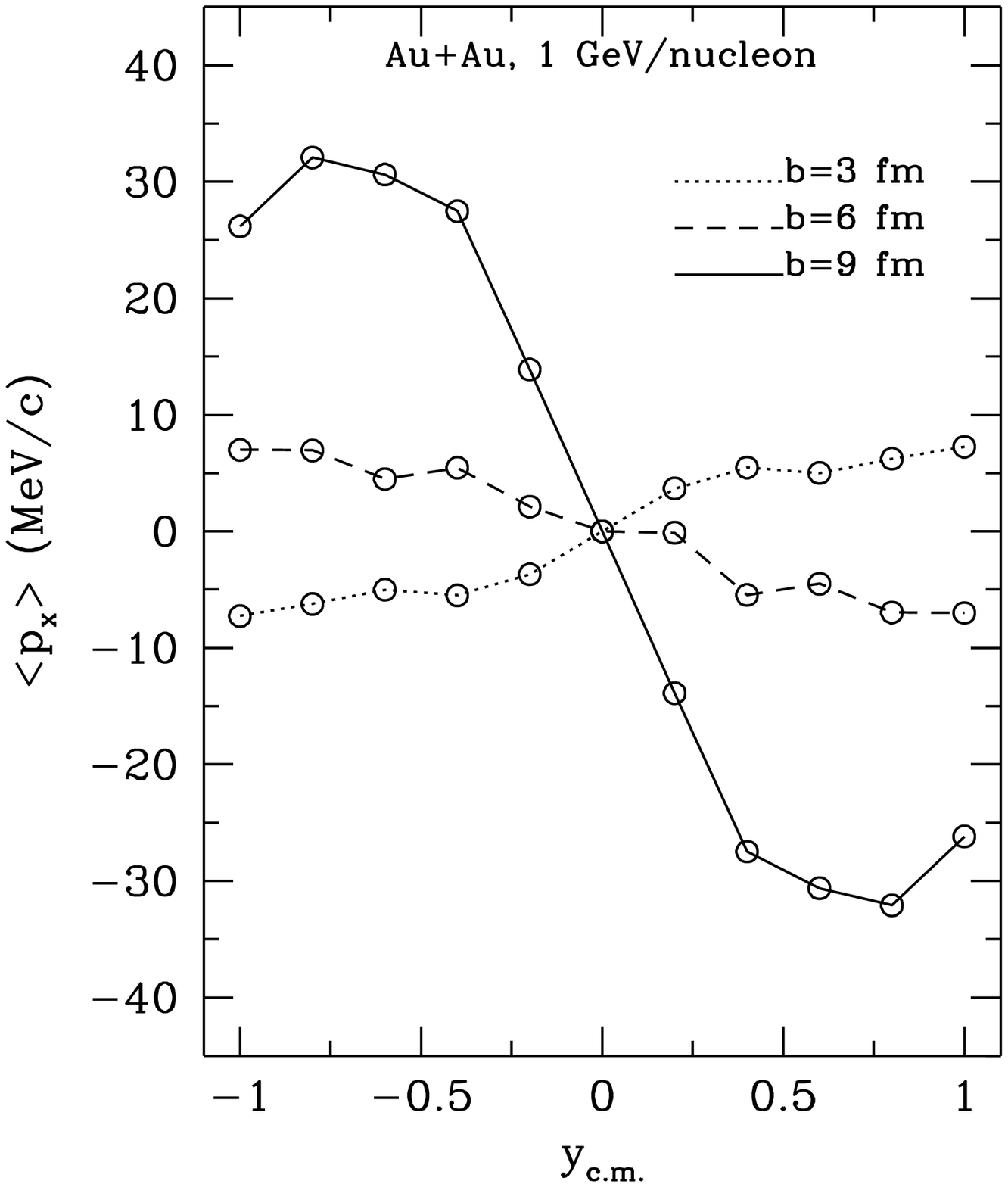,height=4in,width=4in}
\vskip 0.4cm
Fig. 15 ~Pion average in-plane transverse momentum as a function 
of rapidity from Au+Au collisions at 1 GeV/nucleon and impact parameters 
b=3 (dotted), 6 (dashed), and 9 (solid) fm. (from Ref. \cite{LI95C})
\end{figure}

With the confirmation of potential effects on kaon flow in heavy-ion
collisions, it is interesting to note that the attractive antikaon
and antiproton potentials in nuclear medium are expected to change 
the antiflow of these particles caused by their shadowing from nucleons 
as a result of large absorption cross sections. Indeed, it has been 
recently shown that including the attractive potential given by the chiral 
Lagrangian, the strong antikaon antiflow is seen to change into a 
weak flow \cite{LIKO}. Similar medium effects have been shown for 
pion flow in the QMD calculations with an in-medium pion dispersion 
relation \cite{FAE96}.

\begin{figure}
\epsfig{file=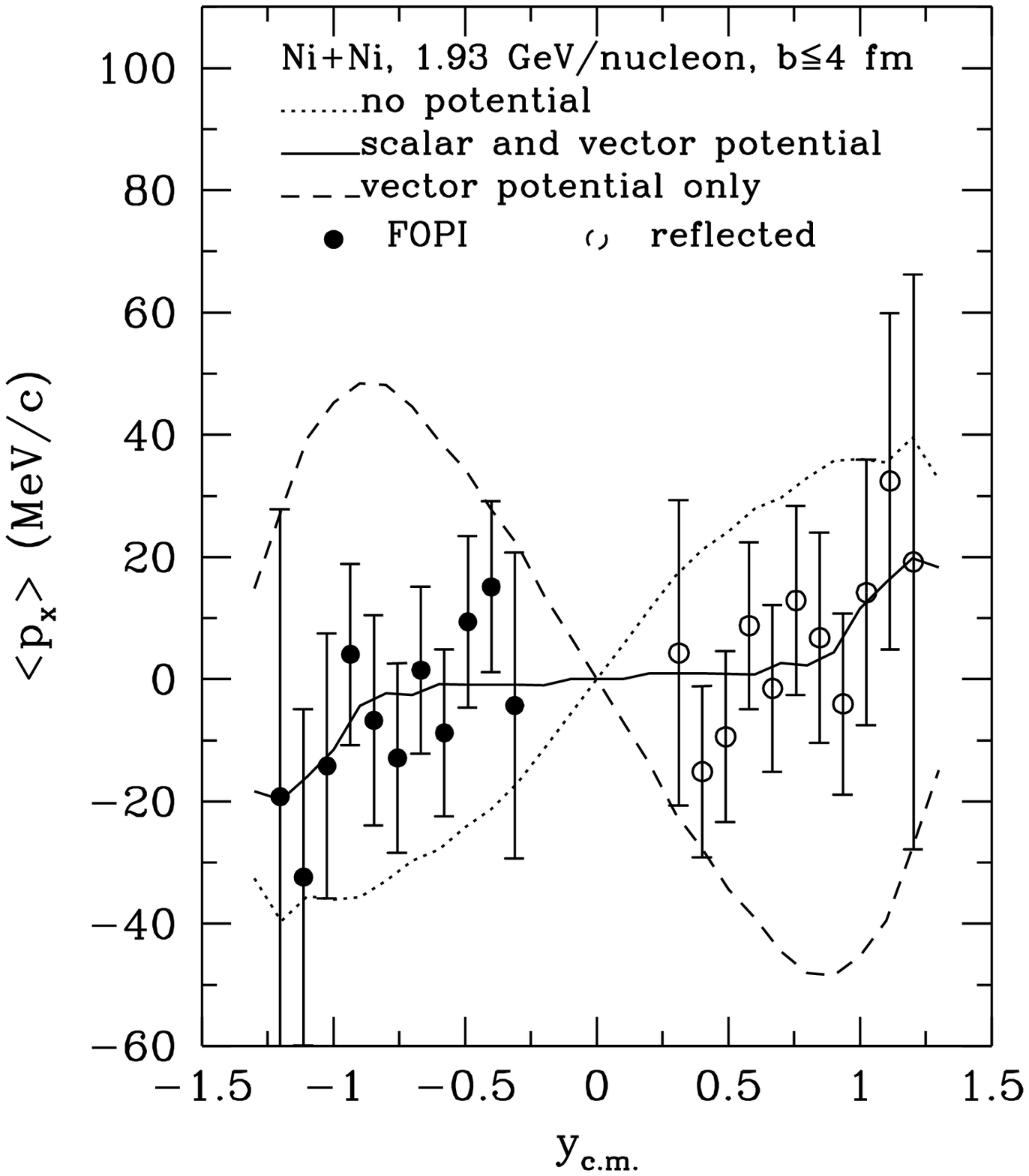,height=4in,width=4in}
\vskip 0.4cm
Fig. 16 ~Kaon average in-plane transverse momentum as a function
of rapidity from Ni+Ni collisions at 1.93 GeV/nucleon. The experimental 
data from the FOPI collaboration \cite{FOPI} are shown by circles.
\end{figure} 
 
\subsection{heavy-ion collisions at AGS and SPS energies}
 
Ultra-relativistic heavy-ion collisions have been carried out
at AGS with a beam energy of about 10 GeV/nucleon and
at SPS with an incident energy of about 200 GeV/nucleon. The
main motivation for carrying out these experiments at increasingly 
higher energies is to create and study in the laboratory the
quark-gluon plasma, which is believed to exist in the early evolution 
of the universe at about one micro second after the Big Bang. At these 
energies, the reaction dynamics becomes much more complex than that at 
SIS energies, as many hadrons, such as eta, kaon, antikaon, rho and 
omega as well as other higher baryon resonances, are abundantly produced, 
and we thus need to include them explicitly. Since the interactions 
among these hadrons are not well understood, various assumptions have 
been introduced to relate the unknown ones to the known ones. Nevertheless, 
cascade-type models have been developed and seem to describe the global 
features of heavy-ion collisions at both AGS and SPS energies reasonably 
well \cite{BALI,ARC,RQMD}. In this subsection, we shall discuss some 
selected topics such as the strangeness enhancement, the difference in 
the slope parameters of $K^+$ and $K^-$ momentum spectra, and the 
enhancement of very low momentum kaons.
  
\subsubsection{Kaon enhancement}
 
In heavy-ion collisions at AGS energies, the $K^+/\pi^+$ ratio has 
been found to be enhanced relative to that from proton-nucleus collisions 
\cite{AGS1,AGS2}.  Different mechanisms have been proposed to explain 
this enhancement. In RQMD \cite{RQMD,SORGE91}, it is attributed to 
meson-baryon interactions in the collision. In ARC \cite{ARC}, it is 
due to interactions between baryon resonances created in the early 
stage of the collision. In a most recent relativistic transport model 
(ART) \cite{BALI}, significant contributions are found not only from 
the baryon-baryon and meson-baryon interactions but also from the
meson-meson interaction. On the other hand, it has been shown 
in Ref. \cite{braun} that the measured kaon yield can be explained 
by a simple fireball model at both thermal and chemical equilibrium.  
Since all models predict a more or less equilibrated system at freeze 
out, it is thus difficult to verify from the kaon yield which models 
correctly describe the kaon production mechanism. Furthermore, in 
these models the energy density reached in the early stage of the 
collision when most kaons are produced have been found to exceed the 
one believed to be for a quark-gluon matter. The conclusions drawn from 
these models are thus expected to change when a more proper treatment 
of the initial stage is included.

\begin{figure}
\epsfig{file=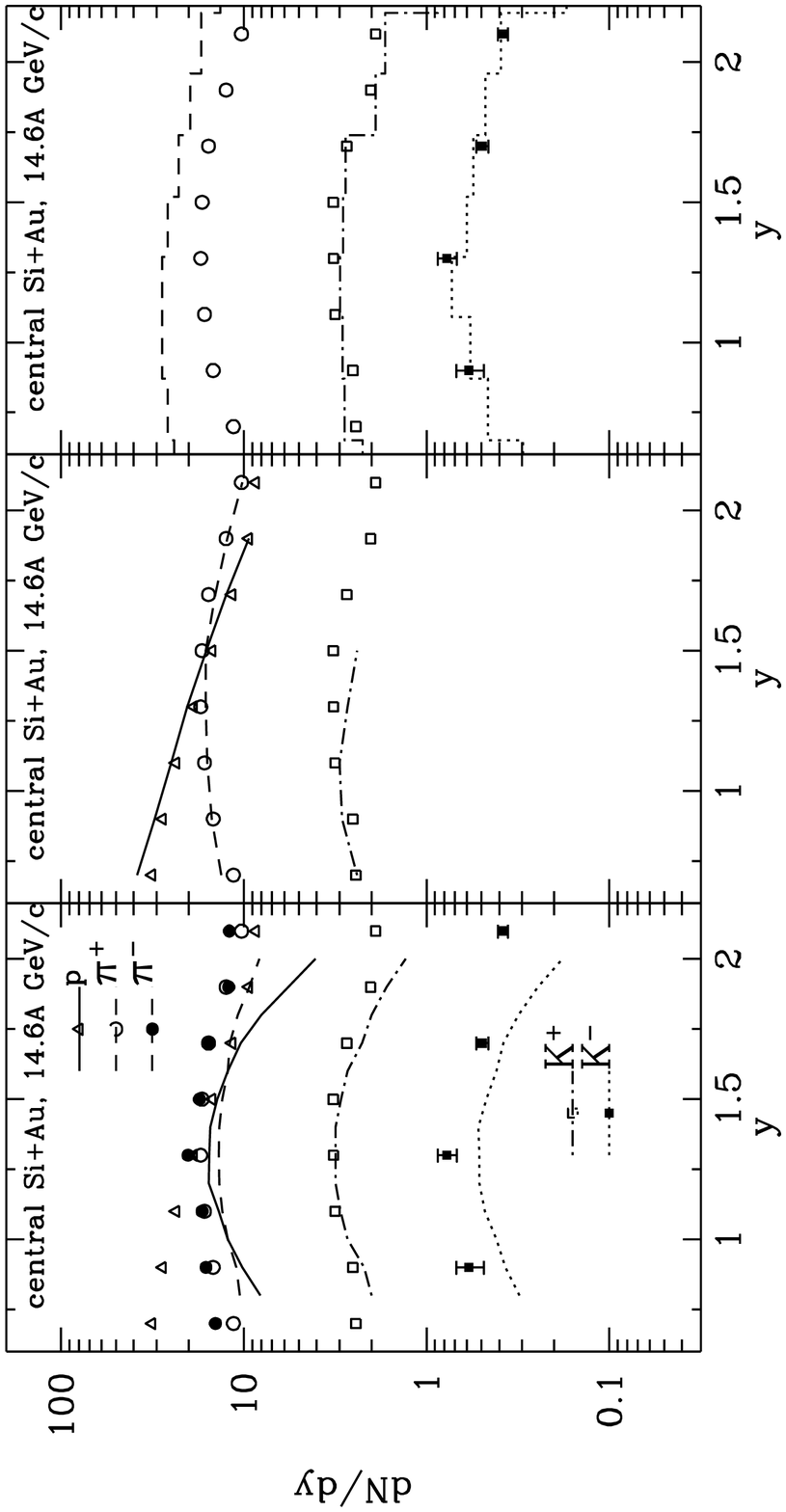,height=4in,width=6in}
\vskip 0.4cm
Fig. 17 ~Particle rapidity distributions in central Si+Au collision
at 14.6 GeV/c/nucleon. The left, middle, and right panels are from
Refs. \cite{FANG93A}, \cite{ARC}, and \cite{SORGE91}, respectively. 
The experimental data from Ref. \cite{AGS2} are also shown.
\end{figure}
 
To understand the enhanced kaon production, the relativistic transport 
model has been used to study the expansion stage of heavy-ion 
collisions at these energies by assuming that a fireball is formed 
in the initial stage \cite{KO91,FANG93A}. The fireball then expands, 
and kaons and antikaons are produced from baryon-baryon, meson-baryon, 
and meson-meson interactions.  Both kaons and antikaons with in-medium 
masses propagate through the hadronic matter under the influence of 
the mean-field potentials and undergo collisions with both nucleons 
and pions. Furthermore, antikaons can be destroyed via the reaction 
$\bar KN\to\pi Y$.  
 
In the left panel of Fig. 17, the particle rapidity distributions from 
Ref. \cite{FANG93A} are compared with the experimental data from Ref. 
\cite{AGS2}. One sees that good agreements with the experimental data 
are obtained for pions, kaons, and antikaons.  The failure of the 
calculated proton distribution at smaller rapidities is due to the 
neglect of protons from the target spectator in the fireball approach. 
In this model study, a significant number of kaons are produced from 
meson-meson interactions which become important as a result of reduced 
total kaon-antikaon masses in dense matter \cite{BRO92}. 
 
In the middle panel of Fig. 17, the results from the ARC model of Ref. 
\cite{ARC} are compared with the experimental data for proton, $\pi^+$ 
and $K^+$. Medium effects are neglected in this cascade-type calculation. 
Kaons are found to be mainly (about 70\%) produced from the interactions 
between baryon resonances. The results from the RQMD calculation 
\cite{SORGE91} for $\pi^+$, $K^+$, and $K^-$ are shown in the right panel
of Fig. 17.  Again, the enhanced kaon production can be well described. 
In the RQMD, kaon production from baryon-meson collisions is found to be 
important. Similar results have been obtained in the ART calculation of 
Ref. \cite{BALI}. The successful description of the kaon data by 
cascade-type calculations, however, does not rule out the mechanism 
proposed in Ref. \cite{KO91}. If the mass of $K{\bar K}$ decreases as 
a result of chiral symmetry restoration, more kaons are produced,
but they will be destroyed by reverse processes to maintain chemical
equilibrium. One thus needs to look at other observables, such as the
phi meson yield as it is sensitive to the change of $K{\bar K}$ energy
as a result of the small difference, about 30 MeV, between the phi meson
mass and the sum of $K$ and $\bar K$ masses. The dilepton spectra from 
kaon-antikaon annihilation will also be useful, as it shows directly 
the invariant mass of the kaon-antikaon pair through the phi meson peak.
 
\subsubsection{Difference in the slope parameters of kaon and
antikaon transverse mass spectra}

In Ref. \cite{FANG93A}, the transverse mass spectra of particles have also
been calculated in the fireball model, and they are shown in Fig. 18. 
All particles have essentially exponential distributions. The slope 
parameters for kaons and antikaons show the difference expected from 
mean-field effects, i.e. the effective temperature of antikaons is 
lower than that of kaons. This can be understood as follows:  Consider
antikaons with high initial energies in the fireball that is formed in 
the collision. They move relatively fast and escape thus from the 
fireball while its size has increased only slightly.  But these antikaons 
must use up much of their kinetic energies to climb out of the deep 
potential well so that the measured kinetic energies are substantially 
smaller.  On the other hand, low energy antikaons stay in the fireball 
and escape later in time when the fireball size is large and the
mean-field potential becomes negligible small. These antikaons therefore 
do not loose much energy. The net effect is that the apparent temperature 
of antikaons after freeze out is lower than the initial temperature.  
For kaons, the potential is only slightly repulsive because of 
cancellation between the attractive scalar field and the repulsive 
vector field. The change of kaon apparent temperature is thus very small.  
If antikaons and kaons have similar initial temperatures, then one expects 
to see a lower final temperature for antikaons than for kaons.

\begin{figure}
\epsfig{file=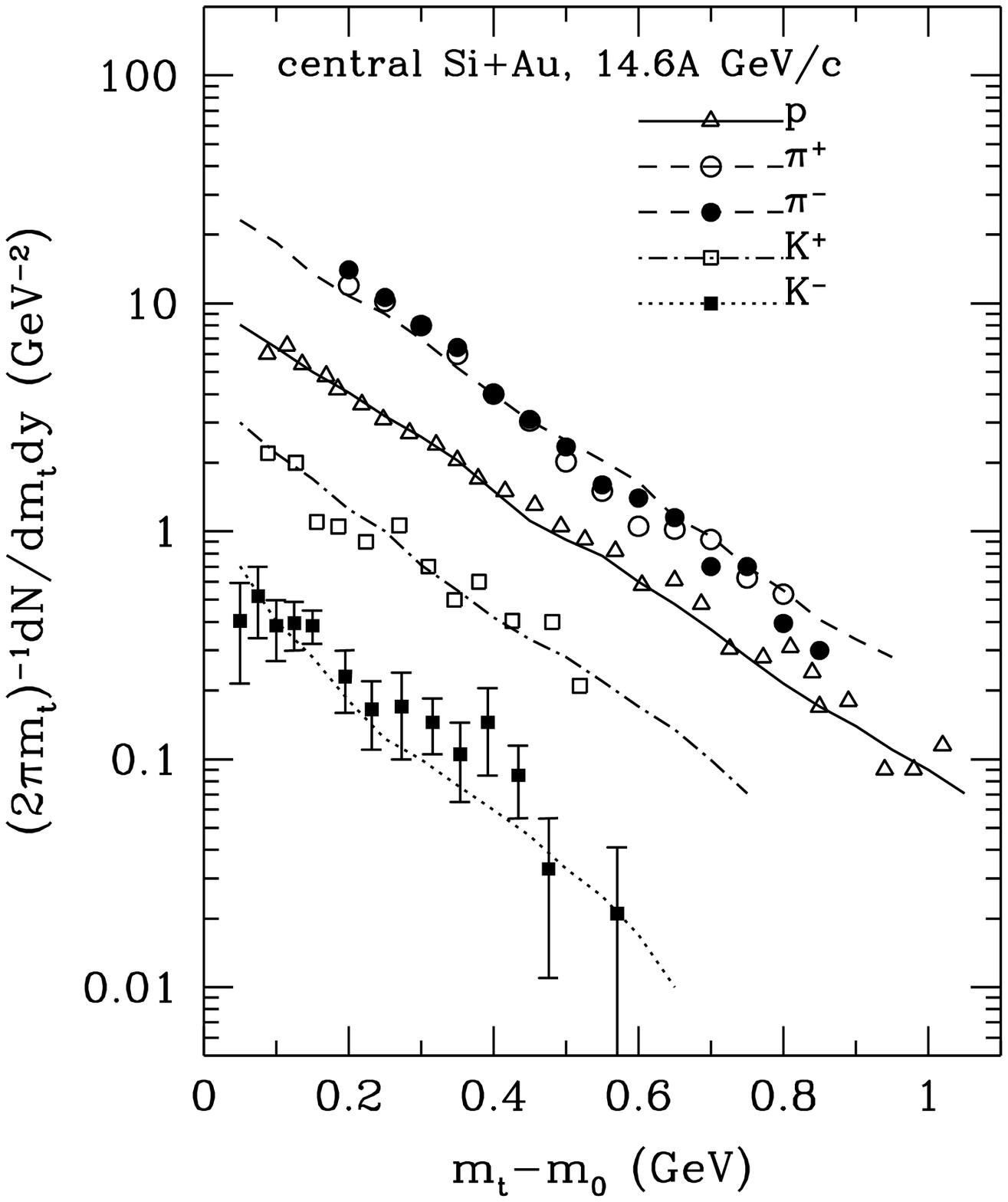,height=4in,width=4in}
\vskip 0.4cm
Fig. 18 ~Particle transverse energy spectra from central Si+Au
collisions. (from Ref. \cite{FANG93A})
\end{figure}
 
Similar effects are seen for antiprotons. The deep attractive mean-field 
potential for antiprotons makes their apparent temperature much lower 
than that for protons \cite{KOCH91}. All these results seem to be 
supported by the experimental observations.
 
We note that the above discussions are for Si+Au collisions. For Au+Au
collisions at AGS energies, the initial compression is expected to be
much more appreciable than in Si+Au collisions, so a larger radial 
expansion may appear and thus reduces the effects of mean-field potential 
on particle spectra \cite{KOCH93}.

\subsubsection{cool kaons}
 
Another interesting preliminary experimental observation in heavy-ion
collisions at AGS energies is the enhancement of low transverse mass ($m_t$)
$K^+$ and $K^-$ in Si+Pb collisions at 14.6 (GeV/c)/nucleon by the E814 
collaboration \cite{stac94}. The spectra of these extremely cold kaons 
can be characterized by an inverse slope parameter (temperature) 
as low as 15 MeV, which is about one order-of-magnitude smaller
than the temperature of normal kaons measured in earlier AGS
experiments \cite{AGS1}.  These low $m_t$ kaons cannot be obtained in 
conventional models such as RQMD \cite{RQMD}, ARC \cite{ARC}, and ART 
\cite{BALI}. Although this experimental result has not been unambiguously 
confirmed by other experiments \cite{ahmad}, it has already generated some 
interesting theoretical explanations.

\begin{figure}
\epsfig{file=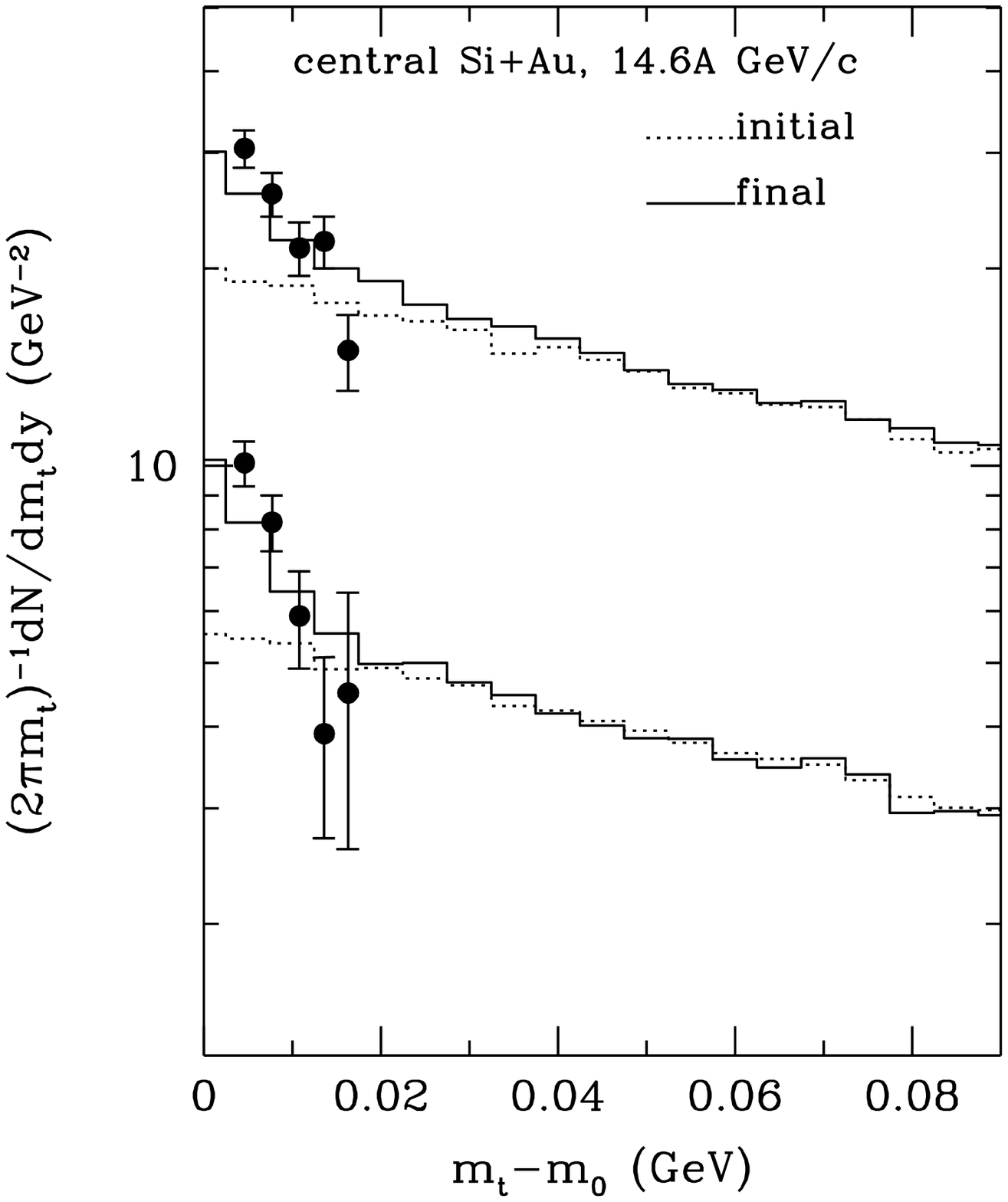,height=4in,width=4in}
\vskip 0.4cm
Fig. 19 ~Initial (dotted) and final (solid) kaon and antikaon
transverse energy spectra. The experimental data are from \cite{stac94}. 
(from Ref. \cite{cool95})
\end{figure} 
 
Based on the Georgi vector limit of the hidden gauge theory, Brown and 
Rho \cite{brow94} have argued that the vector interaction vanishes 
when the chiral symmetry is restored. This seems to be supported by 
calculations based on effective hadronic models for the vector coupling 
constant in medium \cite{xia95}.  If this is indeed the case, then one 
expects that the kaon potential will become attractive at very high 
density and temperature as the scalar attraction becomes dominant.
This idea was first adopted by Koch \cite{koch94}, who considered
the temperature effects, to explain the appearance of these low energy
kaons. In Ref. \cite{cool95}, the density effect is emphasized.
Because of the reduced vector potential, the nuclear equation of 
state becomes softened, leading to the possible formation of a density 
isomer.  As a result, the initial expansion of the nuclear matter is 
relatively slow so that kaons and antikaons can be effectively cooled
by the attractive mean-field potentials. This effect has been studied 
in the relativistic transport model \cite{cool95} by assuming that the 
system is initially in the density isomer. The results are shown in 
Fig. 19 and compared with the preliminary E814 data.  For both kaons 
and antikaons, one obtains in the final transverse mass spectra a cold 
low-$m_t$ component  as observed in the experiment.
 
\subsubsection{strangeness enhancement in heavy-ion collisions at SPS energies}
 
In heavy-ion collisions at SPS energies, enhanced production of
strange particles have also been observed. For example, the antilambda 
yield in the NA35 experiment of S+S at 200 GeV/nucleon \cite{bart90} 
is 1.5 per event and is 115 times greater than that in p-p collisions 
at same energy.   Compared with the 36-fold enhancement of negatively 
charged particles, most of them being negative pions, there is a 
factor of three enhancement of antilambda yield in these collisions.  
This enhancement can be explained if one simply assumes that a 
quark-gluon plasma is formed in the initial stage of the collisions.  
Other explanations have also been proposed.  Aichelin and Werner 
have emphasized the importance of many-body cluster effects \cite{aich91}.
Sorge {\it et al.} \cite{sorg92} have shown that the formation of
a color rope from string excitations can lead to enhanced production 
of antilambdas.  In Ref. \cite{leva92}, this enhancement is explained 
by the lower antilambda production threshold as a result of 
reduced antilambda mass in dense matter.  In a simplified hydrochemical
model, the process $KM\to\bar\Lambda N$, where $M$ denotes either 
a pion or a rho meson, gives an enhanced antilambda production in the
collision. A similar explanation based on the relativistic mean field theory
has been proposed in Ref. \cite{SCHA91}.
 
The reduced phi meson mass in medium has also been shown to lead
to an enhanced phi meson production in CERN heavy-ion collisions 
\cite{sa91,guil91}.  
 
\subsection{dilepton production in heavy-ion collisions}
 
Since dileptons are not subject to the strong final-state interactions 
associated with hadronic observables, they are the most promising probe of
the properties of hot dense matter formed in the initial stage of high 
energy heavy-ion collisions. Dileptons have thus been proposed as 
useful observables for studying medium modifications of the pion 
dispersion relation at both finite density and temperature 
\cite{GALE87,XIA88}, the in-medium properties of vector mesons 
\cite{LI95D,PIS82,KOCH92,RED93,HAT93}, and the phase transition 
from the hadronic matter to the quark-gluon plasma 
\cite{SHUR78,CHIN82,KAJA86,XIA90,ASAK93}.

There are many sources for dilepton production in hadronic matter.
These include the proton-neutron and pion-nucleon bremsstrahlung,
the Dalitz decay of pions, etas, and deltas, the direct decay
of vector mesons, as well as the pion-pion and kaon-antikaon annihilation 
that proceed through the formation of vector mesons. To study the 
in-medium properties of vector meson properties, the pion-pion and 
kaon-antikaon annihilation are most important due to vector dominance 
in the pion and kaon electromagnetic form factors. Neglecting lepton 
masses, the dilepton production cross section from the annihilation 
of two pseudoscalar meson $P$ through a vector meson $V$ is given by
\begin{equation}
\sigma_{PP\to V\to l^+l^-}=a\frac{8\pi\alpha^2k}{3M^3}
\frac{m_V^4}{(M^2-m_V^2)^2+(m_V\Gamma_V)^2}.
\end{equation}
In the above, $M$ is the invariant mass of the lepton pair, $\alpha$ is 
the fine structure constant, and $k$ is the three-momentum of the
pseudoscalar meson in the center-of-mass of the vector meson. 
The mass and width of the vector meson are denoted by $m_V$ and
$\Gamma_V$, respectively. The value of $a$ is 1 and 1/9  for 
$\pi^+\pi^-\to\rho^0\to e^+e^-$ and $K^+K^-\to\phi\to e^+e^-$
($K^0\bar{K^0}\to\phi\to e^+e^-$), respectively.  In this ``form factor" 
approach, the formation time of the dilepton is neglected and the 
effect of the intermediate vector meson is included through the 
electromagnetic form factor. This is a reasonable approach for the 
case without medium effects.
 
With medium-dependent vector meson masses, one needs to adopt the ``dynamical
approach" by including explicitly the formation, propagation and decay of 
the intermediate vector meson so that the change of its properties 
in medium can be included. In this approach, the vector meson formation 
cross section from the meson-meson annihilation is given by
\begin{equation}\label{vpro}
\sigma_{PP\to V}=b\frac{6\pi}{k^2}
\frac{(M\Gamma_V)^2}{(M^2-m_V^2)^2+(m_V\Gamma_V)^2},
\end{equation}
where $b$ is 2 and 1 for $\pi^+\pi^-\to\rho^0$ and $K^+K^-\to\phi$
($K^0\bar{K^0}\to\phi$), respectively.
The decay width of a vector meson of mass $M$ to the dilepton is 
given by \cite{bhaduri}
\begin{equation}\label{vdec}
\Gamma_{V\to l^+l^-}=a\frac{4\pi\alpha^2}
{g_{VPP}^2}\frac{m_V^4}{3M^3},
\end{equation}
where the coupling constants are $g_{\rho\pi\pi}/4\pi\approx 2.9$
and $g_{\phi KK}/4\pi\approx 1.7$, determined by their respective
decay widths. 
 
We note that in the dynamical approach the vector meson mass and 
width in Eqs. (\ref{vpro}) and (\ref{vdec}) are evaluated using the 
local density at the time of its formation and decay, which may be 
different due to the change in density. It has been found in Ref. 
\cite{LI95D} that for vector mesons with large decay widths (e.g., the 
rho meson), the form factor approach gives results qualitatively similar 
to the dynamical approach.  However, for mesons with small decay 
widths (e.g., the phi meson), the dynamic approach is required
as the medium effects are overestimated in the from factor approach.
 
\begin{figure}
\epsfig{file=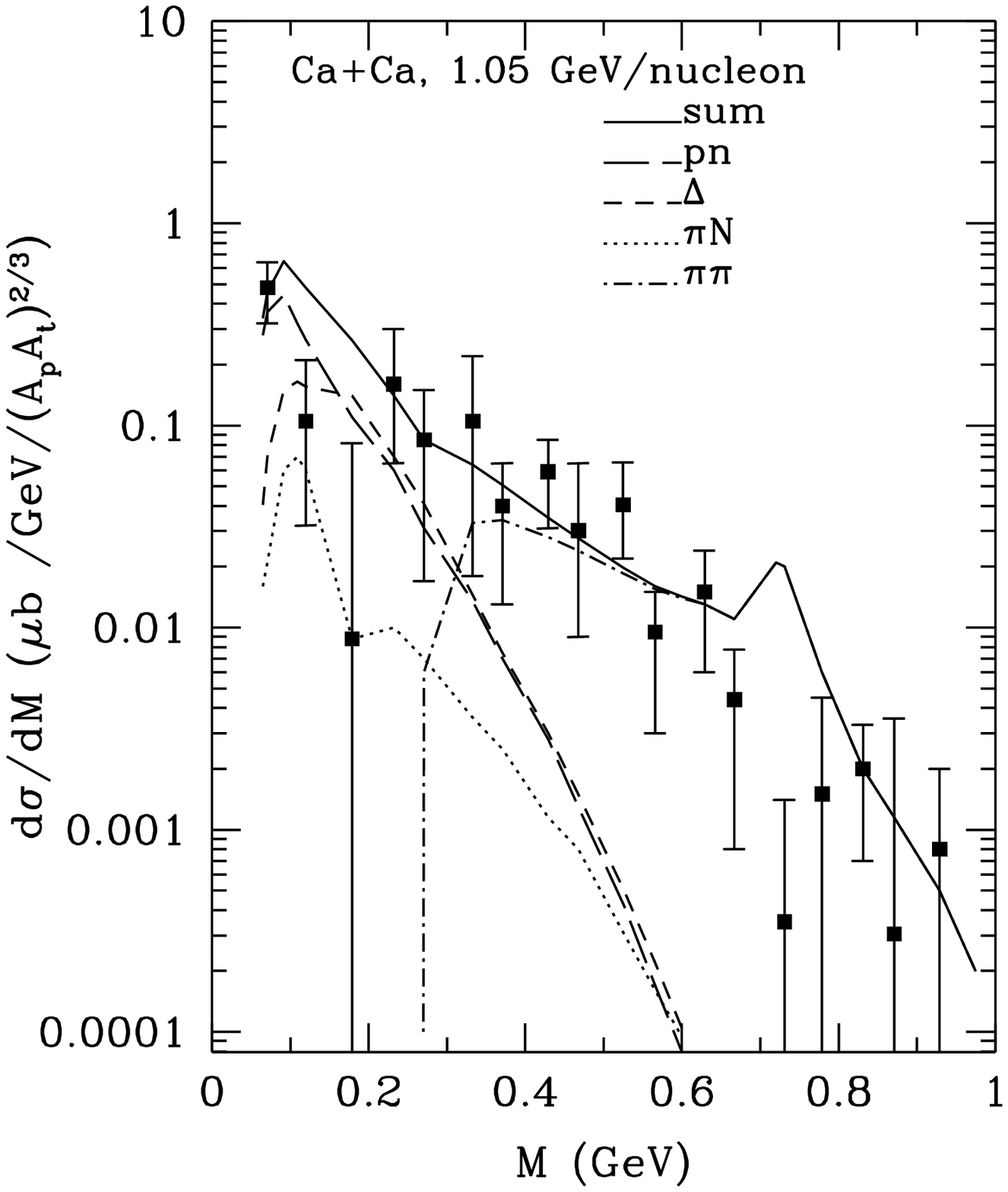,height=4in,width=4in}
\vskip 0.4cm
Fig. 20 ~Dilepton invariant mass spectrum from Ca+Ca collisions 
at 1.05 GeV/nucleon. The experimental data from the DLS collaboration 
\cite{DLS} are given by solid squares while the theoretical total yield 
is given by the thick solid curve. (from Ref. \cite{XIONG90})
\end{figure}

In the following, dilepton production from heavy-ion collisions at 
SIS/GSI, SPS/CERN, and RHIC/BNL is studied in the relativistic 
transport model, and the results are compared with available 
experimental data. 
 
\subsubsection{dileptons from SIS/GSI}
 
Dileptons have been measured from heavy-ion collisions at Bevalac by 
the DLS collaboration \cite{DLS}. The experimental dilepton invariant 
mass spectrum from Ca+Ca collisions at 1.05 GeV/nucleon is shown in
Fig. 20. Based on transport models, perturbative calculations of 
dilepton production, including contributions from proton-neutron 
bremsstrahlung, pion-nucleon interactions, delta and eta decays, and 
pion-pion annihilation, have been carried out in Refs. 
\cite{XIONG90,wolf90,WOLF93}. A typical result from Ref. \cite{XIONG90} 
is shown in Fig. 20. It is seen that dileptons with small invariant 
masses are mainly from proton-neutron bremsstrahlung and delta decays 
while dileptons with large invariant masses are dominated by pion-pion
annihilation. The contribution from eta decays, that was neglected in Ref. 
\cite{XIONG90}, turns out to be very important for invariant masses below
about 500 MeV \cite{WOLF93}. Nevertheless, the eta has a width of only 1.2
keV and decays thus outside of the matter. Its contribution to dilepton
production can in principle be subtracted out if its spectrum is also
measured.  We also see clearly the rho meson contribution to 
dilepton production from pion-pion annihilation as a result of 
rho meson dominance in the pion electromagnetic from factor. 
The data are, however, not accurate enough to extract the in-medium 
rho meson properties.
 
\begin{figure}
\epsfig{file=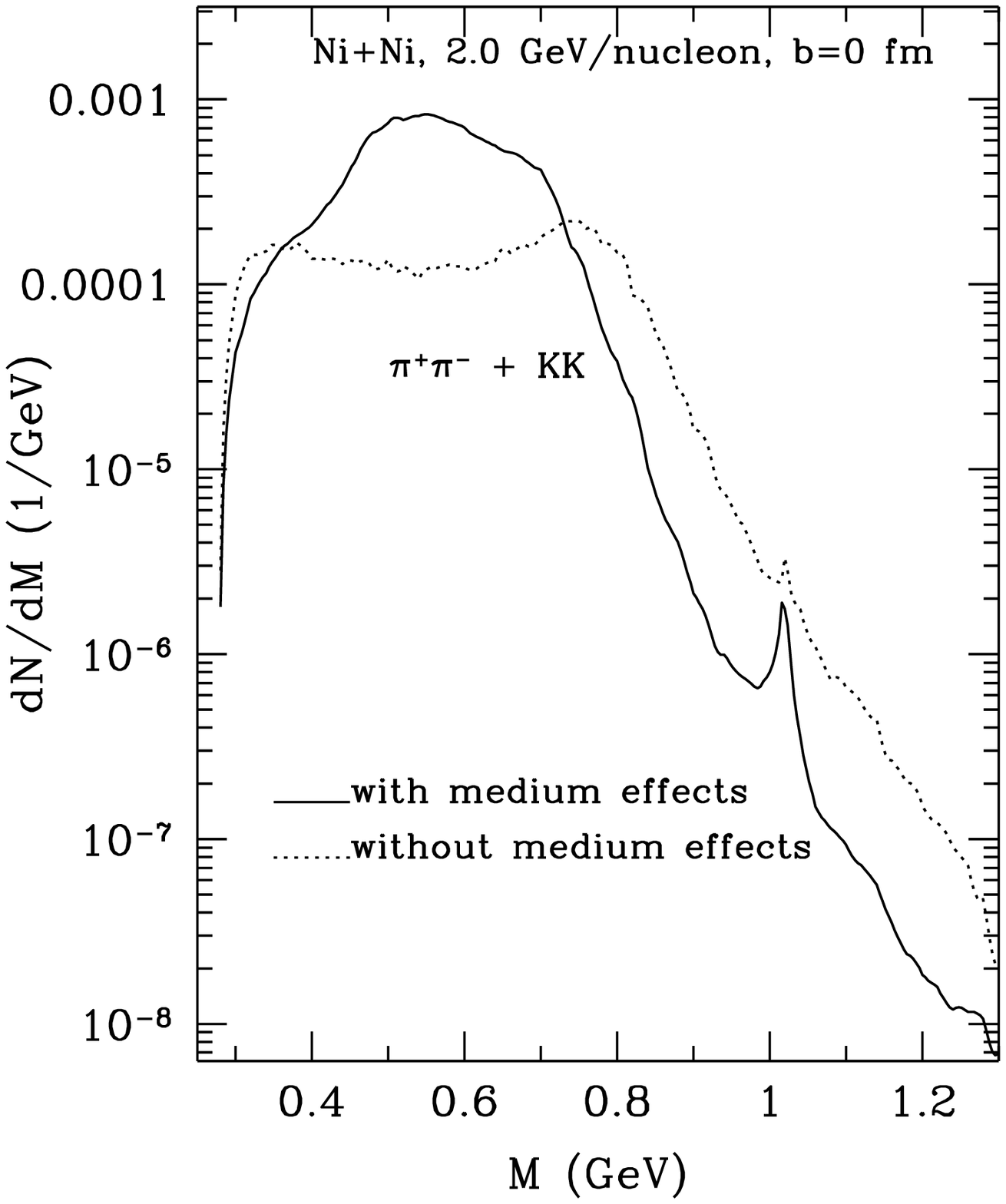,height=4in,width=4in}
\vskip 0.4cm
Fig. 21 ~Dilepton invariant mass spectra from pion-pion and
kaon-antikaon annihilation in Ni+Ni collision at 2 GeV/nucleon and 
impact parameter b=0 fm. The dashed and solid curves are obtained 
using free and in-medium meson masses, respectively. 
(from Ref. \cite{LI95D})
\end{figure}

Future experiments at SIS with the HADES detector \cite{HADES}
will provide data with vastly improved statistics to allow us to study 
more clearly the medium effects on vector meson properties. To see the 
medium effects in these experiments,  the dilepton invariant mass spectra
from Ni+Ni collisions at an incident energy of 2 GeV/nucleon and impact 
parameter of 0 fm with free vector meson masses (dashed curve) and 
in-medium vector meson masses given by Eqs. (\ref{rmass}) and (\ref{pmass})
from the QCD sum rules (solid curve) have been calculated \cite{LI95D} and
are shown in Fig. 21. For pion-pion annihilation, medium effects shift 
the rho meson peak to around 550 MeV, and its height increases by about 
a factor of four.  For kaon-antikaon annihilation, the phi peak shifts 
slightly to a lower invariant mass, and its width is also broadened 
when medium effects are included. In addition, there appears a shoulder 
around 950 MeV. For a complete picture of dilepton production at SIS 
energies, contributions from other channels, such as the Dalitz decay, 
bremsstrahlung, vector mesons produced in initial nucleon-nucleon 
collisions, and vector mesons from decay of high resonances \cite{WINC95}, 
need to be included.
 
For heavy-ion collisions at AGS energies, a similar shift of the rho 
meson strength to lower dilepton invariant masses has been seen in 
RQMD calculations using also the QCD sum-rule results for in-medium 
vector meson masses \cite{HOFF94}. Since a omega meson has a much 
longer lifetime and decays mostly at freeze-out, an interesting 
phenomenon of $\rho$-$\omega$ splitting was seen in the dilepton 
spectra. It would be very useful if dilepton production from heavy-ion 
collisions at AGS energies could be measured.
 
\subsubsection{dileptons from SPS/CERN}
 
For heavy-ion collisions at SPS energies, hot and dense matter is 
also formed in the initial stage of the collisions.  One expects
that medium effects will lead to a shift of the vector meson peaks
in dilepton invariant mass spectra. Experiments from both the CERES
\cite{CERES} and the HELIOS-3 \cite{HELIOS} collaboration have shown
that there is an excess of dileptons over those known and expected
sources which cannot be explained by uncertainties and errors of the
normalization procedures \cite{SUMDI}. In particular, in the CERES 
experiment for central S+Au collisions at 200 GeV/nucleon, a 
significant enhancement of dileptons with invariant masses between 
250 MeV to 1 GeV over that from the proton-nucleus collision has been found.

These experimental data have generated a great deal of interest in the
heavy-ion community. Different models, ranging from schematic estimates
based on a possible enhancement of $\eta$ and $\eta^\prime$ production
\cite{HUANG96,KAPU96}, to hydrodynamical models \cite{GALE96,HAG96},
and relativistic transport models \cite{LI95E,LI95F,CASS95,CASS96,KOCH96},
have been used to study this phenomenon.  Although the contribution from 
pion-pion annihilation, that has not been included in the `cocktail' 
of the CERES collaboration, was found to be important for low-mass 
dileptons, the data in the low mass region are still above the 
theoretical results from these calculations when vector meson
properties in free space are used. This situation is summarized in ig. 22
where the results from Ref. \cite{LI95F} (solid curve) using the 
relativistic transport model with initial conditions determined 
by the RQMD, Ref. \cite{CASS95} (dashed curve) from the Hadron-String 
Dynamics, and Ref. \cite{GALE96} (dotted curve) from the hydrodynamical 
model are shown together with the CERES data. It is seen that in all 
three studies the calculated low mass dileptons are by about a factor 
of 2-4 below the data, and for dileptons around $m_{\rho ,\omega}$ 
they are slightly above the data. It is also of interest to note that 
the three calculations, although with different dynamical models, 
agree more or less with each other, and the differences are less than 
20\%. A strong peak around $m_\phi$ in the results of Ref. \cite{GALE96} 
will become a bump once the mass resolution of the CERES collaboration 
is properly included.
 
\begin{figure}
\epsfig{file=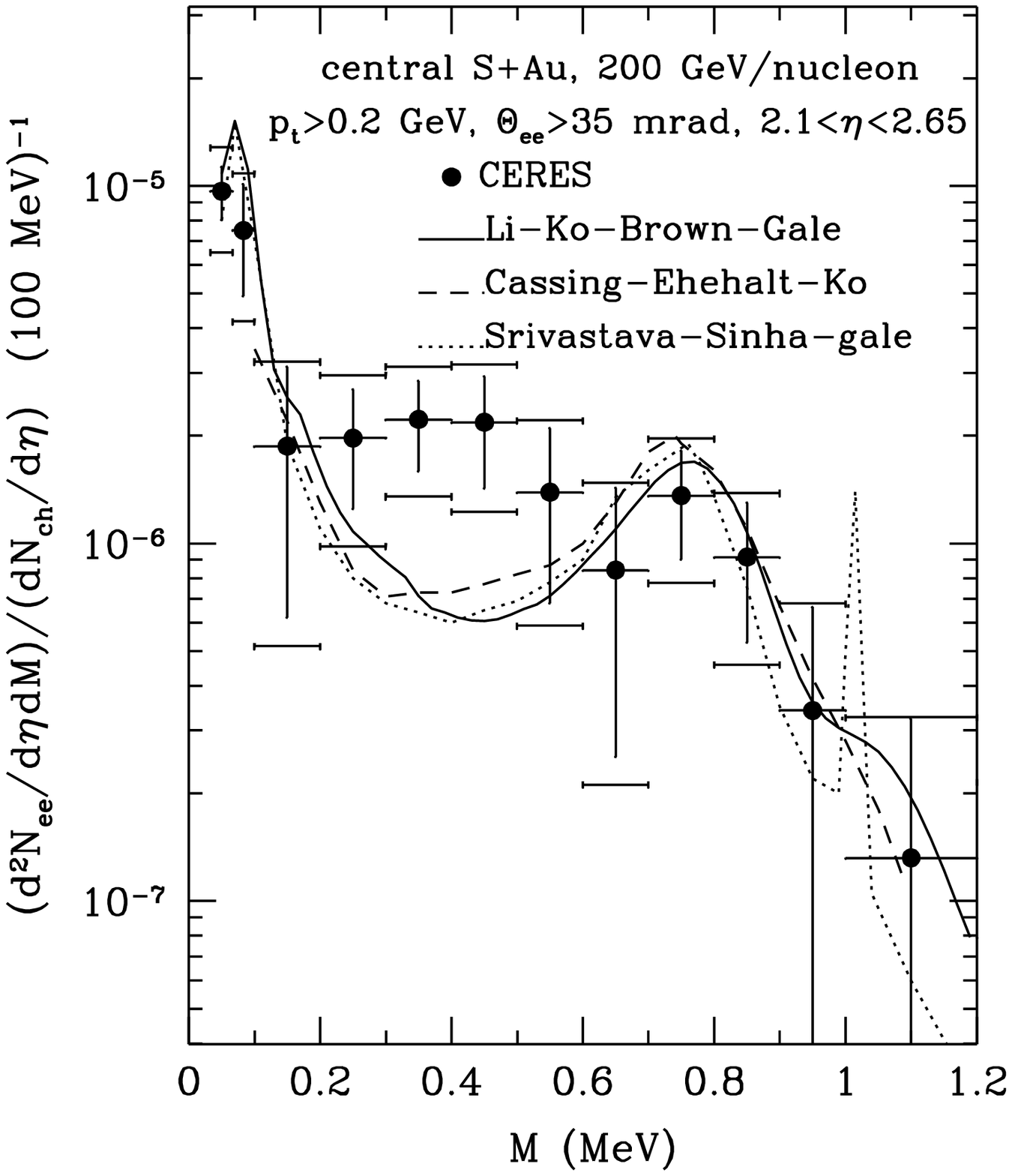,height=4in,width=4in}
\vskip 0.4cm
Fig. 22 ~Dilepton invariant mass spectra in central
S+Au collisions at 200 GeV/nucleon with free meson masses. 
The solid, dashed, and dotted curves are from Refs. \cite{LI95F}, 
\cite{CASS95}, and \cite{GALE96}, respectively. The experimental data
from the CERES collaboration \cite{CERES} are shown by solid circles, 
with bars denoting the statistical errors while the systematic uncertainties
are given by brackets.
\end{figure}
 
The results using in-medium vector meson masses are shown in Fig. 23 
by the solid curve. The agreement with the experimental data is greatly 
improved. The enhancement at low invariant masses is due to both the shift
of primary rho mesons to lower masses and pion-pion annihilation occurring 
in hot dense matter. Since pions have a thermal distribution,
most pion pairs are of low invariant mass. When the rho-meson mass
is reduced, its formation probability from pion-pion annihilation is 
enhanced, thus increasing the production of low-mass dileptons. 
The remaining peak around $m_{\rho ,\omega}$ then comes from the 
decay of omega mesons which have a very small decay width, and therefore 
mostly decay at freeze out when their mass has returned to the free value.  
One notes that dileptons with invariant masses below about 300 MeV are 
mainly from the Dalitz decay of $\pi^0$, $\omega$, and $\eta$.

Dimuon invariant mass spectra have been measured by the HELIOS-3 
collaboration in S+W collisions at 200 GeV/nucleon \cite{HELIOS}. 
The data also show an enhancement of dileptons around $M\approx 0.4-0.6$ 
GeV. This provides another possible indication that the vector (rho) 
meson mass might be reduced in hot dense medium. The comparison of 
theoretical results obtained from the relativistic transport model
with the HELIOS-3 data is shown in Fig. 24 \cite{LI95F}.  With free 
vector meson masses, the theoretical results are below the HELIOS data 
in the mass region from 0.35 to 0.6 GeV by about a factor of two, and 
slightly above the data around $m_{\rho ,\omega}$. Using in-medium vector 
masses, we again see an enhanced dilepton production in the low mass 
region and a reduction around $m_{\rho ,\omega}$. This brings the 
theoretical results in better agreement with the data.  Similar results 
have been obtained in Ref. \cite{CASS96} using the Hadron-String Dynamics.
 
\begin{figure}
\epsfig{file=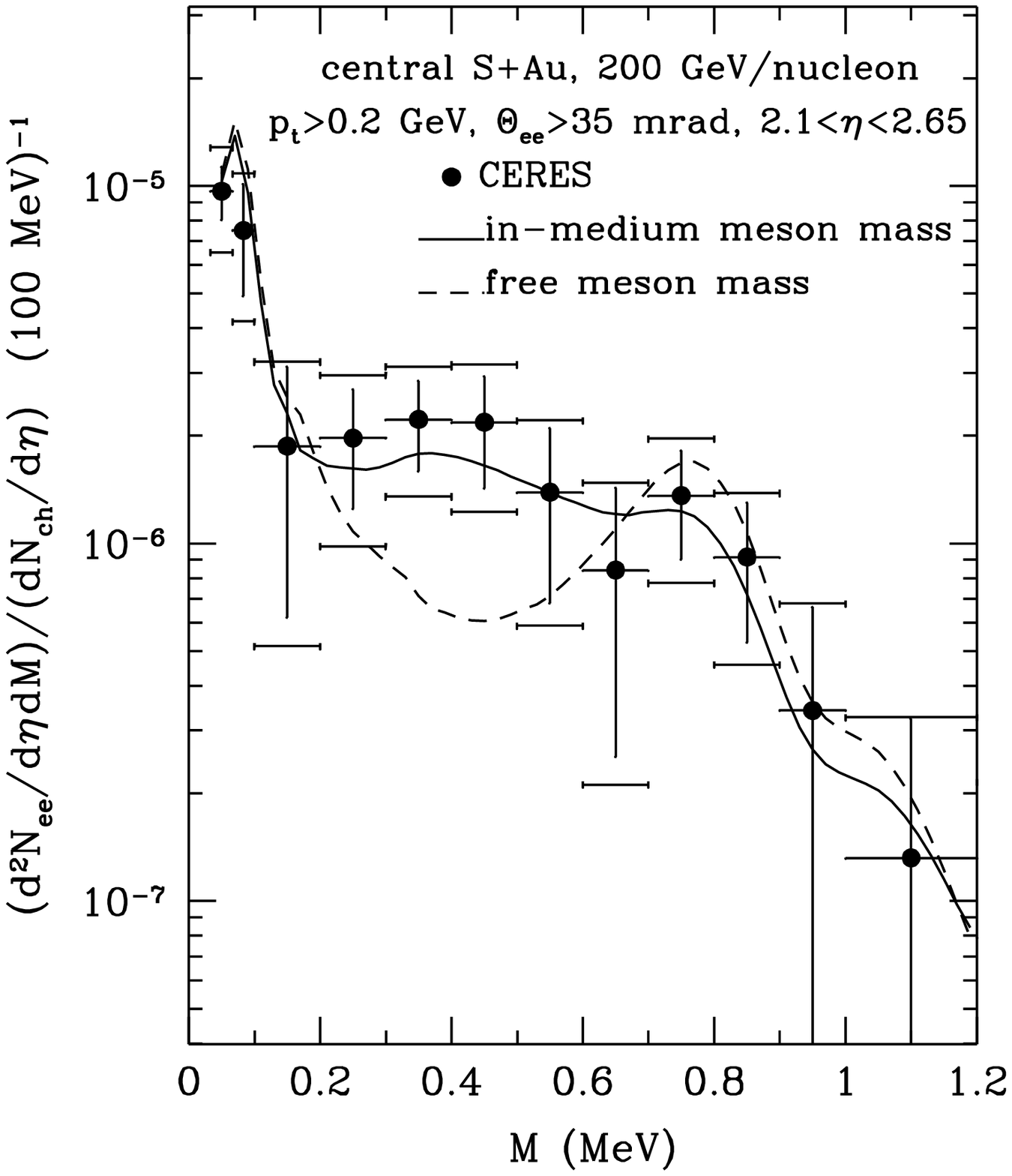,height=4in,width=4in}
\vskip 0.4cm
Fig. 23 ~Same as Fig. 22.  The dotted and solid curves are obtained 
using free and in-medium meson masses, respectively. (from Ref. \cite{LI95F})
\end{figure}
 
The medium effects seen in the HELIOS-3 case are smaller than that in 
the CERES case. This is due to the fact that in the HELIOS-3 experiment 
dileptons are measured at forward rapidity ($3.7<\eta <5.5$) where the 
charged-particle multiplicity is low, while in the CERES experiment 
they are measured in the midrapidity ($2.1<\eta <2.65$) with a higher 
charged-particle multiplicity. It is also of interest to note that the 
theoretical results from both Ref. \cite{LI95F} and Ref. \cite{CASS96} 
are well below the HELIOS-3 data above 1.2 GeV.  This may indicate the 
importance of processes such as $\pi a_1\rightarrow l^+l^-$ 
\cite{XIONG94,SONG94} and the decay of heavier vector mesons such as 
$\omega (1390)$ \cite{GALE94,HAG95}.  Also, in this mass region, the 
contribution from the quark-gluon plasma and the initial Drell-Yan 
processes might become important.

\subsubsection{dileptons from RHIC/BNL}
 
In heavy-ion collisions at the Relativistic Heavy-Ion Collider (RHIC)
being constructed at the Brookhaven National Laboratory, dileptons will
be measured as they are likely to carry the signature for the
quark-gluon plasma expected to be created in the collisions \cite
{SHUR78,CHIN82,KAJA86,ASAK93}. However, there are a number of different
sources of dileptons in heavy-ion collisions at these energies.
For dileptons with low invariant mass, i.e., below about 1 GeV, 
they are dominated by hadronic processes, such as $\pi^0$
and $\eta$ Dalitz decays as well as vector meson ($\rho$, $\omega$,
and $\phi$) direct decays. For intermediate-mass dileptons, i.e.,
between 1 and 2 GeV, dilepton production from heavy meson resonances
\cite{SONG94,GALE94} and charmed meson ($D$ and $D^*$) decays become 
important. It is in this invariant mass regions that it may be possible
to see dileptons emitted from the quark-gluon plasma via quark-antiquark
annihilation if its temperature is sufficient high \cite{xiong}. Above 
2 GeV, dileptons are from Drell-Yan process involving the annihilation 
of quarks and antiquarks in the initial colliding nuclei. In this 
region of dilepton invariant mass, an interesting proposal to see the 
signature of the quark-gluon plasma is the $J/\Psi$ suppression 
\cite{matsui} due to its dissociation in the quark-gluon plasma as a 
result of color screening. Current experiments at CERN/SPS have indeed 
shown that the number of $J/\Psi$ observed versus the Drell-Yan 
background is reduced by about a factor of 2 in heavy-ion collisions 
than in proton-nucleus reactions \cite{na38}. Unfortunately, the data 
can also be explained by hadronic absorption models \cite{gers,vogt,gavin}.

\begin{figure}
\epsfig{file=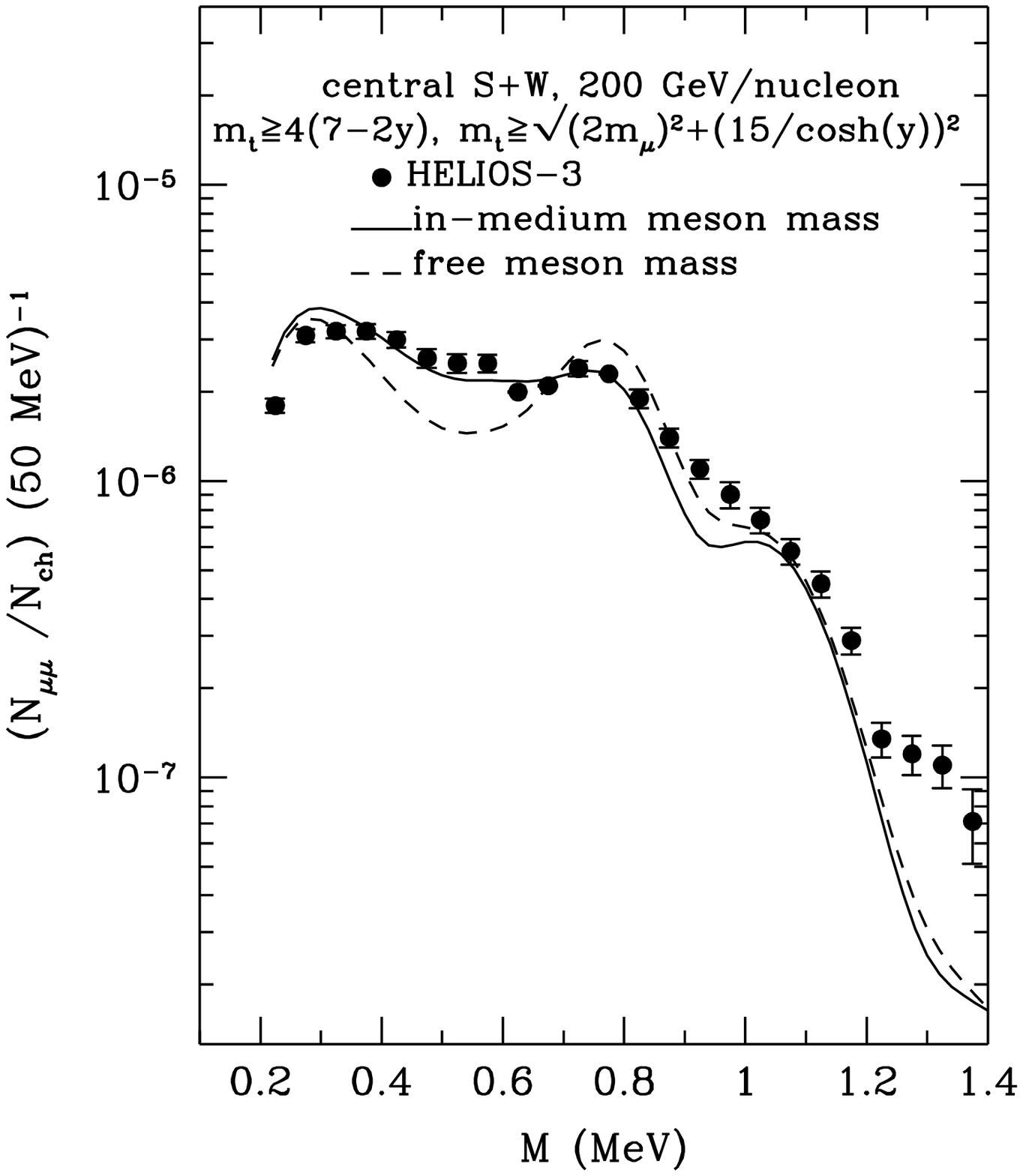,height=4in,width=4in}
\vskip 0.4cm
Fig. 24 ~Dimuon invariant-mass spectrum from pion-pion annihilation
in central S+W collisions at 200 GeV/nucleon. The dotted and solid curves
are obtained with free and in-medium masses, respectively. The experimental 
data from the HELIOS-3 collaboration \cite{HELIOS} are shown by solid circles.  
(from Ref. \cite{LI95F})
\end{figure}
 
On the other hand, the change of phi meson mass in hot matter as 
discussed in Sect. II.C has led to a new signature for identifying 
the quark-gluon plasma to hadronic matter phase transition in 
ultrarelativistic heavy-ion collisions \cite{asko94}.  In a boost 
invariant hydrodynamical calculation with transverse flow and including 
temperature-dependent vector meson masses, it has been found that 
a low mass phi peak at $\sim 880$ MeV besides the normal one at 1,020 MeV
appears in the dilepton spectrum if a first-order phase transition or 
a slow cross-over between the quark-gluon plasma and the hadronic matter 
occurs in the collisions.  Assuming an initial temperature $T_i=250$ MeV,
a critical temperature $T_c=180$ MeV, and a freeze out temperature 
$T_f=120$ MeV, the results are shown by the solid curve in Fig. 25. The
three peaks correspond to the omega, the low-mass phi, and the normal
phi meson, respectively. Because of their reduced mass at finite
temperature according to the QCD sum-rule studies \cite{furns}, 
decays from rho mesons lead to a broad distribution of dileptons
at low masses.
 
The low-mass phi peak is due to the nonnegligible duration time for 
the system to stay near the transition temperature (about 10 fm/c) 
compared with the lifetime of a phi meson in the vacuum ($\sim 45$ fm), 
so the contribution to dileptons from phi meson decays in the mixed 
phase becomes comparable to that from phi meson decays at freeze out.  
Also, its width from collisions with other hadrons remains small 
(about 10 MeV) \cite{kose94,haglin}.
 
Without the formation of the quark-gluon plasma, the low mass phi peak 
is reduced to a shoulder in the dilepton spectrum as shown by the dotted 
curve in Fig. 23, which is obtained by assuming that the initial
temperature of the hadronic matter is just below $T_c$.
 
\begin{figure}
\epsfig{file=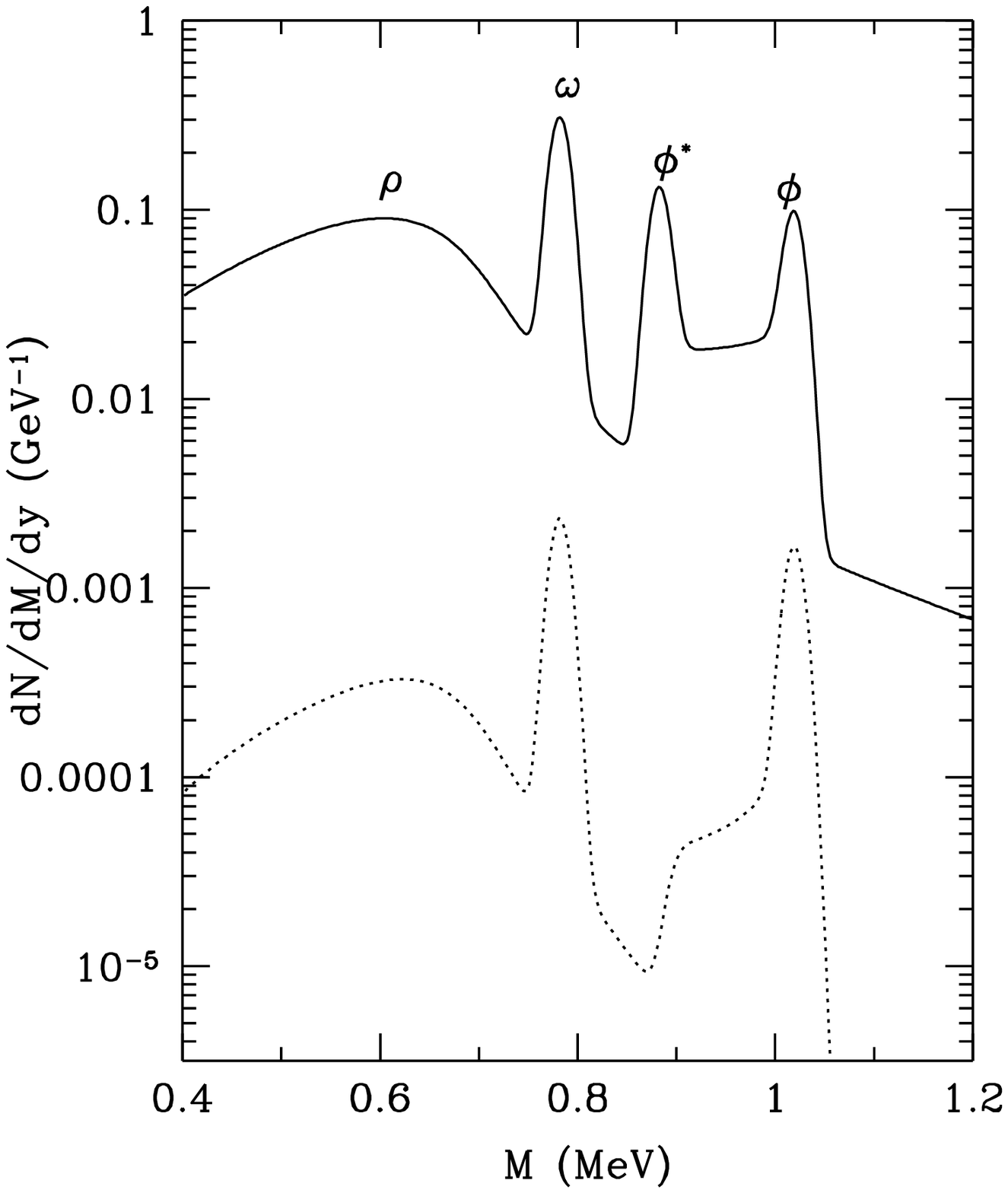,height=4in,width=4in}
\vskip 0.4cm
Fig. 25 ~Dilepton invariant mass spectra from ultrarelativistic
heavy-ion collisions.  Solid and dotted curves correspond to scenarios 
with and without the formation of a quark-gluon plasma, respectively.
(from Ref. \cite{asko94})
\end{figure} 

Since the transverse expansion velocity during the mixed phase is
relatively small, the transverse momentum distribution of the low mass 
phi mesons is largely determined by the temperature of the mixed matter. 
The low mass phi mesons thus also provide information about the 
temperature at which the quark-gluon plasma to hadronic matter 
transition occurs.
 
\section{summary and outlook}
 
Because of the partial restoration of chiral symmetry, the quark 
condensate decreases in hot dense matter. According to the prediction of 
QCD sum rules, hadron masses are then reduced in medium.   As a result,
the thresholds for particle production in medium are reduced, and 
the cross sections for their production are enhanced due to
the increase of phase space in the final states. The dense matter also 
gives rise to strong vector mean-field potentials. With these medium 
effects consistently included in the relativistic transport model, it 
has been shown that for heavy-ion collisions at SIS energies they can 
account for the observed enhancement of kaon, antikaon, and antiproton 
yields, the enhanced soft pions in the transverse direction, and 
particle flows in these collisions.

To place these conclusions on a firmer ground, a number of issues need 
to be addressed in future investigations. In heavy-ion collisions at 
SIS energies, the colliding system consists mainly of nucleons, baryon 
resonances, and pions. Particles can thus be produced from baryon-baryon, 
baryon-meson, and meson-meson collisions. Because of the small 
center-of-mass energy of the pion and the fact that pions materialize 
in the expansion stage of the collisions, hadron production from 
pion-pion interactions can be safely neglected at SIS energies. Kaon 
and antikaon production from pion-baryon collisions have been calculated 
and found to account for about 20\% of their total yield. With improved 
experimental data, these contributions need to be more carefully studied.
Similarly, antiproton production from pion-baryon collisions has not 
been considered and certainly need to be evaluated.
 
For subthreshold particle production, the most important contribution
comes from baryon-baryon interactions, of which interactions involving
baryon resonances are particularly important. One needs thus particle 
production cross sections in baryon-baryon collisions. For proton-proton 
collisions there exist empirical data for kaon, antikaon, and antiproton 
production, mostly at high beam energies \cite{DATA}. The general strategy 
in the transport model is to introduce a parameterization for these data 
and then to extrapolate it to the threshold. The production cross section
near the threshold then depends on the way the parameterization 
and extrapolation are introduced. This is particularly serious 
for antiprotons since there are essentially no experimental data below 
20 GeV. Moreover, no experimental data exist for particle production 
from baryon-baryon collisions involving baryon resonances. Various 
assumptions have been made to relate these cross sections to that
in proton-proton collisions. The validity of these assumptions certainly 
need to be carefully assessed. Also, the role of baryon resonances that 
are higher than the delta needs to be further investigated \cite{SPIE94,BALIN}.
 
For antikaon and antiproton, there is an additional complication
from their large annihilation cross sections. This raises the question
of numerically how to accurately treat the annihilation process in 
transport models. Indeed, different treatment of antiproton annihilation
could lead to almost an order of magnitude difference in the final 
antiproton survival probability \cite{LI94C,antip,MOS94,FAE94,SPIE93}. 
Moreover, the antikaon and antiproton annihilation cross sections
might be modified in medium due to changes in their masses, the screening 
effects from mesons \cite{ARC}, and the effect of Bose-Einstein 
enhancement in the final state \cite{welke}. For antiproton annihilation, 
which leads to about five pions, the latter effect is expected to be 
important in heavy-ion collisions at AGS and SPS energies as the 
abundance of pions is large in these collisions.
 
The medium modification of hadron masses has so far been included in 
transport models only in a minimum way by changing the threshold of 
particle production. In terms of Feynman diagrams, this amounts to change 
only the masses of the initial and final hadrons. In principle, the 
properties of exchanged particles can also be modified \cite{KO91,bbkl}. 
Besides changes of hadron masses, the coupling constants and the form 
factors in the vertices of Feynman diagrams might be modified in medium 
as well.  All these need to be consistently studied.

Finally, off-shell propagation of particles has been neglected in all
transport models. This effect on subthreshold particle production
has been shown to be small \cite{pawelb}. However, off-shell effects
due to correlations in the initial wave function may be important and
need to be further studied.
 
For heavy-ion collisions at AGS energies, medium effects can also 
explain the enhanced $K^+/\pi^+$ ratio, the difference between the 
slope parameters of $K^+$ and $K^-$ transverse kinetic energy spectra, 
the lower apparent temperature of antiprotons than that of protons,
and the cool kaons with low transverse masses.  However, these studies 
have been based on a fireball initial conditions so the initial 
nonequilibrium dynamics is not included. There are already models, 
such as the RQMD \cite{RQMD}, ARC \cite{ARC}, and ART \cite{BALI}, which
can treat the initial stage of heavy-ion collisions at these energies 
as enough inelasticity has been introduced in these models. It will be
useful to extend the relativistic transport model by including these
additional processes so both the collision dynamics and medium
effects can be properly included.

Medium effects are also relevant for heavy-ion collisions at higher 
energies from the SPS at CERN. The reduced particle production thresholds 
in medium have been found to lead to enhanced production of phi mesons, 
antilambdas, and low-mass dileptons. As in the case of heavy-ion collisions 
at AGS energies, the relativistic transport model at its present form can 
only describe the expansion stage of these collisions. In the initial 
nonequilibrium stage, particle production from string fragmentation has 
shown to be essential \cite{RQMD}. It is thus again important to extend 
the relativistic transport model to include consistently in the initial 
stage the string dynamics. Such a step has already being taken by Cassing 
{\it et al.} \cite{HSD}.

We have shown in this review that high energy heavy-ion experiments
offer the possibility to study the properties of hadrons in the
hot dense matter formed in the initial stage of the collisions. This 
study is not only of interest in its own right but also important for 
future heavy-ion experiments at ultrarelativistic energies where the 
quark-gluon plasma is expected to be formed. To find the signatures of 
the quark-gluon plasma, it is essential to have a good understanding of 
the hadronic matter that exists both in the initial and final stages 
of heavy-ion collisions.  Moreover, the modification of hadron properties 
in hot dense matter is expected to give rise to new signals for the 
quark-gluon to hadronic matter transition.  For example, a low mass phi 
peak besides the normal one has been predicted to appear in the dilepton 
invariant mass spectra from ultrarelativistic heavy-ion collisions as a 
result of the decay of phi mesons with reduced in-medium mass during 
the transition.  Furthermore, measurements of the transverse momentum 
distribution of these low mass phi mesons offer a viable means for 
determining the temperature of the transition.

However, experience from heavy-ion collisions at SIS, AGS, and CERN 
energies has shown that to extract useful physics from heavy-ion collisions 
requires transport models which can describe properly the entire collision 
dynamics. The same is expected for heavy-ion collisions at RHIC, so it 
is important to have a model which includes the initial parton cascade 
\cite{geiger}, the evolution of the resulting quark-gluon matter 
\cite{heinz}, the dynamics of hadronization \cite{kapusta}, and the 
final hadronic dynamics. In this respect, the relativistic transport model 
developed extensively for low energy heavy-ion collisions provides a
link of future developments in ultra-relativistic transport model to the
experimental observations.

\bigskip
 
\acknowledgements
 
We are grateful to J. Aichelin, M. Asakawa, G. E. Brown, W. Cassing, 
X. S. Fang, V. Koch, B. A. Li, R. Machleidt, U. Mosel, C. Song, H. Sorge,
and L. Xiong for discussions and/or collaboration. We also thank our 
experimental colleagues, D. Best, P. Braun-Munzinger, A. Dress, 
E. Grosse, I. Kralik, M. Murray, J. Ritman, P. Senger, I. Tserruya, 
T. Ullrich, K. Wolf, and P. Wurm for helpful discussions and/or 
communications. This work was supported in part by the National 
Science Foundation under Grant No. PHY-9509266. C.M.K. also 
acknowledges the support by the Alexander von Humboldt Foundation.

\end{document}